%% file: BH-NegW-Holism_pdf.tex
\documentclass[english,mathscr,psamsfonts,cmex10]{article}
\usepackage{newcent}
\usepackage{avant}
\usepackage{beramono}
\usepackage[T1]{fontenc}
\usepackage[latin9]{inputenc}
\usepackage[a4paper]{geometry}
\usepackage{color}
\usepackage{pifont}
\usepackage{textcomp}
\usepackage{amsmath}
\usepackage{setspace}
\usepackage{amssymb}
\usepackage{esint}
\usepackage[numbers]{natbib}

\makeatletter

\newcommand{\lyxmathsym}[1]{\ifmmode\begingroup\def\b@ld{bold}
  \text{\ifx\math@version\b@ld\bfseries\fi#1}\endgroup\else#1\fi}

\providecommand{\tabularnewline}{\\}

\newcommand{\lyxaddress}[1]{
\par {\raggedright #1
\vspace{1.4em}
\noindent\par}
}

\usepackage{sublabel} 
\usepackage{hyperref}
\usepackage{amssymb}
\usepackage{setspace} 
\usepackage{amsfonts} 
\usepackage{latexsym} 
\usepackage{graphicx} 
\usepackage{multicol} 
\usepackage{epstopdf}
\usepackage{multirow} 
\usepackage{empheq}
\usepackage{flexisym}
\usepackage{mathtools}
\usepackage{mhsetup}
\usepackage{breqn}
\usepackage{mathstyle}
\usepackage{color}
\usepackage{xcolor}
\usepackage{colortbl}
\usepackage{subfig}
\usepackage{subfloat}
\usepackage{fixltx2e}
\usepackage{fancybox} 
\usepackage{euscript} 
\usepackage{exscale} 
\usepackage{pifont} 
\usepackage{array}
\usepackage{afterpage}
\usepackage{bm}
\usepackage{makeidx}
\makeindex
\usepackage{charter}
\usepackage{everysel}
\usepackage{keyval}
\usepackage{ragged2e}
\usepackage{nonfloat}
\usepackage{scalefnt}
\usepackage{ifpdf}
\usepackage{arydshln}
\date{}
\definecolor{Red}{rgb}{1,0.4,0.4}
\definecolor{Green}{rgb}{0.3,1,0.3}

\AtBeginDocument{

}

\makeatother

\usepackage{babel}

\begin{document}

\title{\textsf{\textbf{Gravitational Collapse, Negative World and Complex
Holism}}%
\thanks{\textsf{\textbf{To appear in}}\textsf{\textbf{\emph{ Nonlinear Analysis
B: Real World Applications. }}}\textsf{\textbf{Received : April 22,
2010, Accepted: May 04, 2010}}%
}}

\author{\noindent A. Sengupta%
\thanks{Present address: Institute for Complex Holism, Kolkata%
}}

\maketitle

\lyxaddress{\noindent \begin{center}
Department of Mechanical Engineering\\Indian Institute of Technology
Kanpur, India.\\E-Mail: osegu@iitk.ac.in
\par\end{center}}
\begin{abstract}
Building on the engine-pump paradigm of ChaNoXity, this paper argues
that complex holism --- as the competitive homeostasis of dispersion
and concentration --- is the operating mode of Nature. Specifically,
we show that the negative world $\mathfrak{W}$ is a gravitationally
collapsed black hole that was formed at big-bang time $t=0$ as the
pair $(W,\mathfrak{W})$, with $W$ a real world, and gravity the
unique expression of the maximal multifunctional nonlinearity of the
negative world $\mathfrak{W}$ in the functional reality of $W$.
The temperature of a gravitationally collapsed system does enjoy the
relationship $T\propto1/r$ with its radius, but the entropy follows
the usual volumetric alignment with microstates, reducing to the surface
approximation only at small $r$. It is not clear if quantum non-locality
is merely a linear manifestation of complex holism, with the interaction
of quantum gates in quantum entanglements resulting in distinctive
features from the self-evolved structures of complex holism remaining
an open question for further investigation.
\end{abstract}

\section{\textsf{\large Introduction}}

In a recent two-part discourse \citep{Sengupta2010a}, a rigorous,
scientific, self-contained, and unified formulation of complex holism
has been developed. Science of the last 400 years has essentially
evolved by the reductionist tools of linear mathematics in which a
composite whole is regarded as the sum of its component parts. Increasingly
however, a realization has grown that most of the important manifestations
of nature in such diverse fields as ecology, biology, social, economic
and the management sciences, beside physics and cosmology, display
a holistic behaviour which, simply put, is the philosophy that parts
of any whole cannot exist and be understood except in their relation
to the whole: the system as a whole determines in an important way
how the parts behave. These complex self-organizing systems evolve
on emergent feedback mechanisms and processes that \textquotedblleft{}interact
with themselves and produce themselves from themselves'': they are
\textquotedblleft{}more than the sum of their parts\textquotedblright{}.
Thus society is more than a collection of individuals, life is more
than a mere conglomeration of organs as much as human interactions
are rarely dispassionate. 

Living organisms require both energy and matter to continue living,
are composed of at least one cell, are homeostatic, and evolve; life
organizes matter into increasingly complex forms in apparent violation
of the Second Law of Thermodynamics that forbids order in favour of
discord, instability and lawlessness; infact {}``a living organism
continually increases its entropy and thus tends to approach the dangerous
state of maximum entropy, which is \emph{death}''. However, {}``It
can only keep aloof from it, i.e. stay \emph{alive}, by continually
drawing from its environment \emph{{}``negative entropy''}. It thus
maintains itself stationery at a fairly high level of orderliness
(= fairly low level of entropy) (by) continually sucking orderliness
from its environment'' \citep{Schroedinger1992}. Holism entails
{}``life (to be) a far-from-equilibrium dissipative structure that
maintains its local level of self organization at the cost of increasing
the entropy of the larger global system in which the structure is
imbedded'' \citep{Schneider1994a}, {}``a living individual is defined
within the cybernetic paradigm as a system of inferior negative feedbacks
subordinated to (being at the service of) a superior positive feedback''
\citep{Korzeniewski2001}, {}``life is a balance between the imperatives
of survival and energy degradation'' \citep{Cinquin2002}, and {}``life
is a special complex system of activating mind and restraining body''
\citep{Sengupta2006} identifiable respectively by an anti-thermodynamic
backward and thermodynamic forward arrows.

The linear reductionist nature of present mainstream science raises
many deep-rooted and fundamental questions that apparently defy logical
interpretation within its own framework; as do questions involving
socio-economic, collective (as opposed to individualistic), and biological
relations. The issues raised by this dichotomy have been well known
and appreciated for long enough, leading often to bitter and acrimonious
debate between protagonists of the reductionist and holistic camps:
ChaNoXity \citep{Sengupta2010a} aims at integrating Chaos-Nonlinearity-compleXity
into the unified structure of holism that has been able to shed fresh
insight to these complex manifestations of Nature. The characteristic
features of holism are self-organization and emergence: Self-organization
involves the internal organization of an open system to increase from
numerous nonlinear interactions among the lower-level hierarchical
components \emph{without being guided or managed from outside. }The
rules specifying interactions among the system's components are executed
using only local information, without reference to the global pattern.
\emph{Self-organization} relies on three basic ingredients: positive-negative
feedbacks, exploitation-exploration\emph{,} and multiple interactions.
In \emph{emergence}, global-level coherent structures, patterns and
properties arise from nonlinearly interacting local-level processes.
The structures and patterns cannot be understood or predicted from
the behavior or properties of the components alone: the global patterns
cannot be reduced to individual behaviour. Emergence involves multi-level
systems that interact at both higher and lower level; these emergent
systems in turn exert both upward and downward causal influences. 

Complexity results from the interaction between parts of a system
such that it manifests properties not carried by, or dictated by,
individual components. Thus complexity resides in the interactive
competitive collaboration\emph{}%
\footnote{Competitive collaboration --- as opposed to reductionism --- in the
context of this characterization is to be understood as follows:\emph{
The interde}p\emph{endent parts retain their individual identities,
with each contributing to the whole in its own characteristic fashion
within a framework of dynamically emerging global properties of the
whole}. Although the properties of the whole are generated by the
parts, the individual units acting independently on their own cannot
account for the global behaviour of the total.%
}\emph{ }between the parts; the properties of a system with complexity
are said to {}``emerge, without any guiding hand''. A complex system
is an assembly of many interdependent\emph{ }parts, interacting with
each other through competitive\emph{ }nonlinear\emph{ }collaboration\emph{,
}leading to self\emph{-}organized, emergent holistic behaviour.

What is chaos? Chaos theory describes the behavior of dynamical systems
--- systems whose states evolve with time --- that are highly sensitive
to initial conditions. This sensitivity, expressing itself as an exponential
growth of perturbations in initial conditions, render the evolution
of a chaotic system appear to be random, although these are fully
deterministic systems with no random elements involved. Chaos responsible
for complexity \citep{Sengupta2003} is the eventual outcome of \emph{non-reversible}
iterations of one-dimensional \emph{non-injective} maps; noninjectivity
leads to irreversible nonlinearity and one-dimensionality constrains
the dynamics to evolve with the minimum spatial latitude thereby inducing
emergence of new features as required by complexity. In this sense
chaos is the maximal ill-posed irreversibility of the maximal degeneracy
of multifunctions; features that cannot appear through differential
equations. The mathematics involve topological methods of convergence
of nets and filters%
\footnote{These are generalizations of the usual concept of sequences and, in
what follows, may be read as such. %
} with the multifunctional graphically converged adherent sets effectively
enlarging the functional space in the outward manifestation of Nature.
Chaos therefore is more than just an issue of whether or not it is
possible to make accurate long-term predictions of the system: chaotic
systems are necessarily sensitive to initial conditions and topologically
mixing with dense periodic orbits; this, however, is not sufficient,
and maximal ill-posedness of solutions is a prerequisite for the evolution
of complex structures. 

ChaNoXity involves a new perspective of the dynamical evolution of
Nature based on the irreversible multifunctional multiplicities generated
by the equivalence classes from iteration of noninvertible maps, eventually
leading to chaos of maximal ill-posedness, \citep{Sengupta2003}.
The iterative evolution of difference equations is in sharp contrast
to the smoothness, continuity, and reversible development of differential
equations which cannot lead to the degenerate irreversibility inherent
in the equivalence classes of ill-posed systems. Unlike evolution
of differential equations, difference equations update their progress
at each instant with reference to its immediate predecessor, thereby
satisfying the crucial requirement of adaptability and experience
based learning that constitutes the distinctive feature of complex
systems. Rather than the smooth continuity of differential equations,
Nature takes advantage of jumps, discontinuities, and singularities
to choose from the vast multiplicity of possibilities that rejection
of such regularizing constraints entail. Non-locality and holism,
the natural consequences of this paradigm, are to be compared with
the reductionist determinism of classical Newtonian reversibility
suggesting striking formal correspondence with superpositions, qubits
and entanglement of quantum theory \citep{Sengupta2010a}. Complex
holism is to be understood as complementing mainstream simple reductionism
--- linear science has after all stood the test of the last 400 years
as quantum mechanics is acknowledgedly one of the most successful
yet possibly among the most mysterious of scientific theories. Its
success lies in the capacity to classify and predict the physical
world --- the mystery in what this physical world must be like for
it to be as it is supposed to be.

\section{\textsf{\large ChaNoXity: The New Science of Complex Holism}}

The mathematical structure of ChaNoXity is based on the discrete evolution
of difference equations rather than on the smooth and continuous unfolding
of differential equations. The fundamental goal of chanoxity is to
suggest, justify and institute the existence of an anti-thermodynamic
arrow that allows open systems the privilege of metaphorically {}``sucking
orderliness from the environment'' and thereby survive in the highly
improbable state of being {}``alive''. For an exhaustive account
of the very brief overview recounted below, reference should be made
to \citep{Sengupta2003,Sengupta2010a}.

\subsection{\textsf{Mathematics of ChaNoXity: Nonlinearity, Multiplicity, Non-smoothness}\textmd{\normalsize{}
\citep{Dugundji1966,Eisenberg1971,Munkres1992,Murdeshwar1990}}}

\textsf{\textbf{(A) Topologies.}} (i) If $\sim$ is an equivalence
relation on a set $X$, the class of all saturated sets $[x]_{\sim}=\{y\in X\!:y\sim x\}$
is a topology on $X;$ this \emph{topology of saturated sets} constitutes
the defining topology of chaotic systems.\emph{ }In this topology,
the neighbourhood system at $x$ consists of all supersets of the
equivalence class $[x]_{\sim}$. (ii) For any subset $A$ of the set
$X$, the $A$-\emph{inclusion topology on $X$} comprises $\emptyset$
and every superset of $A$, while the $A$-\emph{exclusion topology
on} $X$ are all subsets of $X-A$. Thus $A$ is open in the inclusion
topology and closed in the exclusion, and in general \emph{every open
set of one is closed in the other. }For a $x\in X$, the $x$-\emph{inclusion}
neighbourhood $\mathcal{N}_{x}$ consists of all non-empty open sets
of $X$ which are the supersets of $\{x\}$, while for a point $y\neq x$,
$\mathcal{N}_{y}$ are the supersets of $\{x,y\}$. In the $x$-\emph{exclusion}
topology, $\mathcal{N}_{x}$ are the non-empty open subsets of $\mathcal{{P}}(X-\{x\})$
that \emph{exclude} $x$. 

\textsl{The possibility of generating different topologies on a set
is of great practical significance in emergent, self-organizing systems
because open sets define convergence properties of nets and continuity
characteristics of functions that nature can play around with to its
best possible advantage.} The topologies introduced above play a key
role in this pogramme.

\textsf{\textbf{\textsl{\small Initial-and-Final Topology}}}\textsf{\textbf{.}}
The topological theory of convergence of nets and filters in terms
of residual and cofinal subsets is fundamental in the development
of this formalism, one of the goals being to understand the Second
Law \emph{{}``dead'' state of maximum entropy}. We consider this
problem as a manifestation of the change of the topologies induced
by a non-biective map $f\!:(X,\mathcal{U})\rightarrow(Y,\mathcal{V})$
to a state of \emph{ininality} of \emph{initial} and \emph{final}
topologies \citep{Murdeshwar1990} of $X$ and $Y$ respectively.
For a continuous $f$ there may be open sets in $X$ that are not
inverse images of open sets of $Y$, just as it is possible for non-open
subsets of $Y$ to contribute to $\mathcal{U}$. When the triple $\{\mathcal{U},f,\mathcal{V}\}$
is tuned in a manner such that neither is possible, the topologies
so generated are the \emph{initial} (smallest/coarsest) and \emph{final}
(largest/finest) topologies on $X$ and $Y$ for which $f\!:X\rightarrow Y$
is continuous. 

For $e\!:X\rightarrow(Y,\mathcal{V})$, the \emph{preimage or} \emph{initial
topology of} $X$ \emph{generated by} \emph{$e$} \emph{and $\mathcal{V}$}
is%
\footnote{For a non-bijective function $f\!:(X,\mathcal{U})\rightarrow(Y,\mathcal{V})$,
\begin{align*}
\mathcal{U}_{\textrm{sat}} & \triangleq\{U\in\mathcal{U}\!:U=f^{-}f(U)\}\\
\mathcal{V}_{\textrm{comp}} & \triangleq\{V\in\mathcal{V}\!:V=ff^{-}(V)=V\cap f(X)\}\end{align*}
Here the {}``inverse'' $f^{-}$ of $f$ is defined by the projective
conditions $ff^{-}f=f$ and $f^{-}ff^{-}=f^{-}$. %
} \begin{equation}
\textrm{IT}\{e;\mathcal{V}\}\triangleq\{U\subseteq X:U=e^{-}(V),V\in\mathcal{V}_{\textrm{comp}}\}\label{Eqn: IT}\end{equation}
and for $q:(X,\mathcal{U})\rightarrow Y$, the \emph{image or} \emph{final
topology of} $\; Y$ \emph{generated by $\mathcal{U}$ and} \emph{$q$}
is \begin{equation}
\textrm{FT}\{\mathcal{U};q\}\triangleq\{V\subseteq Y:q^{-}(V)=U,U\in\mathcal{U}_{\textrm{sat}}\}.\label{Eqn: FT'}\end{equation}

\noindent A bijective \emph{ininal} \emph{function} $f\!:(X,\mathcal{U})\rightarrow(Y,\mathcal{V})$
is a \emph{homeomorphism,} and ininality for functions that are neither
$1:1$ nor onto generalizes homeomorphism; thus \[
U,V\in\textrm{IFT}\{\mathcal{U};f;\mathcal{V}\}\Leftrightarrow\{f(U)\}=\mathcal{V}\mbox{ and }U=f^{-}(V)\]

\noindent reduces to \[
U,V\in\textrm{HOM}\{\mathcal{U};f;\mathcal{V}\}\Leftrightarrow\mathcal{U}=\{f^{-1}(V)\}\mbox{ and }\{f(U)\}=\mathcal{V}\]
for a bijective, open-continuous function. A homeomorphism $f\!:(X,\mathcal{U})\rightarrow(Y,\mathcal{V})$
renders the homeomorphic spaces $(X,\mathcal{U})$ and $(Y,\mathcal{V})$
topologically indistinguishable in as far as their geometrical properties
are concerned. \emph{It is our hypothesis that the driving force behind
the evolution of a system toward a state of dynamical homeostasis
is the attainment of the ininal triple state $(X,f,Y)$ for the system.}
The ininal interaction $f$ between $X$ and $Y$ generates the smallest
possible topology of $f$-saturated sets on $X$ and the largest possible
topology of images of these sets in $Y$ and constitutes the state
of uniformity represented by the maximum entropy of the second law
of thermodynamics. Ininality of $f$ is simply an instance of non-bijective
homeomorphism.\emph{ }

\begin{figure}[!tbh]
\noindent \begin{centering}
{\small 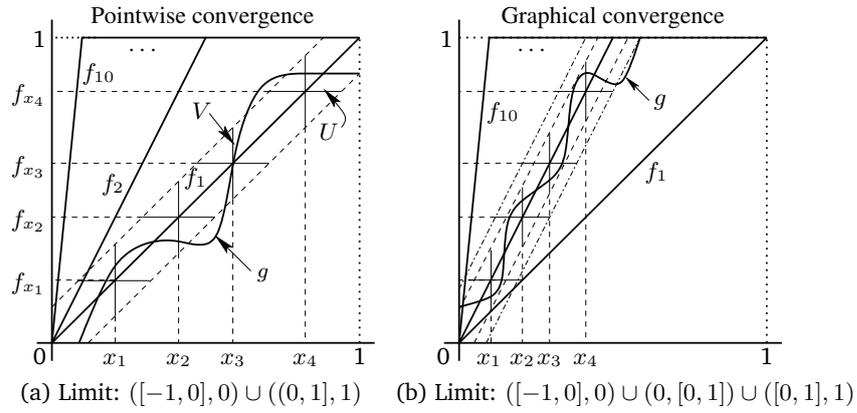}
\par\end{centering}{\small \par}

\caption{{\footnotesize \label{fig: biconv}Pointwise and graphical biconvergence.
Local neighbourhoods of $f_{n}(x)=\protect\begin{cases}
0 & -1\le x\le0\protect\\
nx, & 0<x\le1/n\protect\\
1, & 1/n<x\le1\protect\end{cases}$ at $(x_{i})_{1}^{4}$ with corresponding neighbourhoods $(U_{i})$
and $(V_{i})$ at $(x_{i},f(x_{i}))$. The converged limit in (a)
is a discontinuous function, in (b) it is a multifunction. It is this
extension, from functional to general relations with its various ramifications,
that constitutes the basis of chanoxity. }}

\end{figure}

\textsf{\textbf{(B) Multifunctional Extension of Function Spaces:
Graphical Convergence. }}The multifunctional extension is the{\large{}
}smallest dense extension $\textrm{Multi}(X)$ of the function space
$\textrm{map}(X)$. The main tool in obtaining the space $\textrm{Multi}(X)$
from $\textrm{map}(X)$ is a generalization of pointwise convergence
of continuous functions to (discontinuous) functions \citep{Sengupta2003}
by a process of \emph{graphical convergence of a net of functions}
illustrated in Fig. \ref{fig: biconv}. This defines neighbourhoods
of $f\!\in\!\textrm{map}(X,Y)$ to consist of those functional relations
in $\textrm{Multi}(X,Y)$ whose images at any point $x\in X$ lies
not only arbitrarily close to $f(x)$ (as in the usual case of topology
of \emph{pointwise convergence}), but whose inverse images at $y=f(x)\in Y$
contain points arbitrarily close to $x$. Thus the graph of $f$ is
not only arbitrarily close to $f(x)$ at $x$ in $V$, but must also
be such that $f^{-}(y)$ has at least branch in $U$ about $x$ such
that $f$ is constrained to cling to $f$ as the number of points
on the graph of $f$ increases. Unlike for simple pointwise convergence,
no gaps in the graph of the converged multi is permitted\emph{ not
only on the domain of $f$, but on its range as well. }

\noindent {\small }%
{\small \par}

The usual topological treatment of pointwise convergence of functions
is generalized to generate the boundary%
\footnote{\noindent \label{fn: boundary}The \emph{boundary of $A$ in $X$}
is the set of points $x\in X$ such that every neighbourhood $N$
of $x$ intersects both $A$ and is complement $X-A$: \begin{equation}
{\textstyle \textrm{Bdy}(A)\triangleq\{x\in X\!:(\forall N\in\mathcal{N}_{x})((N\bigcap A\neq\emptyset)\wedge(N\,{\textstyle \bigcap}\,(X-A)\neq\emptyset))\}}\label{Eqn: Def: Boundary}\end{equation}
 with $\mathcal{N}_{x}$ the neighbourhood system at $x$.%
} $\textrm{Multi}_{\parallel}(X,Y)$ between $\textrm{map}(X,Y)$ and
$\textrm{multi}(X,Y)$\begin{align*}
\mbox{Multi}(X,Y) & =\mbox{map}(X,Y)\,{\textstyle \bigcup}\,\mbox{Multi}_{\parallel}(X,Y)\,{\textstyle \bigcup\,\mbox{multi}(X,Y)},\end{align*}

\noindent observe that the boundary of $\textrm{map}(X,Y)$ in the
topology of pointwise biconvergence is a {}``line parallel to the
$Y$-axis''. 

Let $f\!:(X,\mathcal{U})\rightarrow(Y,\mathcal{V})$ be the iterative
evolutionary unfolding of a noninjective map function in $\mbox{Multi}(X)$
and $P(f)$ the number of injective branches of $f$. Denote by \[
F={f\in\mbox{Multi}(X)\!:f\mbox{ is a noninjective function on}X}\subseteq\mbox{Multi}(X)\]
 the basic collection of noninjective functions in $\mbox{Multi}(X)$.
For every $\alpha$ in some directed set $\mathbb{{D}}$, let $F$
have the extension property \[
(\forall f_{\alpha}\in F)(\exists f_{\beta}\in F)\!:P(f_{\alpha})\le P(f_{\beta}).\]
Let a partial order $\preceq$ on $\mbox{Multi}(X)$ be defi{}ned,
for $f_{\alpha},f_{\beta}\in\mbox{map}(X)\subseteq\mbox{Multi}(X)$
by \begin{equation}
P(f_{\alpha})\le P(f_{\beta})\Leftrightarrow f_{\alpha}\preceq f_{\beta},(4)\label{eq: order1}\end{equation}
 with $P(f):=1$ for the smallest $f$, defi{}ne a partially ordered
subset $(F,\preceq)$ of $\mbox{Multi}(X)$. This is actually a preorder
on $\mbox{Multi}(X)$ in which functions with the same number of injective
branches are equivalent. The existence of a maximal non-functional
element in this process, obtained as the set theoretic \textquotedblleft{}limit\textquotedblright{}
of the net of functions with increasing nonlinearity, does not imply
that it belongs to the functional chain as a fi{}xed point. The net
defi{}nes a corresponding net of increasingly multivalued functions
ordered inversely by the relation \begin{equation}
f_{\alpha}\preceq f_{\beta}\Leftrightarrow f_{\beta}^{-}\preceq f_{\alpha}^{-}.\label{eq: order}\end{equation}
from which it follows that \citep{Sengupta2003} \smallskip{}

\textsf{\textbf{Chaotic map.}} Let $A$ be a non-empty closed set
of a compact Hausdorff space $(X,\mathcal{U}).$ A function $f\in\mathrm{Multi}(X)$
is \emph{maximally non-injective} or \emph{chaotic }on $A$ w.r.t.
to $\preceq$ if (a) for any $f_{i}$ there exists an $f_{j}$ satisfying
$f_{i}\!\preceq\! f_{j}$ $\forall i\!<\! j\!\in\!\mathbb{N}$, (b)
the dense set $\mathcal{D}_{+}\!:=\!\{x\!:(f_{\nu}(x))_{\nu\in\mathrm{Cof}(\mathbb{D})}\mbox{ converges in }(X,\mathcal{{U}})\}$
of isolated singletons is countable.\smallskip{}
\footnote{The \emph{residual} and \emph{cofinal} subsets \begin{align}
\textrm{Res}(\mathbb{D}) & =\{\mathbb{R}_{\alpha}\in\mathcal{P}(\mathbb{D})\!:\mathbb{R}_{\alpha}=\{\beta\in\mathbb{D}\textrm{ for all }\beta\succeq\alpha\in\mathbb{D}\}\}.\label{Eqn: residual}\\
\textrm{Cof}(\mathbb{D}) & =\{\mathbb{C}_{\alpha}\in\mathcal{P}(\mathbb{D})\!:\mathbb{C}_{\alpha}=\{\beta\in\mathbb{D}\textrm{ for some }\beta\succeq\alpha\in\mathbb{D}\}\}\label{Eqn: cofinal}\end{align}
of a directed set $\mathbb{D}$ are the basic ingredients of the topological
theory of convergence of a net of functions. %
}

The convergence implicit in the above defi{}nition is \emph{graphical
convergence} of the graphs of $(f_{i})_{i}$ in $\mbox{Multi}(X)$
to allow for the extension of $\mbox{map}(X)$ to this more general
class of relations. The existence and relevance of this extended class
of one-to-many correspondences is fundamental to our consideration
of chaos, complexity, and holism, and the bidirectionality of convergence
with respect to the orthogonal axes of both $(f_{i})$ and its inverse
$(f_{i}^{\lyxmathsym{\textminus}})$. 

The collective macroscopic cooperation between $\mbox{map}(X)$ and
its extension $\mbox{Multi}(X)$ generates the equivalence classes
through fi{}xed points and periodic cycles of $f$. As all points
in a class are equivalent under $f$, a net or sequence converging
to any must necessarily converge to every other in the set. This implies
that the cooperation between $\mbox{map}(X)$ and $\mbox{Multi}(X)$
conspires to alter the topology of $X$ to large equivalence classes.
This dispersion throughout the domain of $f$ of initial localizations
suggests increase in entropy/disorder with increasing chaoticity;
\emph{complete chaos therefore implies the second law condition of
maximum entropy enlarging the function space to multifunctions.}

\textsf{\textbf{(C) The Negative World $\mathfrak{W}$. Motivation:
Competitive Collaboration.}} Of the axioms defining a vector space
$V$, that of the additive\emph{ }inverse stipulating $\forall u\in V,\,\exists(-u)\in V\! u+(-u)=0$,
comprises the crux of \emph{competitive collaboration}. This\emph{
participatory} existence $\mathbb{R}_{-}$ of $\mathbb{R}_{+}$ inducing
a\emph{ reverse arro}w in $\mathbb{R}$,\emph{ competing collaborativel}y
with the forward arrow in $\mathbb{R}_{+}$, serves to \emph{complete}
the structure of $\mathbb{R}$. 

In a parallel vein, let $W$ be a set such that for every $w\in W$
there exists a negative element $\mathfrak{w\in\mathfrak{W}}$ with
the property that \sublabon{equation}\begin{equation}
\mathfrak{W}\triangleq\{\mathfrak{w}\!:\{w\}\,{\textstyle \bigoplus}\,\{\mathfrak{w}\}=\emptyset\}\label{eq: matter-neg(a)}\end{equation}
defines the negative, or exclusion, set of $W$%
\footnote{Notice that this definition is meaningless if restricted to $W$ or
$\mathfrak{W}$ alone; it makes sense, in the manner defined here,
only in relation to the pair $(W,\mathfrak{W})$. %
}. Hence for all $A\subseteq W$ there is a neg(ative) set $\mathfrak{A}\subseteq\mathfrak{W}$
associated with (generated by) $A$ that satisfies

\begin{eqnarray}
A\,{\textstyle \bigoplus}\,\mathfrak{G} & \triangleq & A-G,\quad G\leftrightarrow\mathfrak{G}\nonumber \\
A\,{\textstyle \bigoplus}\,\mathfrak{A} & = & \emptyset.\label{eq: matter-neg(b)}\end{eqnarray}
\sublaboff{equation}The pair $(A,\mathfrak{A})$ act as relative
discipliners of each other in the evolving dissipation and tension,
``undoing'', ``controlling'', ``stabilizing'' the other. The exclusion
topology of large equivalence classes in\emph{ }$\mbox{Multi}(X)$\emph{
}successfully competes with the normal inclusion topology of\emph{
}$\mbox{map}(X)$\emph{ }to generate a state of dynamic homeostasis
in\emph{ }$W$\emph{ that permits out-of-equilibrium complex composites
of a system and its environment to coexist despite the privileged
omnipresence of the Second Law. }The evolutionary process ceases when
the opposing influences in $W$ and its moderator $\mathfrak{W}$
balance in dynamic equilibrium by the generation of the ininal triple. 

\sublabon{figure}%
\begin{figure}[!tbh]
\noindent \begin{centering}
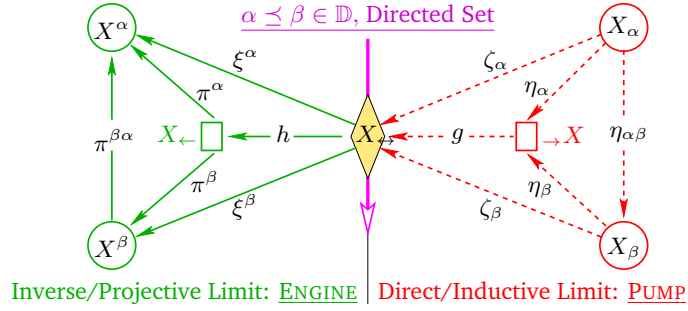
\par\end{centering}

\begin{spacing}{0.85}
\noindent \caption{{\small \label{fig: direct-inverse}Direct and inverse limits of direct
and inverse systems $(X_{\iota},\eta_{\iota\kappa})$, $(X^{\iota},\pi^{\kappa\iota})$.
Induced homeostasis is attained between the two adversaries by the
respective arrows opposing each other as shown in the next figure}{\footnotesize{}
}{\small where expansion to the atmosphere is indicated by decreasingly
nested subsets. }}
\end{spacing}

\end{figure}
\begin{figure}[!tbh]
\noindent \begin{centering}
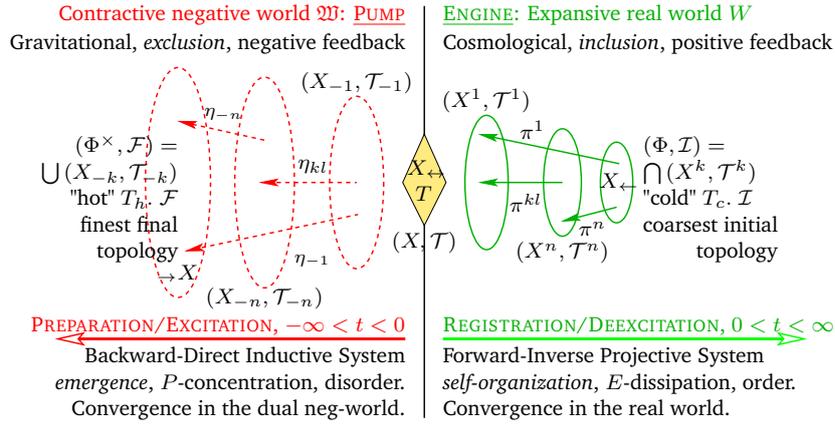
\par\end{centering}

\begin{spacing}{0.85}
\noindent \caption{\label{fig: direct-inv}{\small{} Intrinsic arrows of time based on
inverse-direct limits of inverse-direct systems. Intrinsic irreversibility
follows since the dissipative forward-inverse arrow is the }\emph{\small natural}{\small{}
arrow in $\mathbb{R}_{+}$ equipped with the usual inclusion topology,
while the backward-direct positive arrow of $\mathbb{R}_{-}$ manifests
itself as a dual \textquotedblleft{}negative\textquotedblright{} exclusion
topology in $\mathbb{R}_{+}$. Notice that although $E$ and $P$
are born in $[T_{h},T]$ and $[T,T_{c}]$ respectively, they operate
in the domain of the other in the true spirit of competitive-collaboration.
The entropy increases on contraction since the position uncertainty
decreases faster than the increase of momentum uncertainty}}
\end{spacing}

\end{figure}

\sublaboff{figure}

\textsf{\textbf{(D) Inverse And Direct Limits }}This abstract conceptual
foundation of the existence of a complimentary negative world $\mathfrak{W}$
for every real $W$ permits participatory competitive collaboration
between the two to generate self-organizing complex structures as
summarized in Figs. \ref{fig: direct-inverse} and \ref{fig: direct-inv},
see Refs. \citep{Dugundji1966,Sengupta2010a} for the details. Briefly,
the mathematical goal of chanoxity of establishing the existence of
a disordering arrow for every dissipative ordering eventuality of
large maximal equivalence classes of open sets through the attainment
of the ininal topology, is additionally corroborated by the existence
of these complimentary limits%
\footnote{For a family of sets $(X_{\kappa\in\mathbb{D}})$ the disjoint union
is the set $\coprod_{\alpha\in\mathbb{D}}X_{\alpha}\triangleq\bigcup_{\alpha\in\mathbb{D}}\{(x,\alpha):x\in X_{\alpha}\}$
of ordered pairs, with each $X_{\alpha}$ being canonically embedded
in the union as the pairwise disjoint $\{(x,\alpha):x\in X_{\alpha}\}$,
even when $X_{\alpha}\cap X_{\beta}\neq\emptyset$. If $\{X_{k}\}_{k\in\mathbb{Z}_{+}}$
is an increasing family of subsets of $X$, and $\eta_{mn}\!:X_{m}\rightarrow X_{n}$
is the inclusion map for $m\leq n$, then the direct limit is $\bigcup X_{k}$\textcolor{blue}{.} 

For $\{X^{k}\}_{k\in\mathbb{Z}_{+}}$ a decreasing family of subsets
of $X$ with $\pi^{nm}\!:X^{n}\rightarrow X^{m}$ the inclusion map,
the inverse limit is $\bigcap X^{k}$.%
}, \citep{Dugundji1966}, possessing the following salient features. 

For a given direction $\mathbb{D}$, the connecting maps $\pi$ and
$\eta$ between the family of subsets $\{X^{\alpha}\}$ and $\{X_{\alpha}\}$
are oriented in opposition, the respective inverse and direct limits
of the systems being $X_{\leftarrow}$ and $_{\rightarrow}X$%
\footnote{These limits are conventionally denoted $\underleftarrow{\lim}$ and
$\underrightarrow{\lim}$ respectively%
}. The \emph{mathematical existence} of these opposing limits, applicable
to the problem under consideration, validates the arguments above
and bestows on the gravitational disordering arrow with the sanction
of analytic logic. Accordingly in Fig. \ref{fig: direct-inv}, reversal
of the direction of $\mathbb{D}$ to generate the forward and backward
arrows completes the picture; observe the significant interchange
of the relative positions of the two diagrams defining the homeostatic
equilibrium $X_{\leftrightarrow}(T)$. If any of the two were to be
absent, the remaining would operate within the full gradient $T_{h}\lyxmathsym{\textminus}T_{c}$
and would be evolutionally harmless and impotent; in the homeostatic
competitive case, however, the homeostatic condition $T$ is generated
and defi{}ned by the nonlinear competitive collaboration of these
two opposites as will be seen below.

The inverse and direct limits are thus generated by opposing directional
arrows whose existence follow from very general mathematical principles;
thus for example existence of the union of a family of nested sets
implies the existence of their intersection, and conversely. As a
concrete example, Fig. \ref{fig: direct-inv} specializes to \emph{rigged
Hilbert spaces }$\Phi\subset\mathcal{H}\subset\Phi^{\times}$\[
\overset{{\color{red}\mbox{Entropy-increasing Competitive}}}{{\color{red}\overleftarrow{\Phi^{\times}\triangleq{\displaystyle \cup}_{k}\mathcal{H}_{-k}\,\supset\,\cdots\,\supset\,\mathcal{H}_{-i}\,\supset\,\cdots\,\supset\,\mathcal{H}_{-1}}}}\supset\mathcal{{\color{yellow}H}}\supset\underset{{\color{green}{\color{green}\mbox{Entropy-decreasing, Collaborative}}}}{{\color{green}\underrightarrow{\mathcal{H}^{1}\,\supset\,\cdots\,\supset\,\mathcal{H}^{i}\,\supset\,\cdots\,\supset\,{\displaystyle \cap}_{k}\mathcal{H}^{k}\triangleq\Phi}}}\]
with \emph{$\Phi$} the space of physical states prepared in actual
experiments, and $\Phi^{\times}$ are antilinear functionals on $\Phi$
that associates with each state a real number interpreted as the result
of measurements on the state. Mathematically, the space of test functions
$\Phi$ and the space of distributions $\Phi^{\times}$ represent
defi{}nite and well- understood examples of the inverse and direct
limits that enlarge the Hilbert space $\mathcal{H}$ to the rigged
Hilbert space $(\Phi,\mathcal{H},\Phi^{\times})$, with $\mathcal{H}$
the homeostatic condition. Observe how without the mutual participation
of either of these $\leftrightarrow$ adversaries, emergence of the
real number is impossible.

\subsection{\textsf{Physics of ChaNoXity: Stand-off between Individualism and
Collectivism }}

\noindent %
\begin{figure}[!tbh]
\noindent \begin{centering}
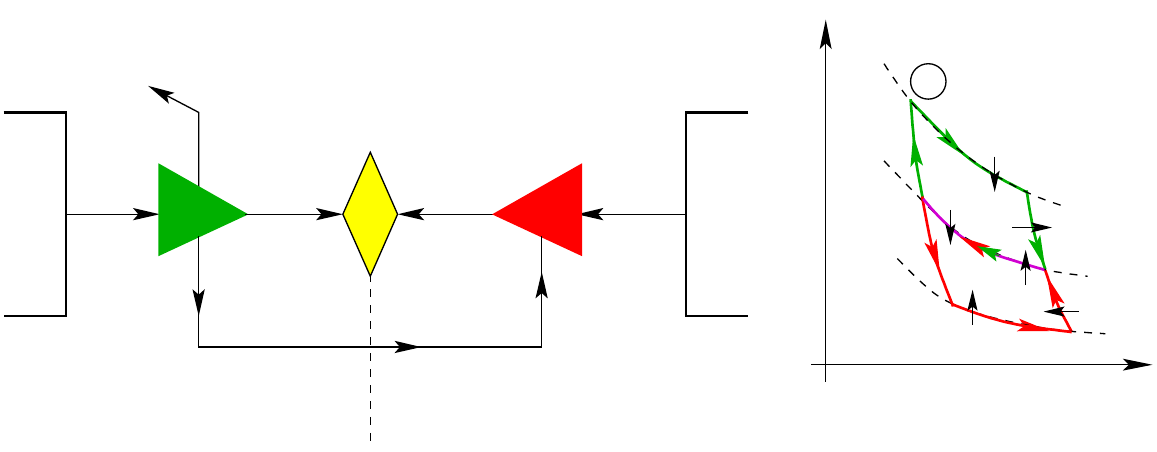
\par\end{centering}

\caption{\label{fig: engine-pump}{\small Reduction of the dynamics of opposites
to an equivalent engine-pump thermodynamic system; $W_{\textrm{rev}}=Q_{h}(1\lyxmathsym{\textminus}T_{c}/T_{h})$,
$W(T)=Q_{h}(1\lyxmathsym{\textminus}T/T_{h})$. The collaborative
confrontation of $Q$ and $q$, as refl{}ected by their inverse relationship
in Eq. (13) below with $\alpha$ the adaptability of these opposites
in defi{}ning the complex$\alpha$ system, bestows on $Q$ the possible
interpretation of a \textquotedblleft{}demand\textquotedblright{}
that is met by the \textquotedblleft{}supply\textquotedblright{} $q$
in a bidirectional feedback loop that sustains, and is sustained by
each other, in the context of the whole.}}

\end{figure}

\noindent Assume that a complex adaptive system is distinguished by
the complete utilization of a fraction $W(T):=[1-\iota(T)]W_{\textrm{rev}}=(1-T/T_{h})Q_{h}$
of the work output of an imaginary reversible engine $(T_{h},E,T_{c})$
with output $W_{\textrm{{rev}}}=(1-T_{c}/T_{h})Q_{h}$ to self-generate
the pump $P$ in competitive collaboration with $E$, Fig. \ref{fig: engine-pump}.
The \emph{irreversibility factor} \sublabon{equation} \begin{equation}
\iota(T)\triangleq\frac{W_{\textrm{rev}}-W(T)}{W_{\textrm{rev}}}\in[0,1]\label{eq: irreversibility(a)}\end{equation}
accounts for that part $\iota W_{\textrm{rev}}$ of available energy
$W_{\textrm{rev}}$ that cannot be gainfully utilized but must be
degraded in increasing the entropy of the universe. Hence \begin{equation}
\iota(T)=\left(\frac{T_{\textrm{R}}}{W_{\textrm{rev}}}\right)S(T)\label{eq: irreversibility(b)}\end{equation}
yields the effective entropy \begin{equation}
S(T)=\frac{W_{\textrm{rev}}-W(T)}{T_{\textrm{R}}}.\label{eq: gen-entropy}\end{equation}
\sublaboff{equation}with reference to the induced temperature $T$.
The self-induced pump decreases the temperature-gradient $T_{h}-T_{c}$
to $T_{h}-T$, $T_{c}\leq T<T_{h}$, generating dynamic stability
to the system. 

Let $\iota$ be obtained from \begin{align}
W_{E}:=Q_{h}\left(1-\frac{T}{T_{h}}\right) & \triangleq W_{P}\nonumber \\
 & =Q_{h}(1-\iota)\left(1-\frac{T_{c}}{T_{h}}\right);\label{eq: irrev(T)}\end{align}
hence \sublabon{equation} \begin{equation}
{\displaystyle \iota(T)=\frac{T-T_{c}}{T_{h}-T_{c}}}\label{eq: iota}\end{equation}
is formally similar to the quality \begin{equation}
x(v)=\frac{v-v_{f}}{v_{g}-v_{f}}\label{eq: quality}\end{equation}

\noindent \sublaboff{equation}of a two-phase liquid-gas mixture with
temperatute $T$ playing the role of specific volume $v$, where $T_{h}-T_{c}$
represents the internal energy that is divided into the non-entropic
$T_{h}-T$ Helmholtz free (available) energy $A$ internally utilized
to generate the pump $P$ and a reduced $T-T_{c}$ entropic dissipation
by $E$, with respect to the induced equilibrium temperature $T$. 

The generated pump is a realization of the energy available for useful,
work arising from reduction of the original gradient $T_{h}-T_{c}$
to $T-T_{c}$. The irreversibility $\iota(T)$ is adapted by the engine-pump
system such that the induced instability of $P$ balances the imposed
stabilizing effort of $E$ to the best possible advantage of the system
and the environment. Hence a measure of the energy that is unavailable
to the system and must necessarily be dissipated to the environment
is the \emph{generalized entropy}\sublabon{equation} \begin{eqnarray}
TS(T)=\iota(T)W_{\textrm{rev}} & \!\!\!=\!\!\! & W_{\textrm{rev}}-W(T)\label{eq: free-energy(a)}\\
 & \!\!\!=\!\!\! & U-A(T)\label{eq: free-energy(b)}\end{eqnarray}
\sublaboff{equation}which the system attains by adapting itself \emph{internally}
to a state of optimal competitive collaboration.

Figure \ref{fig: engine-pump} represents the essence of competitive
collaboration: the dispersion of $E$ is proportional to the domain
$T-T_{c}$ of $P$, and the concentration of $P$ depends on $T_{h}-T$
of $E$. Thus an increase in $\iota$ can occur only at the expense
of $P$ which opposes this tendency; reciprocally a decrease in $\iota$
is resisted by $E$. The induced pump $P$ prevents the entire internal
resource $T_{h}-T_{c}$ from dispersion at $\iota=1$ by defining
some $\iota<1$ for a homeostatic temperature $T_{c}<T<T_{h}$, with
$E$ and $P$ interacting with each other in the spirit of competitive
collaboration at the induced interface $T$.%
\begin{figure}[!tbh]
\noindent \begin{centering}
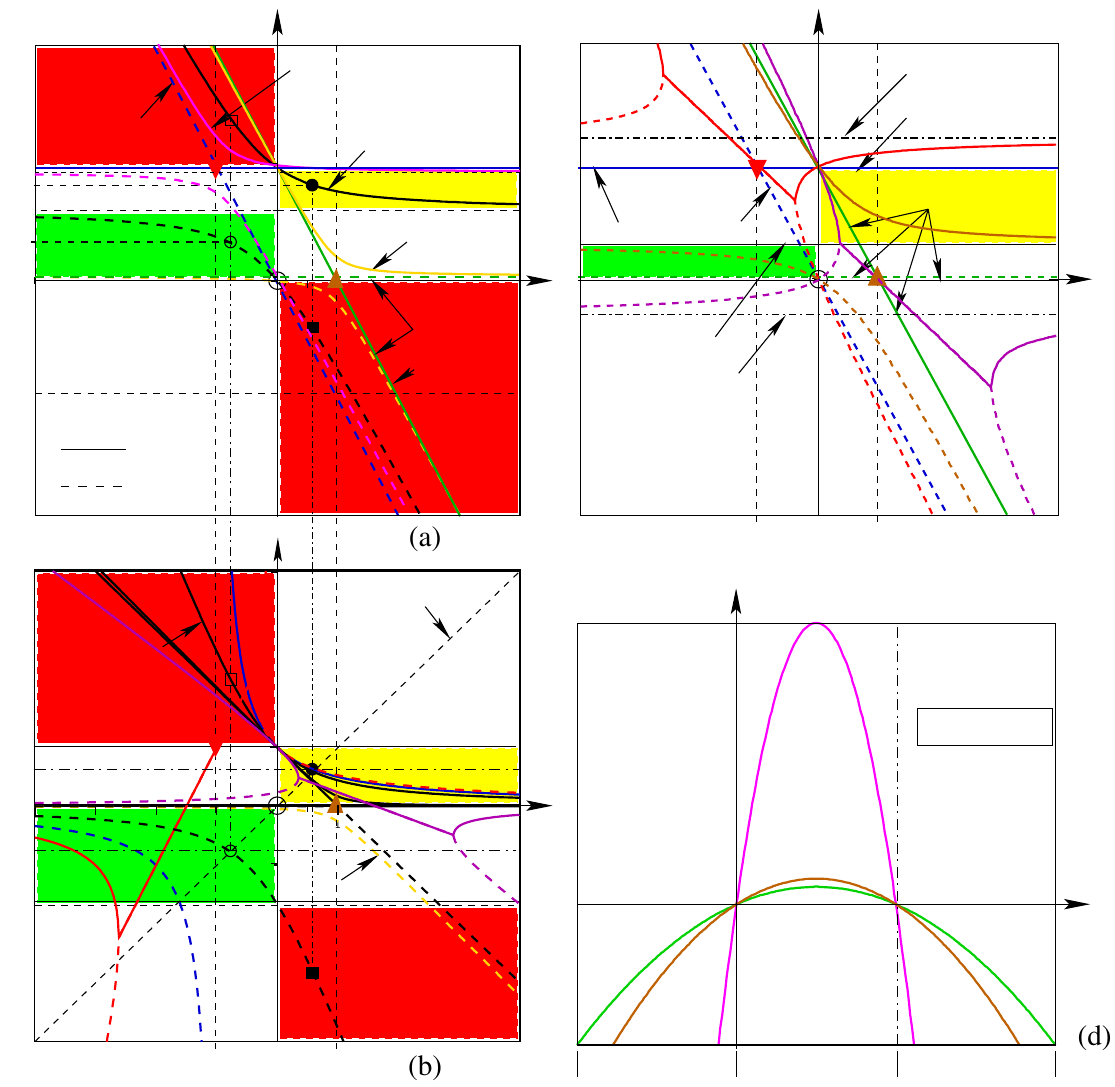
\par\end{centering}

\begin{spacing}{0.85}
\caption{\label{fig: Engine-Pump} {\small The interactive {}``participatory
universe'', $T_{h}=480\mbox{K}$; $\iota_{c}=-T_{c}/(T_{h}-T_{c})$.
The straight lines connecting the $T<T_{c}$ and $T>T_{h}$ segments
in (b) and (c) correspond to complex roots. The colour-code is as
in Fig. \ref{fig: engine-pump}}}
\end{spacing}

\end{figure}

Defining the equilibrium steady-state representing $X_{\leftrightarrow}$
of homeostatic $E\textrm{-}P$ adaptability $\alpha:=\eta_{E}\zeta_{P}$,
the \emph{equation of state of the participatory universe }for $q(T)=(1-\iota)Q(T_{h}-T_{c})/(T-T_{c})=[(1-\iota)/\iota]Q$
with $Q(T)\triangleq Q_{h}-W(T)=Q_{h}-[1-\iota(T)]W_{\textrm{{rev}}}=Q_{h}(T/T_{h})$

{\small \begin{align}
\alpha(T) & =\left(\frac{T_{h}-T}{T_{h}}\right)\left({\displaystyle \frac{T}{T-T_{c}}}\right)\triangleq\frac{q(T)}{Q_{h}}=\frac{q(T)}{Q(T)}\left(\frac{T}{T_{h}}\right)\label{eq: alpha}\end{align}
}in the form $pv=f(T)$ where $p\equiv\zeta_{P}=0$ at $T=0$ and
$v\equiv\eta_{E}$, be the product of the efficiency of a reversible
engine and the coefficient of performance of a reversible pump. Fig.
\ref{fig: Engine-Pump} for $T_{h}=480^{\circ}\textrm{K}$ and $T_{c}=300^{\circ}\textrm{K}$
shows that the engine-pump duality has the significant property of
supporting two different states \sublabon{equation}\begin{eqnarray}
{\color{magenta}{\color{magenta}{\normalcolor T_{\pm}(\alpha)}}} & = & \frac{1}{2}\left[(1-\alpha)T_{h}\pm\sqrt{(1-\alpha)^{2}T_{h}^{2}+4\alpha T_{c}T_{h}}\right]\label{eq: T_pm(a)}\\
 & = & \begin{cases}
((1-\alpha)T_{h},\,0)=(0,0)_{\alpha=1}, & T_{c}=0\\
(T_{h},\,-\alpha T_{h})=(T_{h},T_{h})_{\alpha=-1}, & T_{c}=T_{h}\end{cases}\label{eq: T_pm(b)}\end{eqnarray}
\sublaboff{equation}

\noindent for any value of $\alpha$. 

Fig. \ref{fig: Engine-Pump}(b) suggests that the balancing condition\begin{equation}
\iota(T)=\alpha(T)\label{eq: iota_alpha}\end{equation}
defining the most appropriate equilibrium criterion \sublabon{equation}\begin{eqnarray}
T_{\pm} & = & \frac{T_{h}(T_{h}+T_{c})\pm(T_{h}-T_{c})\sqrt{T_{h}^{2}+4T_{c}T_{h}}}{2(2T_{h}-T_{c})}\label{eq: T_plus(a)}\\
 & = & \begin{cases}
(0.5T_{h},0), & T_{c}=0\\
(T_{h},T_{h}), & T_{c}=T_{h}\end{cases}\label{eq: T_plus(b)}\end{eqnarray}
\sublaboff{equation}directly determines the irreversibility of the
interaction because the tendency to revert back to the original condition
(small $\iota$: predominance of pump $P$) implies large $E$-$P$
adaptability $\alpha$ inviting $E$-opposition and the homeostasy
of Eq. (\ref{eq: iota_alpha}); see Fig. \ref{fig: economy}(a). Note
that at $T_{c}=0$, $T_{\text{\textminus}}=T_{c}$ while at $T_{c}=T_{h}$,
$T_{+}=T_{\text{\textminus}}=T_{c}$ .

A complex system can hence be represented as \vspace{-0.2in}

\begin{center}
\begin{equation}
\begin{array}{c}
{\color{red}\underset{\textrm{Individualism: Competitive "capital"}\,(\uparrow)}{\underbrace{\begin{array}{r}
\underleftarrow{\mbox{\textsc{Gravitational, Backward-Direct}}}\\
P\mbox{-synthesis of }\textrm{concentration, disorder,}\\
\textrm{ \textrm{entropy increasing}, \textrm{\emph{bottom-up emergence}}\ensuremath{\,_{\rightarrow}}}C\end{array}}}}\;{\displaystyle \bigoplus}\;{\color{green}\underset{\textrm{Collectivism: Collaborative "culture"}\,(\downarrow)}{\underbrace{\begin{array}{l}
\underrightarrow{\mbox{\textsc{Cosmological, Forward-Inverse}}}\\
E\textrm{-analysis of dispersion, order, entropy}\\
\textrm{decreasing, \textrm{\emph{top-down self-organization }}}C_{\leftarrow}\end{array}}}}\\
\\\Longleftrightarrow{\color{yellow}\textrm{Synthetic cohabitation of opposites }C_{\leftrightarrow}},\end{array}\label{eq: chanoxity}\end{equation}

\par\end{center}

\noindent where ${\textstyle \bigoplus}$ denotes a non-reductionist
sum of a top-down engine and its complimentary bottom-up pump that
behaves in an organized collective manner with properties that cannot
be identifi{}ed with any of the individual components but arise from
the structure as a whole: these systems cannot dismantle into their
parts without destroying themselves.%
\footnote{The definition of \emph{cybernetics} as the study of systems and processes
that {}``interact with themselves and produce themselves from themselves''
by Louis Kauffman remarkably captures this spirit. %
} Analytic methods cannot simplify them as such techniques do not account
for char- acteristics that belong to no single component but relate
to the whole with all their interactions. Complexity is a dynamical,
interactive and interdependent hierarchical homeostasis of $P$-emergent,
disordering instability of competitive backward \textquotedblleft{}capital\textquotedblright{}%
\footnote{\label{fn: capital}\textquotedblright{}A factor of production which
is not wanted for itself but for its ability to help in producing
other goods; any form of wealth capable of being employed in the production
of more wealth.\textquotedblright{} Wikipedia, the free encyclopedia. 

Capital on its own therefore, is as impotent as sperm without egg.%
} feedback in cohabitation with the adaptive, $E$-organized, ordering
stability of collaborative forward feedback of \textquotedblleft{}culture\textquotedblright{}%
\footnote{\label{fn: culture}\textquotedblright{}The set of shared attitudes,
values, goals, and practices that characterizes an institution, organization
or group; an integrated pattern of human knowledge, belief, and behavior
that depends upon the capacity for symbolic thought and social learning.\textquotedblright{}
Culture, according to Edward Tylor, \textquotedblleft{}is that complex
whole of knowledge, belief, art, morals, law, custom, and any other
capabilities and habits acquired by man as a member of society\textquotedblright{}.
Wikipedia, the free encyclopedia. 

Culture on its own therefore, is as insatiated as egg without sperm.%
} generating non-reductionist holism beyond the sum of its constituents.
\begin{table}[!tbh]
\noindent \begin{centering}
{\renewcommand{\arraystretch}{1.3}\begin{tabular}{|c|c|}
\hline 
\textcolor{green}{$\underrightarrow{\mbox{Forward-Inverse Projective arrow}}$} & \textcolor{red}{$\underleftarrow{\mbox{Backward-DirectInductivearrow}}$}\tabularnewline
\textcolor{green}{(Natural direction in $W$ )} & \textcolor{red}{(Natural direction in $\mathfrak{W}$)}\tabularnewline
\hline
\hline 
Top-down Engine $E$ & Bottom-up Pump $P$\tabularnewline
\hline 
Dissipative: Self-organization & Concentrative: Emergence\tabularnewline
\hline 
Order: Entropy decreasing  & Disorder: Entropy increasing\tabularnewline
\hline 
Collective: Collaborative & Individualistic: Competitive\tabularnewline
\hline
\hline 
\textcolor{green}{$\underrightarrow{\mbox{"Culture"}}$} & \textcolor{red}{$\underleftarrow{\mbox{"Capital"}}$}\tabularnewline
\hline
\end{tabular}}\medskip{}

\par\end{centering}

\caption{{\small Holistic adverseries {}``capital'' and {}``culture'' of
Nature.}}

\end{table}

\noindent Summarizing the evolutionary tension between \textquotedblleft{}capital\textquotedblright{}
and \textquotedblleft{}culture\textquotedblright{} of Table 1, leads
to the 

\noindent \textbf{Definition. Complexity. \label{Definition. Complexity.}}An
open thermodynamic system of many interdependent and interacting parts
is complex if it lives in synthetic competitive cohabitation with
its self-induced negative dual in a state of homeostatic, hierarchical,
two-phase equilibrium of collective top-down, dissipative, self-organizing,
entropy-decreasing engine and individualistic bottom-up, concentrative,
emergent, entropy-increasing pump dual, coordinated and mediated by
the environment (\textquotedblleft{}universe\textquotedblright{}).%
\footnote{\textbf{Succinctly:} Homeostasis of the holistic adverseries of emergent
\textquotedblleft{}capital\textquotedblright{} and self-organizing
\textquotedblleft{}culture\textquotedblright{} defines a complex system.%
}

\subsubsection{\textsf{\label{sub: Complexity:-A-Two-Phase}Complexity: Two-Phase
Mixture of Bottom-Up Individualistic Competition and Top-Down Collective
Collaboration: Critical and Triple Points }}

Consider Fig. \ref{fig: Engine-Pump} for the dual-pair $(W,\mathfrak{W})$
in relation to the formalization represented by Eqs. (\ref{eq: iota},\emph{
b}). Fig. \ref{fig: Engine-Pump}(a, b) defi{}nes four disjoint regions
(I), (II), (III), (IV) characterized by the product \sublabon{equation}
\begin{eqnarray}
\iota\alpha(T) & = & \frac{T(T_{h}\text{\textminus}T)}{T_{h}(T_{h}\text{\textminus}T_{c})}\nonumber \\
 & \gtrless & 0\label{eq: iota-alpha(a)}\end{eqnarray}
with positive values defi{}ning $W$ in (I) and (III) and $\iota\alpha<0$
representing $W$ in (II) and (IV), refer Fig. \ref{fig: Engine-Pump}(d).
The signifi{}cant point is the full specifi{}cation of these regions
in terms of the product and the direct linkages of region (I) with
(II) through $T_{+}$ and of (III) with (IV) by $T_{\lyxmathsym{\textminus}}$
. Considering $T_{c}$ as a variable with $T_{h}$ given, produces
the bounds of Eqs. (\ref{eq: T_pm(b)}) and (\ref{eq: T_plus(b)})
with the rather remarkable property that in the operational range
\begin{equation}
0<T_{c}<T_{h},\end{equation}
\sublaboff{equation}$T_{+}$ and $T_{\lyxmathsym{\textminus}}$ are
composed of bifurcated components of $(T_{+}=(1\lyxmathsym{\textminus}\alpha)T_{h},T_{\lyxmathsym{\textminus}}=0)$
at $T_{c}:=0$ and $(T_{+}=T_{h},T_{\lyxmathsym{\textminus}}=\lyxmathsym{\textminus}\alpha T_{h})$,
at $T_{c}:=T_{h}$ ; thus $T_{\pm}$ in this region are holistic expressions
of themselves at the limiting values of $0$ and $T_{h}$; see Fig.
\ref{fig: Engine-Pump}(a). %
\begin{figure}[!tbh]
\noindent \begin{centering}
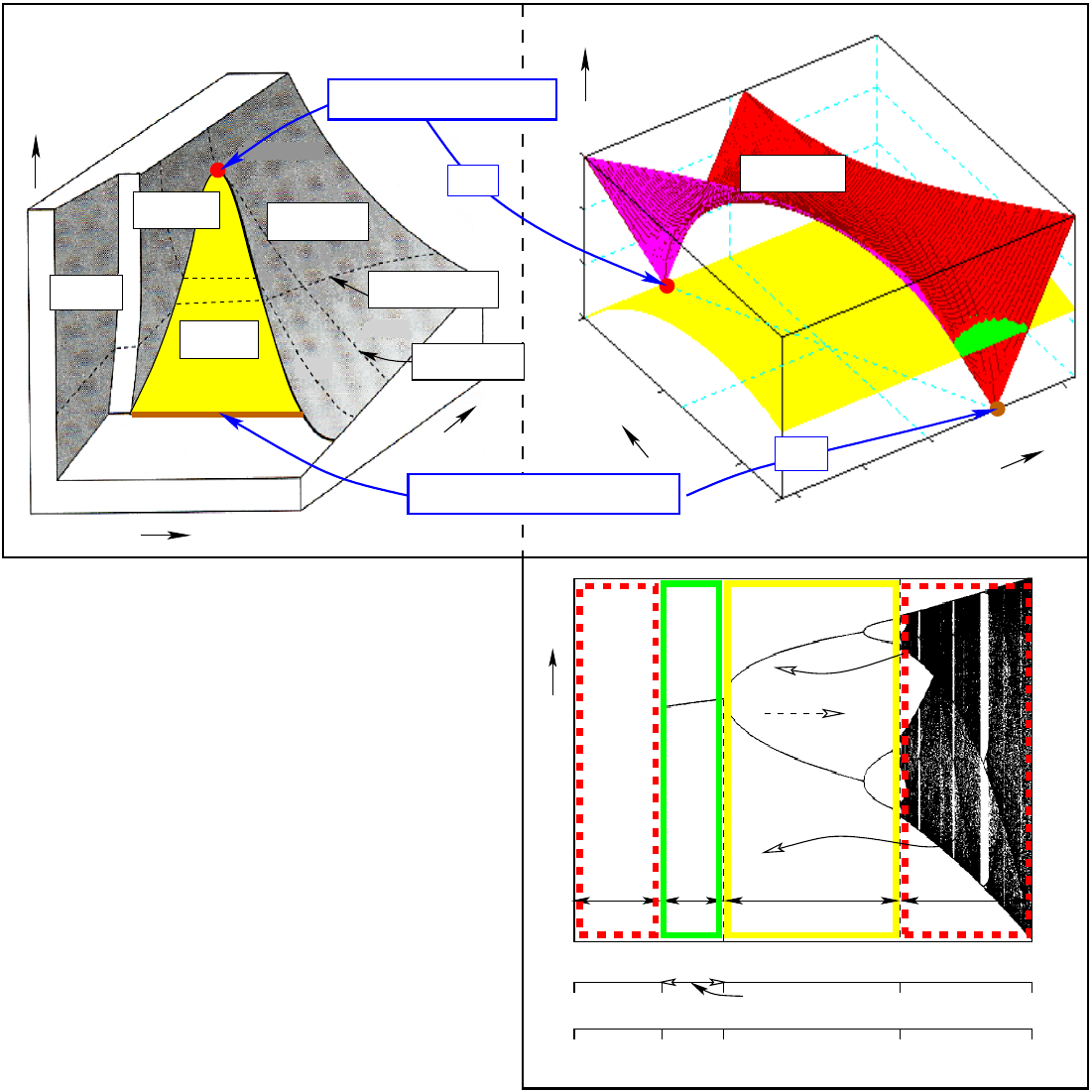
\par\end{centering}

{\small \caption{{\small \label{fig: 2-phase}The 2-phase complex $\iota=\alpha$ region,
(b) with }\emph{\small critical}{\small{} }\emph{\small point}{\small{}
$T_{c}=T_{h}$ at $\alpha_{-}=\lyxmathsym{\textminus}1$, yielding
to $\alpha$-dependent $\alpha=\eta\zeta$ at low $T_{c}$. The }\emph{\small triple}{\small{}
}\emph{\small point}{\small{} $\alpha_{+}=1$, $T_{c}=0$ is approachable
only through this route. Compared to the normal transition of (a),
self-organization in (b) occurs for $\alpha=Pv=\mbox{const}$, with
$P,v,T$ varying according to Eq. (\ref{eq: alpha}). $T_{+}-T_{-}:=\sqrt{T_{h}^{2}+4T_{c}T_{h}}\left(\frac{T_{h}-T_{c}}{2T_{h}-T_{c}}\right)$
at $\iota=\alpha$ is taken as an indicator of fi{}rst-order-second-order
transition because of Eqs. (\ref{eq: T_pm(b)}) and (\ref{eq: T_plus(b)});
(ii) and (iv) of (c) are the signatures of (II) and (IV) in $W$.
The colour convention in (b) follows that of Fig. \ref{fig: engine-pump}. }}
}
\end{figure}

The non-trivial range \[
T_{h}\le T_{c}\le0\]
--- meaningful only for $T_{h}\rightarrow+\infty$ --- is graphed
in Fig. \ref{fig: Engine-Pump}(c) and (d). Of fundamental importance
is the fact that the roots of Eq. (\ref{eq: T_pm(a)}) form \emph{continuous}
\emph{curves} in these regions, bifurcating as individual holistic
components at $\alpha_{\pm}=\pm1$: notice how at these values the
continuous curves change character in disengaging from each other
to form separate linear entities before {}``collaborating'' once
again in generating the profiles $T_{\pm}$ in the operating range.
These adaptations of the engine-pump are substantive in the sense
that the specific $\alpha$-values denote physical changes in the
global behaviour of the system; they mark the \emph{critical} and
\emph{triple} \emph{points} of Fig. \ref{fig: 2-phase}. The two-phase
complex surface denoted by $\alpha=\iota$ is to be distinguished
from the general $Pv$ region $\alpha=\eta\zeta$. Since the ideal
participatory universe satisfies a more involved \emph{nonlinear}
equation (\ref{eq: alpha}) compared to the simple \emph{linear} relationship
of an ideal gas, diagram \ref{fig: 2-phase}(b) is more involved than
the corresponding (a), with the transition at the triple point $\alpha_{+}=1$
displaying very definite distinctive features. While (b) clearly establishes
that the triple point cannot be accessed from the $\iota=\alpha$
surface and requires a detour through the general $\alpha=\eta\zeta$,
it also offers a fresh insight on the origin of the insular nature
of the absolute zero $T=0$. 

Equation (\ref{eq: alpha}) and Fig. \ref{fig: 2-phase}(b) show that
the 2-phase individualistic-collective {}``liquid-vapour'' region
$\iota=\alpha$ is distinguished by the imposed constancy of $\alpha$
--- and hence of the product $Pv$ --- just as $P$ and $T$ separately
remain constant in Fig. \ref{fig: 2-phase}(a). At the critical point
$v_{f}=v_{g}$ for passage to second order phase transition, $T_{c}=T_{h}$
requires $T_{+}=T_{-}$ which according to Eqs. (\ref{eq: T_pm(b)})
and (\ref{eq: T_plus(b)}) can happen only at $\alpha_{-}=-1$ corresponding
to the $(P_{\textrm{cr}},T_{\textrm{cr}})$ of figure (a). At the
other unique adaptability of $\alpha_{+}=1$ at $T_{c}\rightarrow0$,
the system passes into region (IV) from (III) just as (I) passes into
(II) as $T_{c}\rightarrow T_{h}$ at $\alpha_{-}$. Observe from Eq.
(\ref{eq: T_pm(a)}) that \sublabon{equation} \begin{equation}
(T_{c}\rightarrow0)\Longleftrightarrow(T_{h}\rightarrow\infty)\label{eq: reciprocal_a}\end{equation}
allows the self-organizing complex phase-mixture of collaboration
and competition to maintain its state $T$ as the condition of homeostatic
equilibrium%
\footnote{Does the melting of the arctic icebergs and the recent severe blizards
in Europe and USA indicate the veracity of Eq. \ref{eq: reciprocal_a}?%
} when $(T_{+},T_{-})=(0,0)_{\alpha_{+}=1}$. Simultaneously however,
because $T_{c}<T_{h}$,\begin{equation}
(T_{c}\rightarrow T_{h})\Longrightarrow(T_{h}\rightarrow\infty)\label{eq: reciprocal_b}\end{equation}
 implies from Eq. (\ref{eq: T_pm(b)}) that $(T_{+},T_{-})=(T_{h},T_{h})_{\alpha_{-}=-1}$
is also true. Hence 

\begin{equation}
(\alpha_{+})_{T_{c}=0}\sim(\alpha_{-})_{T_{c}=T_{h}}\label{eq: recirocal_c}\end{equation}
generates the equivalence \begin{equation}
(T_{+}-T_{-})_{\alpha_{+},\, T_{c}=0}=(T_{+}-T_{-})_{\alpha_{-},\, T_{c}=T_{h}}\label{eq: reciprocal_d}\end{equation}
\sublaboff{equation}providing an interpretation of the simultaneous
validity of Eqs. (\ref{eq: reciprocal_a},\emph{ b}). The limiting
consideration (\ref{eq: reciprocal_a}) leaves us with two regions:
(I) characterized by $\iota\alpha>0$ of the complex real world $W$
and (IV) of $\iota\alpha<0$ of the negative world $\mathfrak{W}$.
The three phases of matter of solid, liquid and gas of our perception
manifests only in $W$, the negative world not admitting this distinction
is a miscible concentrate in all proportions. The reciprocal implications
(\ref{eq: reciprocal_a}-\emph{d}) at the big-bang degenerate singularity
$\alpha_{+}=+1$ at $t=0$ \citep{Sengupta2010a}, instantaneously
causes the birth of the $(W,\mathfrak{W})$ duality at some \emph{unique
admissible} value of $\alpha$ for $0<T_{c}<T_{h}$ and complexity
criterion $\iota=\alpha$, breaking the equivalence $\alpha_{_{+}}\sim\alpha_{-}$
of Eq. (\ref{eq: recirocal_c}). 

Figure \ref{fig: 2-phase}(c) identifies the complex $W$ on the bifurcation
diagram of the logistic map $\lambda x(1-x)$ that we now turn to.

\subsection{\textsf{Philosophy of ChaNoXity: Both adversaries win, both lose }}

\subsubsection{\textsf{\large The Logistic Map }$\lambda\mathbf{x(1-x)}$\textsf{\large :
An Elementary Nonlinear Qubit}}

A correspondence between the dynamics of the engine-pump system and
the logistic map $\lambda x(1-x)$, with the competitive, backward-direct
iterates $f^{i}(x)$ corresponding to the {}``pump'' $\mathfrak{W}$
and the collaborative, forward-inverse iterates $f^{-i}(x)$ to the
{}``engine'' $W$, is summarized in Table \ref{tab: world_neg-world}.
The two-phase complex region (I), $\lambda\in(3,\lambda_{*})$, $T_{+}\in(T_{c},T_{h})$,
$\iota\in(0,1)$, is the outward manifestation of the tension between
the regions (I), (III) on the one hand and (II), (IV) on the other:
observe from Eq. (\ref{eq: T_pm(a)}) and Fig. \ref{fig: Engine-Pump}
that at the environment $T_{c}=(0,T_{h})$ the two worlds merge at
$\alpha_{\pm}=\pm1$ bifurcating as individual components for $0<T_{c}<T_{h}$.
The logistic map --- and its possible generalizations --- with its
rising and falling branches denoted $\left(\uparrow\right)$ and $\left(\downarrow\right)$,
see Fig. \ref{fig: 2-4-8}, constitutes a perfect example of an elementary
\emph{nonlinear} \emph{qubit, }not represented as a (complex) linear
combination: nonlinear combinations of the branches generate the evolving
structures, as do the computational base $(1\;0)^{\textrm{T}}$ and
$(0\;1)^{\textrm{T}}$ for the linear qubit. This qubit can be prepared
efficiently by its defining nonlinear, non-invertible, functional
representation, made to interact with the environment\emph{ }through
discrete non-unitary time evolutionary iterations, with the final
(homeostatic) equilibrium {}``measured'' and recorded through its
resulting complex structures. 

The effective power law $f(x)=x^{1-\chi}$\citep{Sengupta2006} for\sublabon{equation} 

\begin{eqnarray}
\chi & \!\!\!=\!\!\! & 1-\frac{\ln\left\langle f(x)\right\rangle }{\ln\left\langle x\right\rangle },\qquad0\le\chi\le1,\label{eq: chi_1}\\
\left\langle x\right\rangle  & \!\!\!\triangleq\!\!\! & 2^{N}\overset{\lambda=\lambda_{*}}{\longrightarrow}\infty\label{eq: chi_2}\\
\left\langle f(x)\right\rangle  & \!\!\!\triangleq\!\!\! & 2f_{1}+{\textstyle \sum_{j=1}^{N}\sum_{i=1}^{2^{j-1}}}f_{i,i+2^{j-1}},\; N=1,2,\cdots,\nonumber \\
 & \!\!\!=\!\!\! & \{[(2f_{1}+f_{12})+f_{13}+f_{24}]+f_{15}+f_{26}+f_{37}+f_{48}\}\label{eq: chi_3}\end{eqnarray}
\sublaboff{equation}and the hierarchical levels $(N=1)$, $[N=2]$,
$\{N=3\}$, with $\left\langle x\right\rangle $ the $2^{N}$ microstates
of the basic unstable fixed points resulting from the $N+1$ macrostates
$\{f^{i}\}_{i=0}^{N}$ constituting the net feedback $\left\langle f(x)\right\rangle $,
bestows the complex system with its composite holism. Hence\begin{equation}
\chi_{N}=1-\frac{1}{N\ln2}\,\ln\left[2f_{1}+\sum_{j=1}^{N}\sum_{i=1}^{2^{j-1}}f_{i,i+2^{j-1}}\right]\label{eq: chi_N(c)}\end{equation}
is the measure of chanoxity, where $f_{i}=f^{i}(0.5)$, $f_{i,j}=|f^{i}(0.5)-f^{j}(0.5)|$,
$i<j$, and 

\begin{equation}
\chi=\iota=\alpha,\qquad\lambda\in(3,\lambda_{*}:=3.5699456)\label{eq: iota.chi.alpha}\end{equation}
in Regions (I) and (III) can be taken as the definite assignment of
thermodynamical purview to the dynamics of the logistic map for $\iota\alpha=\chi^{2}$,
$\chi$ being the \emph{measure of chanoxity}, Eq. (\ref{eq: chi_1}).
\begin{table}[!tbh]
\begin{centering}
\renewcommand{\arraystretch}{1.5}\begin{tabular}{c|c|c|c|}
\cline{2-4} 
\cline{2-4} & $\mathnormal{\iota;T;\alpha}$ & $\lambda$; $\chi$ & $x_{\textrm{fp}}$\tabularnewline
\cline{2-4} 
\cline{2-4}\textsc{ } & \cellcolor{Red}$(-\infty,\,\iota_{c}];(-\infty,0];[\infty,0)$ & \cellcolor{Red}$(0,1],(1,2]$; $0$ & \cellcolor{Red}$(\bullet,-),(\circ,-)$\tabularnewline
\cdashline{2-4}\textsc{ } & \multicolumn{3}{c|}{\cellcolor{Red}\textsc{$\iota\alpha<0:$ Multifunctional Simple }$\mathfrak{W}$
\textbf{(IV: Individualistic)}}\tabularnewline
\cline{2-4} 
\cline{2-4}\textsc{ } & \cellcolor{Green}\textsc{ }$(\iota_{c},\,0);(0,T_{c}];(0,-\infty)$ & \cellcolor{Green}\textsc{ }$(2,3)$; $0$ & \cellcolor{Green}\textsc{ }$(\circ,\bullet)$\tabularnewline
\cdashline{2-4}\textsc{ } & \multicolumn{3}{c|}{\cellcolor{Green}\textsc{ $\iota\alpha>0:$ }\textbf{\textsc{Functional
Simple}} $W$\textbf{ (III: Collective)}}\tabularnewline
\cline{2-4} 
\cline{2-4}\textsc{ } & \cellcolor{yellow}$(0,1);(T_{c},T_{h});(\infty,0)$ & \cellcolor{yellow}\textbf{$[3,\lambda_{*})$}; $[0,1)$ & \cellcolor{yellow}\textbf{$\mathbf{\mathnormal{(\circ,\bullet/\circ)}}$}\tabularnewline
\cdashline{2-4}\textsc{ } & \multicolumn{3}{c|}{\cellcolor{yellow}\textbf{\textsc{ $\iota\alpha>0:$ Functional Complex}}\textbf{
$W$ (I: Individualistic~$\oplus$~Collective)}}\tabularnewline
\cline{2-4} 
\cline{2-4}\textsc{ } & \cellcolor{Red}$[1,\infty);[T_{h},\infty);[0,-\infty)$ & $[\lambda_{*},4),\,4$; $\{0,1\}$\cellcolor{Red} & \cellcolor{Red}$(\circ,\circ)$\tabularnewline
\cdashline{2-4}\textsc{ } & \multicolumn{3}{c|}{\cellcolor{Red}\textsc{ $\iota\alpha<0:$ }\textbf{\textsc{Multifunctional
Chaotic}}\textbf{ $\mathfrak{W}$ (II: Individualistic)}}\tabularnewline
\cline{2-4} 
\end{tabular}\bigskip{}

\par\end{centering}

\caption{{\small \label{tab: world_neg-world}Emergence of the {}``Participatory
Universe'', for $0<T_{c}<T_{h}$ in $W$; $\iota_{c}=-T_{c}/(T_{h}-T_{c})$:
putting dynamics and thermodynamics together. Region I of complex
homeostacy in competition and collaboration is the holistic cohabitation
of the opposites individualism and collectivism.}}

\end{table}

\begin{singlespace}
Table \ref{tab: world_neg-world} demonstrates that the dynamics of
the logistic map undergoes a discontinuous transition from the monotonically
increasing $0\le\chi<1$ in $3\le\lambda<\lambda_{*}$ of region (I)
to a disjoint world\textbf{ }at $\chi=0$ in the fully chaotic $\lambda_{*}\le\lambda<4$
of (II), thereby\emph{ reducing chaos to effective linear simplicity.}
Eq. (\ref{eq: iota.chi.alpha}), Fig. \ref{fig: Engine-Pump}(a),
(b) demonstrate that the boundary $\mbox{Multi}_{\parallel}$ between
$W\!:=\mbox{map}$ and $\mathfrak{W\!}:=\mbox{Multi}$ of the chaotic
region II, $\lambda\in[\lambda,4)$, occurs for $\chi=0$, $\iota=\alpha\neq0$
\emph{in equivalence with }(III), for $T_{c}>0$. As $T_{c}\rightarrow0$,
the boundary degenerates a point at $\alpha=1$. 
\end{singlespace}

According to Table \ref{tab: world_neg-world}, $\chi=0$ of (I) and
$\chi=1$ of (II) establishes the one-one correspondences $(\lambda\in(2,3),\,\lambda\in[\lambda_{*},4))$
at the boundary $\mbox{Multi}_{\left\Vert \right.}$, and $(\lambda=\lambda_{*},\,\lambda\ge4)$
at $T_{c}=T_{h}\rightarrow\infty$, Eq. (\ref{eq: reciprocal_a}),
of a boundaryless transition between these complimentary dual worlds.
Hence $T_{c}\ge T_{h}$ is to be interpreted to imply $-\infty<T_{c}\le0$
of negative temperatures that define $\mathfrak{W}$. 

Observe that quantum mechanics comprises the linear boundary between
$W$and $\mathfrak{W}$.

\subsubsection*{\noindent \textsf{Index of Complexity }}

\noindent Equation (\ref{eq: iota_alpha}) for $\iota=\alpha$ leads
to \begin{align}
\iota_{\pm} & =\frac{T_{h}-2T_{c}\pm\sqrt{T_{h}^{2}+4T_{h}T_{c}}}{4T_{h}-2T_{c}}\label{Eqn: iota_pm}\\
 & =\begin{cases}
(0.5,0), & T_{c}=0\\
\pm\frac{1}{2}\left(\sqrt{5}\mp1\right), & T_{c}=T_{h}\end{cases}\nonumber \end{align}
at temperatures $T_{\pm}$ of Eq. (\ref{eq: T_plus(a)}) denoted as
$T_{\bullet}$ and $T_{\circ}$ in Fig. \ref{fig: Engine-Pump}(a)\textsf{.}
The complexity $\sigma$ of a system is expected to depend on both
the irreversibility $\iota$ and the interaction $\alpha$; thus the
definition 

\noindent \renewcommand{\arraystretch}{1.25}\begin{equation}
\sigma_{\pm}\triangleq\frac{-1}{\ln2}\left\{ \!\!\!\begin{array}{l}
{\displaystyle \tilde{\iota}_{-}}[\iota_{+}\ln\iota_{+}+(1-\iota_{+})\ln(1-\iota_{+})]\\
{\displaystyle \iota_{+}}[\tilde{\iota}_{-}\ln\tilde{\iota}_{-}+(1-\tilde{\iota}_{-})\ln(1-\tilde{\iota}_{-})]\end{array}\right.\label{eq: complex1}\end{equation}
with $\tilde{\iota}_{-}=\iota_{-}/\iota_{c}\in[0,1]$, ensures the
expected two-state, logistic-like, non-linear qubit $(\left\uparrow \right\downarrow )$
signature at $T_{+}$ and $T_{-}$. 

Unification of thermodynamic and logistic qubit $(\left\uparrow \right\downarrow )$
dynamics of the self-induced engine-pump system is achieved by extending
(\ref{eq: iota_alpha}) to (\ref{eq: iota.chi.alpha}). This identification
of thermodynamic and dynamic properties in the evolution of a complex
system by associating its dynamical degree $\chi:=1-\frac{\ln\left\langle f(x)\right\rangle }{\ln\left\langle x\right\rangle }$
linearly increasing with $\lambda$ with the thermodynamic competitive-collaboration
$\alpha$, focuses on the distinction between $\chi$ and complexity
$\sigma_{+}$ representing a homeostatic balance between dispersion
and concentration.

\subsubsection{\textsf{\large Quantum Mechanics: A Linear Representation of Chaos}}

\begin{flushright}
\textsl{\large $\bullet$}\textsl{ Bell's inequalities represent,
first of all, an experimental test of the consistency of quantum mechanics.
Many experiments have been performed in order to check Bell's inequalities;
the most famous involved EPR pairs of photons and was performed by
Aspect and co-workers in 1982. This experiment displayed an unambiguous
violation of CHSH inequality and an excellent agreement with quantum
mechanics. More recently, other experiments have come closer to the
requirements of the ideal EPR scheme and again impressive agreement
with the predictions of quantum mechanics has always been found. If,
for the sake of argument, we assume that the present results will
not be contradicted by future experiments with high-efficiency detectors,
we must conclude that}\textsc{ }\textsl{Nature does not support the
EPR point of view}\textsc{.} \textsl{In summary,} \textsl{the world
is not locally realistic.\\}\textsl{\large $\bullet$}\textsl{ These
profound results show us that entanglement is a fundamentally new
resource, beyond the realm of classical physics, and that it is possible
to experimentally manipulate entangled states. A major goal of quantum
information science is to exploit this resource to perform computation
and communication tasks beyond classical capabilities. Violation of
Bell's inequalities is a typical feature of entangled states}\textsc{.
$\quad$\citet{Benenti2004}}\bigskip{}

\par\end{flushright}

\noindent Is quantum mechanics indeed a general theory that applies
to everything from subatomic particles to galaxies as it is generally
believed to be, i.e., is Nature indeed governed by entanglements of
linear superposition in Hilbert space or is it an expression of the
nonlinear holism of emergence, self-organization, and complexity that
we have constructed above? What is clear is that some basic structure
of holistic {}``entanglement'' is involved in the expressions of
Nature; what is not so clear and is the subject of our present concern
is whether this is linear quantum mechanical or nonlinear, self-organizing-emergent,
and complex. 

Composite systems\textsf{\large{} }in QM are described by tensor products
of vector spaces, a natural way of putting linear spaces together
to form larger spaces. If $V$, $W$ are spaces of dimensions $n$,
$m$, $A\!:V_{1}\rightarrow V_{2}$, $B\!:W_{1}\rightarrow W_{2}$
are linear operators, then $C:=\sum_{i}\alpha_{i}A_{i}\otimes B_{i}$
on the $nm$-dimensional linear space $V\otimes W$ defined by $C(\left|v\right\rangle \otimes\left|w\right\rangle )=\sum_{i}\alpha_{i}(A_{i}\left|v\right\rangle \otimes B_{i}\left|w\right\rangle )$,
together with the bi-linearity of tensor products, endows $V\otimes W$
with standard properties of Hilbert spaces inherited from its components,
with the state space of the composite being the tensor product of
the spaces of the components. 

In quantum mechanics, the basic unit of classical information of the
\emph{b}(inary)(dig)\emph{it} of either {}``on $\left|\uparrow\right\rangle $''
of {}``off $\left|\downarrow\right\rangle $'', is replaced by the
\emph{qubit} of a normalized vector in two\emph{-}dimensional complex
Hilbert space spanned by the orthonormal vectors $\left|\uparrow\right\rangle :=(1\;0)^{\textrm{T}},\left|\downarrow\right\rangle :=(0\;1)^{\textrm{T}}$.
The qubit differs from a classical bit in that it can exist either
as $\left|\uparrow\right\rangle $ or as $\left|\downarrow\right\rangle $
or as a superposition $\alpha\left|\uparrow\right\rangle +\beta\left|\downarrow\right\rangle $
$(\left|\alpha\right|^{2}+\left|\beta\right|^{2}=1)$ of both. The
distinguishing feature in the quantum case is a consequence of the
linear superposition principle that allows the quantum system to be
in any of the \textsf{$2^{N}$ }basic states simultaneously, leading
to the non-classical manifestations of interference, non-locality
and entanglement. 

\begin{figure}[!tbh]
\begin{centering}
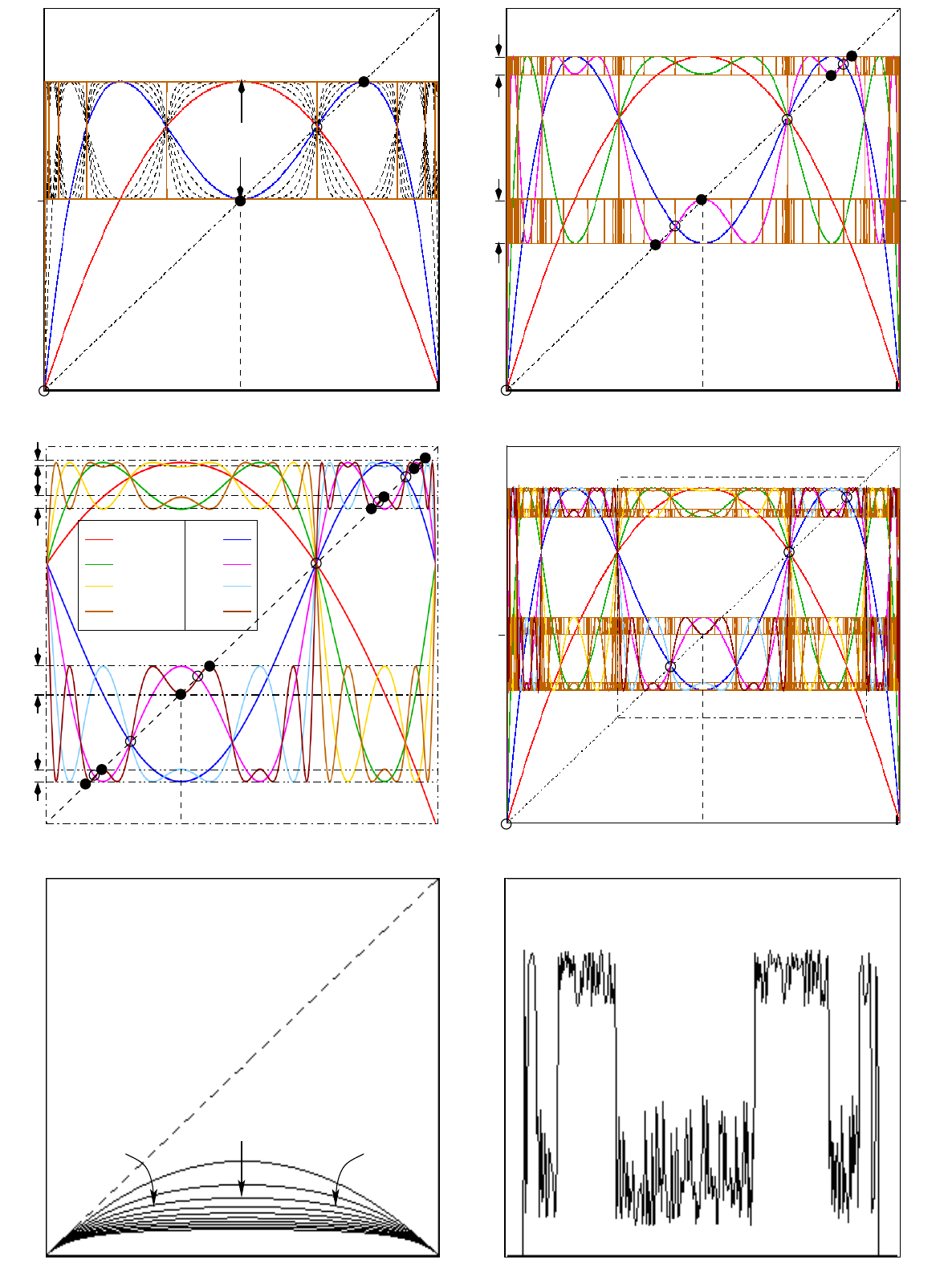
\par\end{centering}

\caption{\label{fig: 2-4-8}Complex entangled holism (a)-(d), generated by
the logistic nonlinear qubit $f(x)=\lambda x(1\lyxmathsym{\textminus}x)$.
The effective nonlinearity $0\le\chi\le1$ of the representation $f(x)=x^{1\text{\textminus}\chi}$
increases with $\lambda$, as the system becomes more holistic with
an larger number of interacting parts of unstable fi{}xed points shown
unfi{}lled, the stable fi{}lled points being the interacting, interdependent,
components of the evolved pattern. The resulting holistic patterns
of one, two, and three hierarchical levels are entangled manifestations
of these observables, none of which can be independently manipulated
outside of the collaborative whole. Direct iterates $f^{i}(x)$, (f),
of increasing entropy, individualism, concentration comprises the
backward gravitational arrow, the inverse iterates $f^{\lyxmathsym{\textminus}i}(x)$,
(e), of decreasing entropy, collectivism and dispersion is its opposite,
with homeostasis of the graphically converged multifunctions comprising
the dynamic equilibrium.}

\end{figure}

Entanglement is the new quantum resource that distinguishes it so
fundamentally from the classical in the sense that with the qubit,
the degeneracy $2^{N}$ of composite entangled states is hugely larger
than the $2N$ possibilities for classical systems. An immediate consequence
of this is that for physically separated and entangled $S$ and $E$
in state $(\left|\left\uparrow \right\downarrow \right\rangle +\left\downarrow \right\uparrow )/\sqrt{2}$
for example, a measurement of $\left|\uparrow\right\rangle $ on $S$
reduces/collapses the entangled state to the separable $\left|\left\uparrow \right\downarrow \right\rangle $
so that a subsequent measurement on $E$ in the same basis always
yields the predictable result $\left|\downarrow\right\rangle $; if
$\left|\downarrow\right\rangle $ occurs in $S$ then $E$ will be
guaranteed to return the corresponding reciprocal value $\left|\uparrow\right\rangle $.
System $\left|E\right\rangle $ has accordingly been altered by local
operations on $\left|S\right\rangle $, with a measurement on the
second qubit always yielding a predictable complimentary result from
measurements on the first. In the linear setting of quantum mechanics,
multipartite systems modeled in $2^{N}$-dimensional tensor products
$\mathcal{H}_{1}\otimes\cdots\otimes\mathcal{H}_{N}$ of 2-dimensional
spin states, correspond to the $2^{N}$ {}``dimensional space''
of unstable fixed points in the evolution of the logistic nonlinear
qubit. This formal equivalence illustrated in Fig. \ref{fig: 2-4-8}
while clearly demonstrating how holism emerges in $2^{N}$-cycle complex
systems for increasing complexity with increasing $\lambda$ --- the
emergent $2^{N}$-cycle are {}``entangled'' in the basic $(\uparrow)$
and $(\downarrow)$ components as the system self-organizes to the
graphically converged multifunctional limits indicated by the brown
lines: the parts surrendering their individuality to the holism of
the periodic cycles also focuses on the significant differences between
complex holism and quantum non-locality.

The converged holistic behaviour of complex {}``entanglement'' reflects
the fact that the subsystems have combined nonlinearly to form an
emergent, self-organized structure of the $2^{1}$, $2^{2}$ and $2^{3}$
cycles in Fig. \ref{fig: 2-4-8}(a), (b) and (c)\textbf{\small{} }that
cannot be decoupled without destroying the entire assembly; contrast
with the quantum entanglement and the notion of partial tracing for
obtaining properties of individual components from the whole. Unlike
the quantum case, the complex evolutions are not linearly superposed
reductionist entities but appear as emergent, self-organized holistic
wholes. In this sense complex holism represents a stronger form of
{}``entanglement'' than Bell's nonlocality: \emph{linear systems
cannot be chaotic, hence complex, and therefore holistic.} While quantum
non-locality is a paradoxical manifestation of linear tensor products,
complex holism is a natural consequence of the nonlinearity of emergence
and self-organization\emph{.} %

Nature uses the $2^{N\rightarrow\infty}$ multiplicities of chaos
as an intermediate step in attaining states that would otherwise be
inaccessible to it. Well-posedness of a system is an obviously inefficient
way of expressing a multitude of possibilities as this requires a
different input for every possible output. The countably many outputs
arising from the non-injectivity of $f$ for a given input is interpreted
to define complexity because\emph{ }in a nonlinear system each of
these possibilities constitute a experimental result in itself that
may not be combined in any definite predetermined manner\emph{.} This
multiplicity of possibilities that have no predetermined combinatorial
property is the basis of the diversity of Nature.

The reduced density matrix plays a key role in \emph{decoherence},
a mechanism by which open quantum systems\emph{ }interact with their
environment leading to spontaneous suppression of interference and
appearance of classicality, involving transition from the quantum
world of superpositions to the definiteness of the classical objectivity.
Partial tracing over the environment of the total density operator
produces an {}``environment selected'' basis in which the reduced
density is diagonal. This irreversible decay of the off-diagonal terms
is the basis of decoherence that effectively bypasses {}``collapse''
of the state on measurement to one of its eigenstates. This derivation
of the classical world from the quantum is to be compared with nonlinearly-induced
emergence of complex patterns through the multifunctional graphical
convergence route of the type in Fig. \ref{fig: 2-4-8}. Multiplicities
inherent in this mode illustrated by the filled circles, liberated
from the strictures of linear superposition and reductionism, allow
interpretation of objectivity and definiteness as in classical probabilistic
systems through a judicious application of the Axiom of Choice: \emph{To
define a choice function is to conduct an experiment. }Because of
the drive toward ininal topology of maximal equivalence classes of
open sets at chaos, the selection by choice function refers to the
analogue of continuous quantum probability of the Bloch sphere rather
than the discrete or randomized classical probability. Non-local entanglement
and interference, the distinguishing features of this distinction,
are more pronounced and pervasive in nonlinear complexity than in
linear \emph{isolated and closed,} quantum systems, with its origins
in the noninvertible, maximal ill-posedness of the dynamics of the
former compared to the bijective, reversible unitary Schrodinger evolution
of the later. This identifying differentiation of quantum non-locality
and complex holism forms the basis of the following inferences.

Unlike in the quantum-classical transition, complex evolving systems
are in a state of homeostasis with the environment with {}``measurement''
providing a record of such interaction; probing holistic systems for
its parts and components is expected to lead to paradoxes and contradictions.
A complex system represents a state of dynamic stasis between the
opposites of bottom-up pump induced synthesis of concentration, disorder,
and emergence, and top-down engine dominated analysis of dissipation,
order, and self-organization, the pump effectively deceiving the Second
Law through entropy reduction and gradient dissipation. While quantum
non-locality is a natural consequence of quantum entanglement that
endows multi-partite systems with definite properties at the expense
of the individual constituents, the effective power law $f(x)=x^{1-\chi}$
of Eqs. (\ref{eq: chi_1}),(\emph{b}),(\emph{c}) and the discussions
of the effective linearity of the chaotic region in Table \ref{tab: world_neg-world},
suggests the integration of quantum mechanics with chanoxity by identifying
$\left\langle x\right\rangle =2^{N}$ of Eq. (\ref{eq: chi_N(c)})
with the dimension of the resulting Hilbert space leading to the conjecture
that \emph{quantum mechanics is an} \emph{effective linear representation}
\emph{$\chi=0$ of the fully chaotic, maximally illposed }$\mbox{Multi}_{\parallel}$\emph{
boundary $\lambda_{*}\le\lambda<4$} that manifests itself --- in
equivalence with $\lambda\in(2,3)$ --- through a bi-directional,
contextually objective, inducement of $W$ in adapting to the Second
Law of Thermodynamics. Quantum Mechanics resides at the interfacial
boundary of $W$ and $\mathfrak{W}$ thereby possessing both the properties
both of functional objectivity of the former and mutifunctional ubiquity
of $\mathfrak{W}$. The opposites of the (pump) preparation of the
state and the subsequent (engine) measurement collaborate to define
the contextual reality of the present\emph{. }This\emph{ }combined
with the axiom of choice allows the inference that \emph{quantum mechanical
{}``collapse'' of the wave function is a linear objectification
of the measurement choice function, }the {}``measurement'' process
allowing the quantum boundary between the dual worlds of Table \ref{tab: world_neg-world}
to interact with the {}``apparatus'' in $W$ to generate the complex
{}``reality'' of the present %
\footnote{\label{fn: Penrose-1}{}``While the linearity of quantum theory's
unitary process gives that theory a particular elegence, it is that
very linearity (or unitarity) which leads directly to the measurement
paradox. Is it so unreasonable to believe that this linearity might
be an approximation to some more precise (but subtle) nonlinearity?
$\cdots$ Einstein's theory explained these deviations, but the new
theory was by no means obtained by tinkering with the old; it involved
a completely radical change in perspective. This it seems to me is
the kind of change in the structure of quantum mechanics that we must
look towards, if we are to obtain the needed nonlinear theory to replace
the present-day conventional quantum theory'' \citep{Penrose2006}. %
}. 

Possibly the most ambitious projected utility of quantum entanglement
and interference is of \emph{quantum} \emph{computers} \citep{Benenti2004}.
Any two-level quantum system --- like the ground and an excited states
of an ion --- that can be prepared, manipulated, and measured in a
controlled way comprises a qubit, a collection of $N$ qubits with
its $2^{N}$ dimensional wave function in a Hilbert space representing
a quantum computer. \emph{Neglecting its coupling with the environment},
the \emph{unitary} (hence invertible) time evolution of the computer
is governed by the Schrodinger equation, with measurements disrupting
this process. A \emph{quantum} \emph{computation} therefore consists
of three basic steps: (i) preparation of the input state, (ii) implementation
of the desired unitary transformation (quantum \emph{gates})\emph{
}acting on this state, and (iii) measurement of the output. In an
ion-trap quantum computer for example, any linear array of ions constrained
within a trap formed by static and oscillating electric fields is
the quantum register. Ions are prepared in a specific qubit state
by a laser pulses, the linear interaction between qubits being moderated
by the collective vibrations of the trapped collection of ions. 

The significant attributes of the programme for quantum computers
in direct conflict with the defining features of chanoxism are the
following. Isolation from the environment, invertible unitary interactions
and the ability to selectively operate on constituent parts of the
entangled state (of {}``Alice'', for example, who {}``shares an
e-bit with Bob'') that in the ultimate analysis depend on the linear
invertibility of unitary evolution, and superposition of quantum states.
As none of these hold in complex holism, \emph{being externally imposed
classical interactions of the quantum system with its environment}
and not self-generated, it can be hypothesized that \emph{holistic
computation, as the source of its linear quantum realization, is unlikely
to be feasible}: unlike linear superpositions, any of the evolved
holistic multifunctional entities in Fig. \ref{fig: 2-4-8}(a), (b),
(c) cannot be decomposed or altered without adversely affecting the
entire pattern with the inevitable consequence of critical instabilities
impeding any serious, non-trivial, quantum computation.

The labeling of the interdependent, interacting, stable fixed points
in Fig. \ref{fig: 2-4-8} follows the following rule. The interval
$[0,1]$ is divided into two equal parts at $\frac{1}{2}$ with 0
corresponding to $\mbox{L}$ and 1 to $\mbox{R}$. At any stage of
the iterative hierarchy generated by the unstable (unfilled) points
with the $f_{i<j}$ shown, the stable points are labeled left to right
according to the prescription of Table \ref{tab: label} shown for
$\left\langle f(x)\right\rangle =\{[(2f_{1}+f_{12})+f_{13}+f_{24}]+f_{15}+f_{26}+f_{37}+f_{48}\}$,
the mean value of $f$ according to Eq. (\ref{eq: chi_3}). This gives
the symbolic representation \sublabon{equation} \begin{eqnarray}
N=1 &  & (0;1)\label{eq: N.eq.1}\\
N=2 &  & [(01,00);(10,11)]\label{eq: N.eq.2}\\
N=3 &  & \{[(011,010),(000,001)];[(101,100),(110,111)]\}\label{eq: N.eq.3}\end{eqnarray}
\sublaboff{equation}

\noindent for the self-organized, emergent levels corresponding to
$N=1,2,3$.

\noindent %
\begin{table}[!tbh]
\begin{centering}
\begin{tabular}{|c||c|c|}
\hline 
$f^{i}(0.5)$ & $\mbox{L}$ of unstable f.p. & $\mbox{R}$ of unstable f.p.\tabularnewline
\hline
\hline 
convex up & 0 & 1\tabularnewline
\hline 
convex down & 1 & 0\tabularnewline
\hline
\end{tabular}\bigskip{}

\par\end{centering}

\caption{\label{tab: label}Rule for symbolic representation of the stable
fi{}xed points of Fig. \ref{fig: 2-4-8} at each hierarchical level.
Using this convention, these can be labeled left to right in (a),
(b), (d) of the fi{}gure as (0; 1), {[}(01, 00); (10, 11){]}, and
\{{[}(011, 010), (000, 001){]}; {[}(101, 100), (110, 111){]}\} respectively.}

\end{table}

As a specific example for $N=2$, the complex {}``entangled'' holistic
pattern of Fig. \ref{fig: 2-4-8}(b) clearly demostrates that the
four components of Eq. (\ref{eq: N.eq.2}) cannot be decoupled into
Bell states, being itself nonlinearly {}``entangled'' rather than
separated. The various operations historically performed on the respective
qubits of the entangled pair to generate dense coding and teleportation
$(N=3)$ for example, are not meaningful on the nonlinear holistic
entities; in fact it is possibly not significant to ascribe any specific
qubit to the individual members of the strings in Eq. (\ref{eq: N.eq.2}).
These suggestive points of departure between linear quantum nonlocality
generated by external operations and nonlinear self-evolved complex
holism calls for a deeper investigation that we hope to perform subsequently. 

Nevertheless, the phenomenal sucess of linear quantum mechanics to
{}``classify and predict the physical world'' begs a proper perspective.
Our hypothesis is that nature operates in accordance with chanoxity
only in its {}``kitchen'' that forever remains beyond our direct
perception; what we do observe physically is only a linearized, presentable,
table-top version of this complexity, through the quantum linear interface
of $W-\mathfrak{W}$. This boundary between the dual worlds, of course,
carries signatures of both, which seems to explain its legendary observational
success.

\subsubsection{\textsf{\large Black Hole and Gravity: The Negative World} \textsf{\large and
its Thermodynamic Legacy}}

\paragraph{\textsf{(I)\label{sub Defining-Example:}A Defining Example: The
$(W,\mathfrak{W})$, }(\textsf{Top-Down, Bottom-Up}), (\textsf{Particle,
''Wave''}) \textsf{Duality}\textcolor{red}{\emph{ }}}

\begin{figure}[!tbh]
\noindent \begin{centering}
\emph{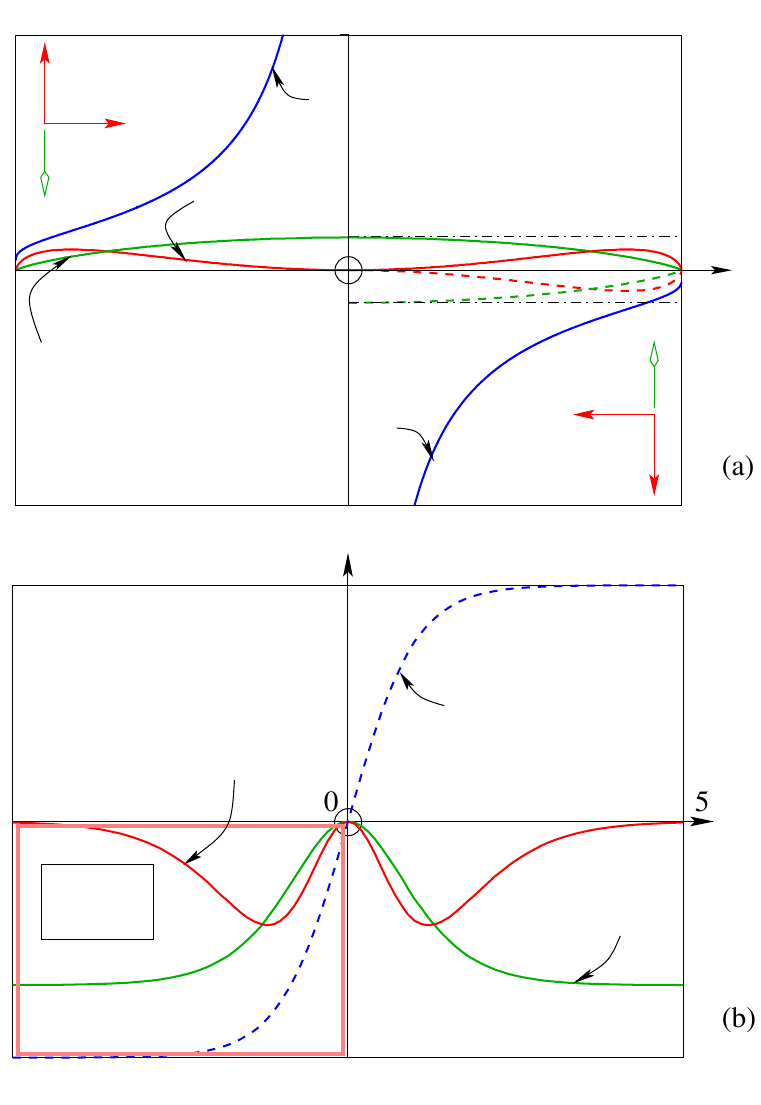}
\par\end{centering}

\begin{spacing}{0.85}
\noindent \caption{{\small \label{fig: S-C-T}(a) Normalized ($N=\mu=B=1=k_{\textrm{B}}$)
negative temperature, specific heat and entropy for a $(\left\uparrow \right\downarrow )$
system. (b) The virial negative world $\mathfrak{W}$ of negative
specific heat and entropy. }}
\end{spacing}

\end{figure}

\noindent Consider the two-state paramagnet of $N$ elementary $(\left\uparrow \right\downarrow )$
dipoles with magnetic field $B$ in the $+z$-direction. Then with
$\mu$ the magnetic moment and\emph{\begin{align}
E & =-N\mu B\tanh\left(\frac{\mu B}{kT}\right)\label{eq: E-T}\end{align}
}the total energy of the system, the corresponding expressions for
temperature, entropy, and specific heat plotted in Fig. \ref{fig: S-C-T}
displays the typical unimodal, two-state, $(\left\uparrow \right\downarrow )$
character of $S$ that admits the following interpretation. In the
normalized ground state energy $E=-1$ of all spins along the $B$-axis,
the number of microstates is $1$ and the entropy $0$. As energy
is added to the system some of the spins flip in the opposite direction
till at $E=0$ the distribution of the $\uparrow$ and $\downarrow$
configurations exactly balance, and the entropy attains the maximum
of $\ln2$. On increasing $E$ further, the spins tend to align against
the applied field till at $E=1$ the entropy is again zero with all
spins opposing the field for a single microstate and \emph{negative}
$T$. The mainstream \citep{Baierlein1999} relying on uni-directionality,
holds that \emph{{}``all negative temperatures are hotter than positive
temperatures.} Moreover, the coldest temperature is just above $0$K
on the positive side, and the hottest temperatures are just below
$0$K on the negative side''\emph{.} This view of\emph{ $\mathbb{R}_{-}$
}of negative temperatures as a set of\emph{ }{}``super positives''
is to be compared with what it really should be: the negative world
$\mathfrak{W}$. %

Bidirectionally, quite a different interpretation based on the \emph{virial
theorem} relates the average kinetic energy of a system to its average
potential energy, $2\mathscr{T}=-\sum_{k=1}^{N}\mathbf{F}_{k}\cdot\mathbf{r}_{k}$
where $\mathbf{F}_{k}$ is the force on the $k^{\textrm{th}}$ particle
at $\mathbf{r}_{k}$. For power law potentials $\mathscr{V}(r)=cr^{n}$,
the theorem takes the simple form \begin{align}
\mathscr{T} & =\frac{n\mathscr{V}}{2},\quad(\mbox{Virial Theorem})\label{eq: virial}\end{align}
which for attractive $c=-1$ gravitational systems $n=-1$, reduces
to \begin{equation}
\mathscr{T}+E=0,\qquad E\mbox{ total energy}.\label{eq: virial1}\end{equation}

\noindent Since the potential energy decreases ($d\mathscr{V}/dr>0$)
faster than increase in kinetic energy ($d\mathscr{T}/dr<0$) the
total energy $E$ decreases with bounding radius $dE/dr>0$. With
$\mathscr{T}\sim NT$, it follows \sublabon{equation} \begin{equation}
\mathscr{V}\sim-NT\Rightarrow T\propto{\displaystyle \frac{1}{r}},\qquad\frac{dT}{dr}<0,\label{eq: T-r}\end{equation}
and the gas gets hotter with shrinking radius. Hence \begin{equation}
C_{V}\triangleq\frac{dE}{dT}=\left(\frac{dE}{dr}\right)\left(\frac{dr}{dT}\right)<0\label{eq: C_V}\end{equation}
and, from $dS(r)\triangleq dE/T=-dT/T$, the entropy \begin{equation}
S(r)\propto\ln r<0,\qquad r\rightarrow0,\label{eq: S-ln(r)}\end{equation}
\sublaboff{equation}becomes negative as $r\rightarrow0$ gravitationally.\emph{
}It is clear from Eq. (\ref{eq: C_V}) and Fig. \ref{fig: S-C-T}
that only $E>0$ with its direction reversed\emph{,} $1<E<0$, can
qualify for the gravitational region of negative specific heat and
entropy. Hence \emph{\[
E(r)=\tanh(r),\qquad-\infty<r<0,\,1<E<0\]
}applies only to the gravity-induced region of negative $T$ and therefore
of negative\emph{ $r$}, and

\noindent \sublabon{equation}\begin{eqnarray}
C_{V}(r) & \triangleq & -r^{2}\mbox{sech}^{2}(r)\le0\label{eq: C_V-S}\\
 & = & -r^{2}+r^{4}-\frac{2}{3}\, r^{6}+\frac{17}{45}\, r^{8}-\cdots,\label{eq: C_V-S1}\\
S(r) & \triangleq & \int\frac{dE}{T}=\int C_{V}\left(\frac{dT}{dr}\right)\left(\frac{dr}{T}\right)=\int r\mbox{sech}^{2}(r)\, dr\nonumber \\
 & = & \ln\cosh(r)-r\tanh(r)\overset{r\rightarrow\infty}{\longrightarrow}-\ln2\le0\label{eq: C_V-S2}\\
 & = & -\frac{1}{2}\, r^{2}+\frac{1}{4}\, r^{4}-\frac{1}{9}\, r^{6}+\frac{17}{360}\, r^{8}-\cdots\label{eq: C_V-S3}\end{eqnarray}
\sublaboff{equation}are both negative and even functions of $r$,
as predicted by the arguments above. In the gravitationally collapsed
region, therefore, entropy is proportional to $r^{2}$ (surface) \emph{rather
than to $r^{3}$ }(volume)\emph{ at small $r$ --- a characteristic
feature of the black hole.} This most noteworthy manifestation of
viriality in the dynamics of a $(\left\uparrow \right\downarrow )$
system, of the natural appearance of negative $r$, can be taken as
a confirmation of the existence of a negative, gravitationally collapsed
world, that in fact constitutes a black hole\emph{.} In this negative
multifunctional dual $\mathfrak{W}$, where {}``anti-second law''%
\footnote{All qualifications are with respect to $W.$%
} requires heat to flow spontaneously from lower to higher temperatures
with positive temperature gradient along increasing temperatures,
the engine and pump exchange their roles \emph{with disordering compresssion
of the system by the environment --- rather than entropy-decreasing
expansion against it as in $W$ --- being the natural direction in
$\mathfrak{W}$}. For an observer in $W$ heat flows from higher \emph{negative
}temperatures to lower\emph{ negative }temperature\emph{.}%
\footnote{Refer Appendix (A) for some additional considerations.%
}

The opposing arrow of $\mathfrak{W}$ translated to $W$, generate
the full curves of Fig. \ref{fig: S-C-T}; hence the entropy, specific
heat and temperature are all positive as seen from $\mathfrak{W}$
$(1<E<0)$ dashed in the figure. In this framework, entropy increases
with energy in $\mathfrak{W}$, rather than negative temperatures
acting {}``as if they are higher than positive temperatures'': the
temperature increases to infinity in $\mathfrak{W}$ with $E\rightarrow0_{+}$
as it does in $W$ for $E\rightarrow0_{-}$ with the \emph{toroidal}
\emph{interconnection} between the complimentary dual worlds through
the equivalences at $E=\pm1$ and $T=\mp\infty$ allowing them to
collaborate competitively as realized by the full curves. The maximum
entropy of $\pm\ln2$ occurs at $E=0$ and the minimum at $E=\pm1$
when all spins are aligned unidirectionally in single microstates.
\emph{This manifestation of $\mathfrak{W}$ in $W$ produces the characteristic
two-state $(\left\uparrow \right\downarrow )$ signature of complexity
and holism through the induced contractive manifestation of gravity}%
\footnote{In GR, gravity is a manifestation of the curvature of spacetime geometry.%
}

The pathological $T_{h}\le T_{c}$ of (II) in the fully-chaotic region
$\lambda\ge\lambda_{*}$, Fig. \ref{fig: Engine-Pump} and Table \ref{tab: world_neg-world}
where no complex patterns are possible, can now be understood with
reference to Eqs. (\ref{eq: reciprocal_a}-\emph{d}) iff $T_{h}=\infty$
when (II) and (IV) merge in the single region of negative temperatures
with its own {}``negative'' dynamics in relation to $W$. Reciprocally
at $T_{c}=0$, region (III) vanishes and with $T_{h}=\infty$ leads
to the two surviving $\alpha\ge0$ portions of Table \ref{tab: world_neg-world},
one for $\iota\alpha<0$ of the multifunctional (IV) of $\mathfrak{W}$
and the other functional $\iota\alpha>0$ of $W$ (I). Since matter
is born only in $W$ as a gravitational materialization of the miscible
mixture $\mathfrak{W}$, the boundary $\mbox{Multi}_{\parallel}(X)$
between the two worlds at $\chi=0$, $\lambda\in[\lambda_{*},4)\sim\lambda\in(2,3)$
is an expression of functional-particle, maximally-multifunctional-{}``wave''
duality that is inaccessible from $W$ because the equivalence at
$E=\pm1$ generates a passage between these antagonistic domains.

\paragraph{\noindent \textsf{(II) Schwarzschild-de Sitter Metric: The Negative
World $\mathfrak{W}$. }}

\begin{figure}[!tbh]
\noindent \begin{centering}
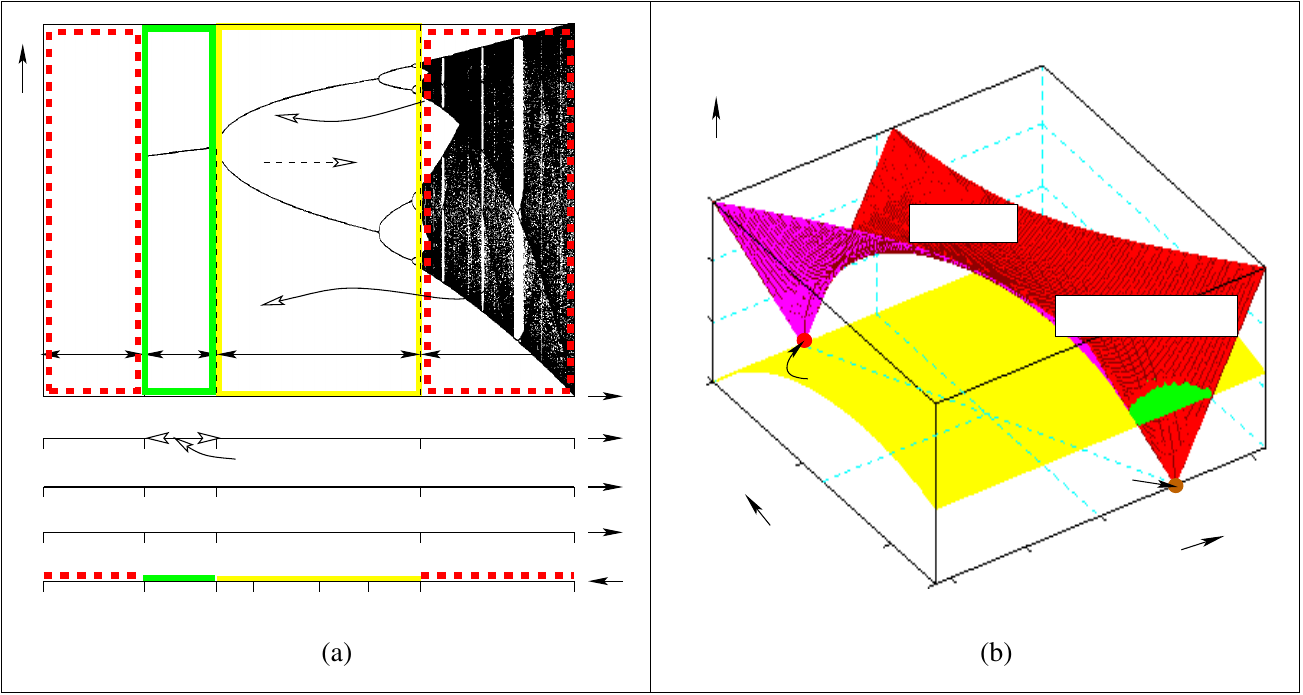
\par\end{centering}

\caption{(a) {\small \label{fig: bifur-phase}Bifurcation profile of the Universe:
Integration of the dynamical and thermodynamical perspectives. The
complex phase (I) is a mixture of concentration and dispersion. $r_{M}={\displaystyle \frac{2GM}{c^{2}},}\, r_{\Lambda}={\displaystyle \sqrt{\frac{3}{\Lambda}}}$
are the Schwarzschild and cosmological radii respectively with $\Lambda=1.3793\times10^{-52}\mbox{ m}^{-2}$,
$G=6.6742\times10^{-11}\mbox{ m}^{3}/\mbox{kg-s}^{2}$. $\iota_{c}=-T_{c}/(T_{h}-T_{c})$.}}

\end{figure}

\noindent The (self-induced) engine-pump system that forms the basis
of our approach has a relativistic analogue in the Schwarzschild-de
Sitter metric\sublabon{equation}\begin{equation}
ds^{2}=-f_{\textrm{SdS}}dt^{2}+f_{\textrm{SdS}}^{-1}dr^{2}+r^{2}d\Omega^{2}\label{eq:Sch-deS}\end{equation}
where \begin{equation}
f_{\textrm{SdS}}=1-\left(\frac{2GM}{c^{2}}\right)\frac{1}{r}-\left(\frac{\Lambda}{3}\right)r^{2},\quad M>0,\,\Lambda>0\label{eq:Sch-deS1}\end{equation}
\sublaboff{equation}for an expansive cosmological opposition to gravitational
compression. The zeros of $f_{\textrm{SdS}}$ give the limiting values
\renewcommand{\arraystretch}{2.25} \begin{equation}
r_{M,\Lambda}=\left\{ \!\!\!\begin{array}{ll}
{\displaystyle \frac{2GM}{c^{2}},} & \quad\Lambda=0:\,\mbox{only gravitational contraction}\\
{\displaystyle \sqrt{\frac{3}{\Lambda}},} & \quad M=0:\,\mbox{only cosmological expansion}\end{array}\right.\label{eq: r_Sch-Cosmo1}\end{equation}
of the Schwarzschild and cosmological radii $r_{M}$ and $r_{\Lambda}$,
respectively. Equation $f_{\textrm{SdS}}=0$ solved for $M>0$ and
$\Lambda>0$\sublabon{equation}\begin{eqnarray}
M(\Lambda) & \!\!\!=\!\!\! & \frac{c^{2}}{2G}\, r\left(1-\frac{\Lambda}{3}\, r^{2}\right)>0\Rightarrow r<\sqrt{\frac{3}{\Lambda}}\triangleq r_{\Lambda}\label{eq: M_Lambda}\\
\Lambda(M) & \!\!\!=\!\!\! & \frac{3}{r^{2}}\left(1-\frac{2GM}{c^{2}}\,\frac{1}{r}\right)>0\Rightarrow r>\frac{2GM}{c^{2}}\triangleq r_{M};\label{eq: M_Lambda-1}\end{eqnarray}

\noindent by imposing convenient bounds on the dynamics of the gravitational-cosmological
tension, implies that \begin{equation}
r_{M}<r<r_{\Lambda},\quad M>0,\Lambda>0\label{eq: M_Lambda-2}\end{equation}
corresponds to region (I) of the complex holistic world, Fig. \ref{fig: bifur-phase}.
Reciprocally, \begin{equation}
r_{\Lambda}<r<r_{M},\quad M<0,\,\Lambda<0\label{eq: M_Lambda-3}\end{equation}
\sublaboff{equation}is an indicator of the negative world $\mathfrak{W}$
of negative temperature, Fig. \ref{fig: bifur-phase} being a detailed
representation of this equivalence: for $\Lambda=1.38\times10^{-52}\,\mbox{m}^{2}$,
the cosmological horizon $r_{\Lambda}\sim1.47\times10^{26}\mbox{\,\ m}$
is of the size of the observable universe, while with $M=3.8\times10^{52}\mbox{ kg}$
as the mass of the observable universe, the gravitational Schwarzschild
radius (event horizon) $r_{M}\sim6.0\times10^{25}\mbox{\,\ m}<r_{\Lambda}$.

The tilting of light cones for an expansionless $\Lambda=0$ universe
at the removable singularity of the Schwarzschild event horizon $r_{M}$
that prevents future directed timelike or null worldlines reaching
$r>r_{M}$ from the interior, corresponds to the passage to $W$ through
the $T_{h}=\infty$, $\alpha_{\lyxmathsym{\textminus}}=\lyxmathsym{\textminus}1$,
Black Hole critical point, Fig \ref{fig: bifur-phase}(a), with $r_{M}>r\rightarrow0$
collapsing gravitationally. At the other extreme of $M=0$ as $r_{\Lambda}<r\rightarrow\infty$
expands without limit, $T\rightarrow0$, denoting crossover to $W$
through the $T_{c}=0$, $\alpha_{+}=1$, Big Bang triple point. The
gravitationally collapsed expression Eq. (\ref{eq: T-r})\sublabon{equation}
\begin{equation}
T=\left(\frac{\hslash c}{k_{\textrm{B}}}\right)\frac{1}{r},\label{eq: Hawking_T}\end{equation}
is the Hawking temperature%
\footnote{With $T$ negative in $\mathfrak{W}$, $r$ must also be so.%
}, while the entropy Eq. (\ref{eq: C_V-S2}), that \emph{for small}
$r$ reduces to \begin{equation}
S=-\left(\frac{c^{3}}{\hslash G}\right)r^{2},\label{eq: Hawking_S}\end{equation}
\sublaboff{equation}is the negative of Hawking-Bekenstein entropy,
but has the usual volumetric dependence of $\ln2$ at full dispersion.
The fully chaotic region $\lambda_{*}\le\lambda\le4$ of the boundary
$\mbox{Multi}_{\parallel}$ between $W$ and $\mathfrak{W}$, in equvalence
with $2<\lambda<3$ of region (III) (see Fig. \ref{fig: bifur-phase}),
is the {}``skin'' of the gravitational black hole $\mathfrak{W}$
at the physical singularity $r=0$, identified as $M<0,\,\lambda<0$,
in Eq. (\ref{eq: M_Lambda-3}). \emph{Gravity as we experience it
in $W$, is the legacy of the thermodynamic arrow of $\mathfrak{W}$,
}see Fig.\emph{ }\ref{fig: S-C-T}\emph{. }The subsequent $T_{c}>0$,
$T_{h}<\infty$ exposition of Fig. \ref{fig: Engine-Pump} is responsible
for the complex structures and patterns of Nature\emph{.} %

The only real root of the cubic $f_{\textrm{{SdS}}}(r)=0$, \sublabon{equation}

\begin{eqnarray}
-r_{\textrm{SdS}} & \!\!\!=\!\!\! & \frac{c^{2}}{\left(3GM\Lambda^{2}c^{4}+\sigma\right)^{1/3}}+\frac{\left(3GM\Lambda^{2}c^{4}+\sigma\right)^{1/3}}{c^{2}\Lambda},\label{eq: SdS roots_negative}\\
r_{\pm} & \!\!\!=\!\!\! & -\frac{r_{\textrm{SdS}}}{2}\pm i\sqrt{3}\left[\frac{c^{2}}{2\left(3GM\Lambda^{2}c^{4}+\sigma\right)^{1/3}}-\frac{\left(3GM\Lambda^{2}c^{4}+\sigma\right)^{1/3}}{2c^{2}\Lambda}\right]\label{eq: SdS roots_complex}\end{eqnarray}
\sublaboff{equation}with $\sigma:=\Lambda c^{4}\sqrt{9G^{2}M^{2}\Lambda^{2}-\Lambda c^{4}}$,
has the significant property of being \emph{negative}; the remaining
complex pair merge to a single real value for\sublabon{equation}
\begin{equation}
\sigma=0\Rightarrow9G^{2}M^{2}\Lambda=c^{4}\label{eq: Nariai-1}\end{equation}
at the Nariai radius \begin{eqnarray}
\rho_{N}=-\frac{r_{\textrm{{SdS}}}}{2} & = & \frac{1}{\sqrt{\Lambda}}\nonumber \\
 & = & \frac{3GM}{c^{2}}\label{eq: Nariai-2}\end{eqnarray}
permitting $f_{\textrm{{SdS}}}$ to be factored as $f_{\textrm{{SdS}}}(r)=-(\Lambda/3r)(r-\rho_{\textrm{N}})^{2}(r+2\rho_{\textrm{N}})$.
In general $f_{\textrm{{SdS}}}$ has two positive real zeros $\rho_{M}$
and $\rho_{\Lambda}$ satisfying \begin{equation}
\begin{array}{r}
r_{M}<\rho_{M}<\rho_{N}={\displaystyle \frac{3GM}{c^{2}}}<\rho_{\Lambda}<r_{\Lambda},\\
\mbox{iff}\quad0<{\displaystyle \frac{3GM}{c^{2}}}\sqrt{\Lambda}<1\end{array}\label{eq: Nariari-3}\end{equation}
\sublaboff{equation}with $\rho_{M}$ monotonically \emph{increasing}
and $\rho_{\Lambda}$ monotonically \emph{decreasing} to the common
value of $\rho_{\textrm{N}}$ as $\Lambda\rightarrow c^{4}/9G^{2}M^{2}$.
Specifically requiring \begin{align*}
\left(\frac{3G}{c^{2}}\right)M\sqrt{\Lambda} & >1\\
 & \Rightarrow M\gtrsim3.8\times10^{52}\mbox{ kg}\end{align*}
as we do, prescribes $M$ of magnitude of the mass of the observable
universe. 

While bypassing the significance of the negative root, the Nariai
solution by equating the cosmological and mass horizons $\rho_{\Lambda}=\rho_{M}$
at the double root, deny the emergent feedback system of the two competitors
of expansion ($\Lambda)$ and compression ($M$) the privilege of
collaborating with each other. Figure \ref{fig: bifur-phase} and
our approach by not insisting on this essentially unique reductionism,
illustrate how gravity effectively moderates the Second Law dictate
of \emph{death }by allowing the system to {}``continually draw from
its environment \emph{negative entropy}''. The opposing effects at
the degenerate singularities $\alpha_{+}=1$ and $\alpha_{-}=\lyxmathsym{\textminus}1$
must be allowed to interact in order to generate the complex structure
for $3GM\Lambda/c^{2}>1$ leading to a negative $\mathfrak{W}$ with
negative $M$. Recall from Fig. \ref{fig: bifur-phase}(a) and Table
\ref{tab: world_neg-world} that component (II) of $\mathfrak{W}$
corresponds to the extreme nonlinearity $\lambda\ge\lambda_{*}$:
this is the region where density of matter and curvature of spacetime
are \textquotedblleft{}so infi{}nitely strong that even light cannot
escape\textquotedblright{} This regularizes the singularity through
collaborative competition of gravitational collapse and de Sitter
expansion in $W$: mutual support of the two opposites generates the
complex holistic structures of Fig. \ref{fig: bifur-phase}%
\footnote{{}``Matter tells spacetime how to bend and spacetime tells matter
how to move'' --- J. A. Wheeler.%
}. The negative real root of Eq. (\ref{eq: SdS roots_negative}) adds
additional justification of the negative world through negative $M$;
observe however that $\Lambda$ is not affected by this negativity
of $r$, Eqs. (\ref{eq: M_Lambda}, \ref{eq: M_Lambda-1}). 

What is the significance of the negative root (\ref{eq: SdS roots_negative})
of negative mass $M$? Our assertion in Sec. \ref{sub: Complexity:-A-Two-Phase}
that the three phases of matter are born only in $W$ at $t=0$ and
have no meaning in $\mathfrak{W}$ is supported by this distribution
of the zeros, $\mathfrak{W}$ being characterized fully by just the
vacuum energy $\Lambda$. $\mathfrak{W}$ induces in $W$ two simultaneous
effects (recall Figs. \ref{fig: direct-inv} and \ref{fig: S-C-T}):
its concentrative, individualistic \textquotedblleft{}capital\textquotedblright{}-ist
arrow of compression induces the expansive collaborative \textquotedblleft{}culture\textquotedblright{}-d
arrow of $W$ with its own dispersion inducing the gravitational attraction
in $W$. This is how Nature\textquoteright{}s holism operates through
unipolar gravity, with the concentration in $\mathfrak{W}$ completing
its bipolarity. Gravity is uniquely distinct from other known interactions
as it straddles $(W,\mathfrak{W})$ in establishing itself, the other
known forms reside within $W$ itself.%
\footnote{\label{fn: Penrose-2}{}``Gravity seems to have a very special status,
different from that of any other field. Rather than sharing in the
thermalization that in the early universe applies to all other fields,
gravity remained aloof, its degrees of freedom lying in wait, so that
the second law could come into play as these begin to be taken up.
Gravity just seems to have been different. However one looks at it,
it is hard to avoid the conclusion that in those circumstances where
quantum and gravitational effects must both come together, gravity
just behaves differently from other fields. For whatever reason, Nature
has imposed a gross temporal asymmetry on the behaviour of gravity
in such circumstances.'' \citep{Penrose2006}%
} It is this unique expression of the maximal multifunctional nonlinearity
of $\mathfrak{W}$ in the functional reality of $W$ that is responsible
for the inducement of {}``neg-entropy'' effects necessary for the
sustenance of life.

\section{\textsf{\large Conclusion: }\textsf{\emph{\large Reality }}\textsf{\large is
not}\textsf{\emph{\large{} Flat }}}

In his remarkable explorations along \emph{The Road to Reality}, Roger
Penrose\textsf{\emph{\large }}%
\footnote{\textsf{\emph{\large \label{fn: Penrose-3}}}{}``The\emph{ }usual
perspective with regard to the proposed marriage betwen these two
theories is that one of them, namely general relativity, must submit
itself to the will of the other. $\cdots$ Indeed the very name 'quantum
gravity' that is normally assigned to the proposed union, carries
the implicit connotation that it is a standard \emph{quantum} (field)
theory that is sought. Yet I would claim that there is observational
evidence that Nature's view of this union is very different from this!
Her design for this union must be what, in our eyes, would be a distinctly
non-standard one, and that an objective state reduction must be one
of its important features.'' \citep{Penrose2006}%
}\textsf{\emph{\large{} }}repeatedly stresses his conviction of {}``powerful
positive reasons to believe that the laws of present-day quantum mechanics
are in need of a fundamental (though presumably subtle) change'',
basing his arguments on the {}``distinctly odd type of way for a
Universe to behave'' in the reversible unitarity of Schrodinger evolution
\textbf{U} being inconsistently paired with irreversible state reduction
\textbf{R. }This leads him to posit that {}``perhaps there is a more
general mathematical equation, or evolution principle, which has both
\textbf{U} and \textbf{R }as limiting approximations'', see footnote
\ref{fn: Penrose-1}. In fact, {}``a gross time-asymmetry (is) a
necessary feature of Nature's quantum-gravity union'': gravity {}``just
behaves differently from other fields''. Specifically, {}``there
is some connection between $\mathbf{R}$ and the Second Law'', with
quantum state reduction being an \emph{objectively real process} arising
from the difference of gravitational self-energy\emph{ $E_{G}$}%
\footnote{\emph{Gravitational self-energy} in a mass distribution is the amount
of (binding) energy gained in assembling the mass from point masses
dispersed at infinity.%
} between different space-time geometries of the quantum states in
superposition. Thus, all observable manifestations in Nature are interpreted
to be \emph{always} gravity induced, quantum superpositions decaying
into one or the other state. 

This philosophical stance is operationally consistent with the foundations
of our theory, recall Sec. \ref{sub Defining-Example:} in particular,
the details being however, conspicuously different. The homeostasy
of top-down-engine and bottom-up-pump endows the state of dynamical
equilibrium with the distinctive characteristic of competitively cohabitating
opposites (Eq. \ref{eq: chanoxity}) in its continual search for life
and order. The reality of the natural world of \emph{not }being in
a {}``flat'' \citep{Friedman2007} state of dispersive maximum entropy
is infact the quest of open systems to stay alive by temporarily impeding
this eventuality through self-organized competitive homeostasis. Hierarchical
top-down-bottom-up complex holism does not support {}``flatness'';
because of its antithetical stance toward self-organization and emergence:
such a world is essentially a dead world. The survival of open living
systems lies in its successfully guarding against this contingency
through the expression of gravity as a realization of the multifunctional
{}``quantum'' $\mathfrak{W}$ on the materially tangible $W$. 

A socially significant remarkable example of this competitive collaboration
is the open source/free software dialectics, developed essentially
by an independent, dispersed community of individuals. Wikipedia as
an exceptional phenomenon of this collaboration, along with Linux
the computer operating system, are noteworthy manifestations of the
power and reality of self-organizing emergent systems. How are these
bottom-up community expressions of {}``peer-reviewed science'' ---
with bugs, security holes, and deviations from standards having to
pass through peer-review evaluation of the system (author) in dynamic
equilibrium of competitive collaboration with the reviewing environment
--- able to {}``outperform a stupendously rich company that can afford
to employ very smart people and give them all the resources they need?
Here is a posible answer: Complexity. Open source is a way of building
complex things'' \citep{Naughton2006}. Note also that {}``the world's
biggest computer company (IBM) decided that its enginners could not
best the work of an ad-hoc open-source collection of geeks (Apache
Web server), so they threw out their own technology and decided to
go with the geeks!'' \citep{Friedman2007}. 

Which brings us to the main issue: Building anything, open-source
or otherwise, requires investment of resources, financial and human.
While the human incentive of open-sourcing for personal recognition
through peer-review is a major deciding factor for the individual
component, {}``collaborating for free in the open-source manner (as)
the best way to assemble the best brains for the job'' guarantees
the collective ingredient needed for emergence of these complex systems
that are far beyond the capacity of any single organization to handle.
The blended model of revenue generation followed by most of the major
open source groups contributes to the financial assets required for
the self-generation of the backward pump as operationally viable,
with the dispersive engine of a readily available market completing
the engine-pump paradigm of chanoxity; \emph{economics infact is about
collectivism to inhibit human selfish individualism and promote evolution
to a state of sustainable homeostatic, collective and societal holism.}
The (social) unit {}``may be the individual or a collective of individuals.
If it is a collective, could its behaviour be deduced from the sum
of the behaviour of its components? Or could its behaviour be governed
by other things than the sum of its components?'' Unlike other customs
in the analysis of social phenomena, the through and through individualistic
character of neoclassical economics based almost entirely on the analysis
of the behaviour of a single individual and his interaction with others
{}``begins and ends with the individual, and sadly, there is barely
any role to anything which is a reflection of the collective. $\cdots$
From the utility maximizing behaviour of individuals we derived the
demand; from the profit maximizing behaviour of firms we derived the
supply. The opposition of forces here is quite clear and well depicted
by the demand and supply analysis (founded on Newtonian mechanics).
Market is where the conflicting forces meet, and the most basic question
is what might influence the outcome of an encounter between a consumer
and a seller?'' \citep{Witztum2005} Further insight into these economic
considerations are considered in Appendix (B). 

The science of collective holism is specifically addressed to issues
such as these leading to an understanding of their true perspective.
\bigskip{}

\begin{center}
\textsf{\textbf{\large APPENDIX}}
\par\end{center}{\large \par}

\begin{center}
\textsf{\textbf{\large (A): Gravity and Entropy}}
\par\end{center}{\large \par}

\begin{figure}[!tbh]
\noindent \begin{centering}
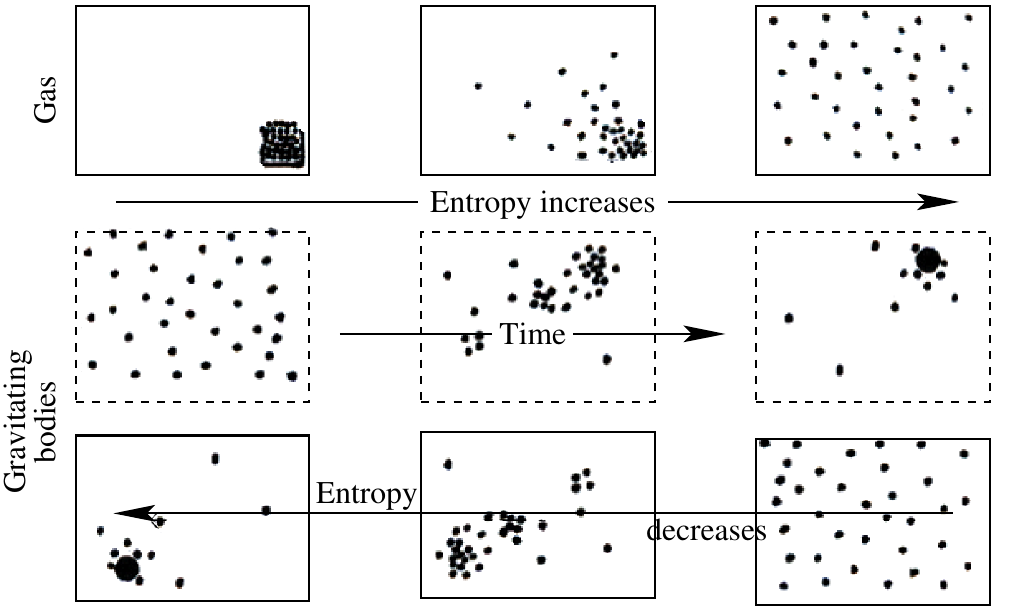$\qquad\quad$
\par\end{centering}

\caption{{\small \label{fig: Penrose}Increasing entropy with increasing time.
(a) For gas in a box, initially tucked in one corner, entropy increases
as the gas spreads throughout the box, finally reaching the uniform
state of thermal equilibrium. (b) With gravity, things tend to be
the other way about. An initial uniformly spread system of gravitating
bodies represent a relatively low entropy, and clumping tends to occur
as the entropy increases. From \citet{Penrose2006}, p. 707.}}

\end{figure}

\noindent Figure \ref{fig: Penrose} adapted from \citet{Penrose2006},
with the accompanying caption reproduced, is a vivid illustration
of the special property of long range unipolar gravity%
\footnote{\label{fn: Penrose-4}{}``Whereas with a gas, the maximum entropy
of thermal equilibrium has the gas uniformly spread throughout the
region in question, with large gravitating bodies maximum entropy
is achieved when all the mass is concentrated in one place --- in
the black hole.'' \citep{Penrose2006}%
}, and further supports our arguments against considering negatives
as {}``super positive''. Panels (a) and (b) are from \citep{Penrose2006}
with the identification of (b) added. Recalling Figs. \ref{fig: direct-inverse},
\ref{fig: direct-inv}, (a) and (c) represent the engine-pump duality
expressed in Fig. \ref{fig: S-C-T}(a).%

\newpage{}

\begin{center}
\textsf{\textbf{\large (B): The Economic Turmoil: }}\textsf{\textbf{\emph{\large Creative
Destruction}}}\textsf{\textbf{\large{} of Economic Holism}}
\par\end{center}{\large \par}

Modern individualistic, neo-classical Western economics, is a static
Newtonian equilibrium theory, where supply by the firm equals the
demand of the consumer. Linear stablity is central to this variant
of economic thinking that has come under severe strain in recent times,
\emph{Economics Needs a Scientific Revolution}\citep{Bouchaud2008}\emph{,
The Eonomy Needs Agent-Based Modelling }\citep{Farmer2009}\emph{,
Meltdown Modelling }\citep{Buchanan2009} reflecting some of the manifestations
of this disillusionment. The linear mathematics of neoclassicalism
is founded in calculus with maximization and contraint-based optimization
being the ground rules, see \citep{Hands2004} for example, that {}``Western
economics became obsessed with'' \citep{Rothschild1995}. These Marshallian
linear static models seeking to maximize utility for the consumer
and profit for firms, as epitomized in Pareto optimality%
\footnote{\textbf{\label{fn: Pareto}Pareto Effi{}ciency.} Given a set of alternative
allocations for a collective of individuals, a change from one allocation
to another that can make at least one individual better off without
making any other worse, is called a Pareto improvement. An allocation
is Pareto optimal when no further Pareto improvements can be made.
The collectivism of Pareto effi{}ciency can be expressed as 

\textbullet{} \emph{A Pareto effi{}cient situation is one in which
any change to make someone better off is impossible without making
somebody else worse off.}%
}, Nash equilibrium%
\footnote{\textbf{\label{fn: Nash}Nash Equilibrium.} Let $(S,f)$ be a game
with $n$ players, where $S$ is the strategy set for player $i$,
$S=S_{1}\times S_{2}\times\cdots\times S_{n}$ is the set of strategy
profi{}les and $f=(f_{1}(x),\cdots,f_{n}(x))$ is the payoff function.
Let $x_{\lyxmathsym{\textminus}i}$ be a strategy profi{}le of all
players except player $i$. When each player $i\in1,\cdots,n$ chooses
strategy $x_{i}$ resulting in strategy profi{}le $x=(x_{1},\cdots,x_{n})$
then player $i$ obtains payoff $f_{i}(x)$: the payoff depends on
the collective strategy of all. The collectivism of Nash equilibrium
is 

\textbullet{} \emph{A strategy profi{}le $x^{*}\in S$ is a Nash equilibrium
if no unilateral deviation in strategy by any single player is profi{}table
for him:} \begin{equation}
\forall i,x_{i}\in S_{i},\, x_{i}\neq x_{i}^{*}:f_{i}(x_{i}^{*},x_{-i}^{*})\ge f_{i}(x_{i},x_{-i}^{*}).\label{eq: Nash}\end{equation}
}, and Prisoner's Dilemma%
\footnote{\label{fn: PD}\textbf{Prisoner's Dilemma.} Two suspects $A$ and
$B$ --- each being interested only in maximizing his own advantage
without any concern for the (collective) well-being of the other ---
are arrested by the police. The police have insuffi{}cient evidence
for conviction, and, having separated the prisoners, visit each of
them to offer the same deal. If one testifi{}es for the prosecution
against the other (\textquotedblleft{}competes\textquotedblright{}
$(\downarrow)$ with the other) and the other remains silent (\textquotedblleft{}collaborates\textquotedblright{}
$(\uparrow)$ with the other), the betrayer goes free $(T)$ and the
cooperating accomplice receives the full 10-year sentence $(S)$.
If both cooperate, they are sentenced to only six months $(R)$ in
jail for a minor charge. If each competes with the other, both receives
a fi{}ve-year sentence $(P)$. Each prisoner must choose to compete
with the other or to cooperate. How should the prisoners act?

\textbullet{} The \emph{Prisoner's Dilemma} can be summarized as follows,
with $(\uparrow),(\downarrow)$ denoting {}``collaboration'', {}``competition''
of one with the other: \renewcommand{\arraystretch}{1.4}

\noindent \begin{center}
\begin{tabular}{|c||c|c|}
\hline 
$A\downarrow\setminus B\rightarrow$ & Collaborate $(\uparrow)$  & Compete $(\downarrow)$ \tabularnewline
\hline
\hline 
$(\uparrow)$ &  (6 mo ($R$), 6mo ($R$)) &  (10 ye ($S$), Free ($T$)) \tabularnewline
\hline 
$(\downarrow)$ & (Free ($T$), 10 ye ($S$))  &  (5 ye ($P$), 5 ye ($P$)) \tabularnewline
\hline
\end{tabular}
\par\end{center}

The Nash equilibrium of this game, which is not Pareto optimal $(\left\uparrow \right\uparrow )$,
is $(\left\downarrow \right\downarrow )$ of 5 years each: competition
dominates cooperation with competitors having a higher fi{}tness than
cooperators, compare Eq. \ref{eq: chanoxity} and Def. \ref{Definition. Complexity.}. 

The pay-off matrix of benefi{}ts received by the parties defi{}nes
a Prisoner\textquoteright{}s Dilemma when $T>R>P>S$. 

In the \emph{Iterated Prisoner\textquoteright{}s Dilemma}, when additionally
$2R>S+T$, the participants have to choose their mutual strategy repeatedly
with memory of their previous encounters, each getting an opportunity
to {}``punish'' the other for earlier non-collaboration. Cooperation
may then arise as an equilibrium outcome, the incentive to defect
being overcome by the threat of punishment leading to the possibility
of a cooperative outcome. As the number of iterations increase, the
Nash equilibrium tends to the Pareto optimum, the likelihood of cooperation
increases, and a collective state of competitive-collaborating homeostasy
emerges. %
} for example, work as might well be expected with reasonable justification,
as long as its canonized axioms of linearity of people with rational
preferences acting independently with full and relevent information
make sense. This framework of rationality of economic agents of individuals
or company working to maximize own profits, of the {}``invisible
hand'' transforming this profit-seeking motive to collective societal
benefaction, and of market efficiency of prices faithfully reflecting
all known information about assests \citep{Bouchaud2008}, can at
best be relevent under severely restrictive conditions: {}``the supposed
omniscience and perfect efficacy of a free market with hindsight looks
more like propaganda against communism than plausible science. In
reality, markets are not efficient, humans tend to be over-focused
in the short-term and blind in the long-term, and errors get amplified,
ultimately leading to collective irrationality, panic and crashes.
Free markets are wild markets. Surprisingly, classical economics has
no framework through which to understand 'wild' markets'' (\citet{Bouchaud2008}).
These {}``perfect world'' models are meaningful under {}``linear''
conditions only: {}``these successfully forecast a few quarters ahead
as long as things stay more or less the same, but fail in the face
of great change'' (Farmer and Foley\citep{Bouchaud2008}), {}``as
long as the influences on the economy are independent of each other,
and the past remains a reliable guide to the future. But the recent
financial collapse was a systemic meltdown, in which interwined breakdowns
$\cdots$ conspired to destabilize the system as a whole. We have
had a massive failure of the dominant economic model'' (Buchanan\citep{Bouchaud2008}). 

These authors advocate an agent-based computer-modelling of economics
({}``the meltdown has shown that regulatory policies have to cope
with far-from-equilibrium situations''), for simulating the interdependence
and interactions of autonomous individuals with a view to assessing
their effects on the system as a whole: the complex behaviour of adaptive
system emerges from interactions among the components of the system
and between the system and the environment. Individual agents are
typically characterized as boundedly rational, presumed to be acting
in what they perceive as their own interests such as economic benefit
or social status, employing heuristics or simple decision-making rules.
The computer keeps track of multiple agent interactions, monitoring
a far wider range of nonlinear intercourse than conventional equilibrium
models are capable of; {}``because the agent can learn from and respond
to emerging market behaviour, they often shift their strategies, leading
other agents to change their behaviour in turn. As a result prices
don't settle down into a stable equilibrium, as standard economic
theory predicts'' (Buchanan\citep{Bouchaud2008}). %
\begin{figure}[!tbh]
\noindent \begin{centering}
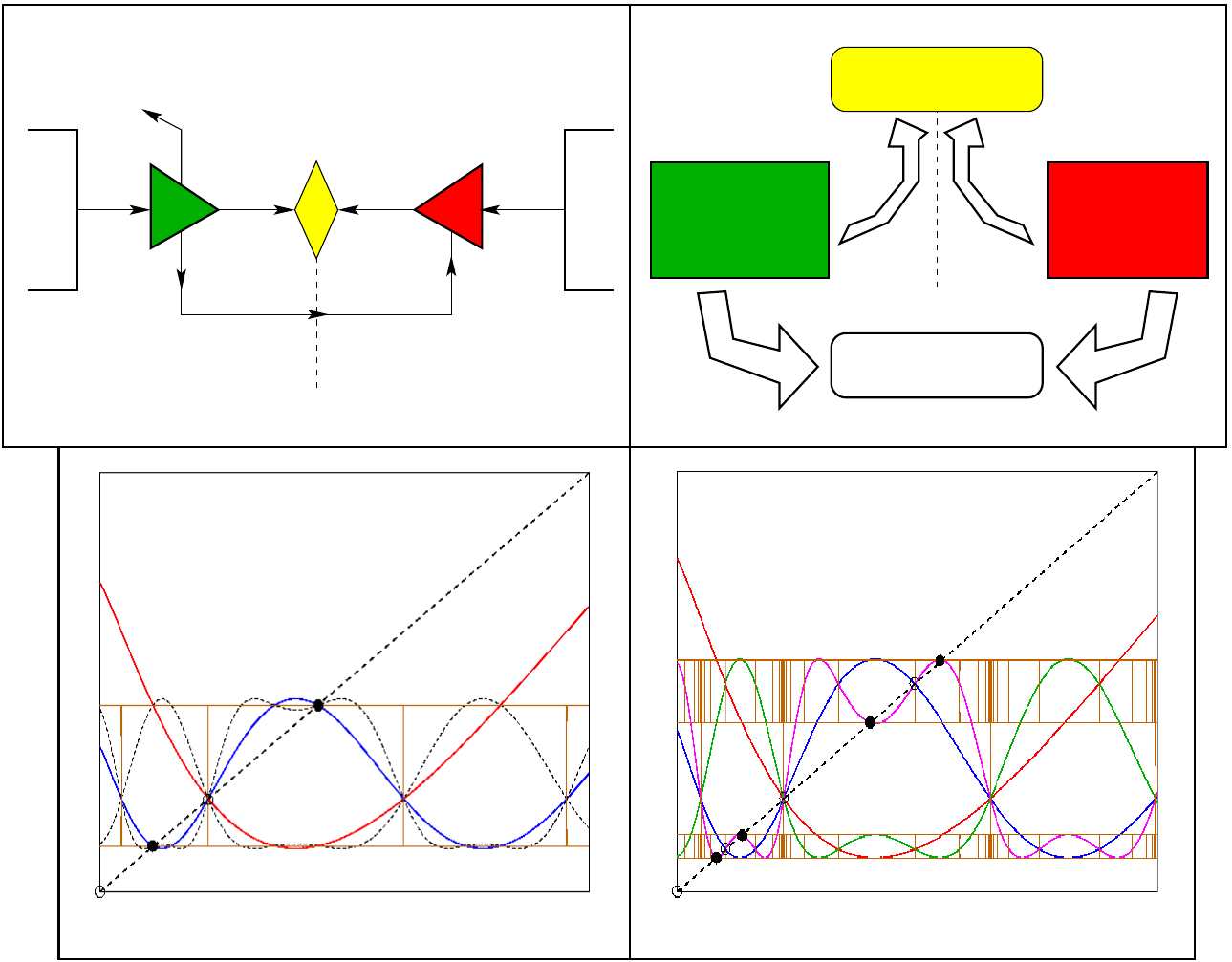
\par\end{centering}

\caption{\label{fig: economy}Economy as a complex system, $U$ is the \textquotedblleft{}universe\textquotedblright{}.
(b) of neo-classical economics is adapted from \citet{Witztum2005}.
According to this point of view, economics as the principal instrument
of collective interaction in society, is to be distinguished from
the exclusively individualistic stance of neo-classicalism. The Samuelson
tatonnement of (c) and (d), to be compared with \ref{fig: 2-4-8}(a)
and (b), show the emergence of economic complexity for nonlinear demand
and supply profi{}les $D(p)=\frac{8.0}{1.1+p}-1.75$, $S(p)=10p^{1.5}e^{-p}$
respectively with $p$ the commodity price.}

\end{figure}

This cellular automata%
\footnote{Cellular automata (CA) are simple models of spatially extended decentralized
systems comprising a number of individual component cells interacting
with each other through local communications, with the state of a
cell at any instant depending on the states of its neighbours. The
division of CA into four classes \citep{Wolfram2002} corresponding
to the attractors of dynamical systems --- Class 1: Stable Fixed Point,
Class 2: Stable Limit Cycle, Class 3: Chaotic, Class 4: Complex ---
renders them attractive tools for graphical visualization of evolution
like the emergence of altruistic or cooperative behaviour in Prisoner's
Dilemma \citep{Nowak1992} from classical Darwinian competition of
second-law dispersion. %
} generated computer-graphics evolution of the economy strongly resembles
the engine-pump realism of chanoxity as summarized in Fig. \ref{fig: economy}.
The competitive collaboration of the engine and its self-generated
pump is identified as the tension between the consumer with its dispersive
collective spending engine (collaborative {}``culture'') in conflict
with the individualistic resource generating pump (competitive {}``capital'')
in mutual feedback cycles, attaining market homeostasis not through
linear optimization and equilibrium of intersecting supply-demand
profiles, but through nonlinear feedback loops that generate entangled
holistic structures like those of Fig. \ref{fig: 2-4-8}. Supply and
demand in human society are rarely independent: aggressive advertising
for example can completely dominate the individual behaviour of these
attributes. To take this into account, the interactive feedback between
the opposites of engine consumption and pump production can be modelled
as a \emph{product} of the supply and demand factors that now, unlike
in its static manifestation of neo-classicalism, will evolve in time
to generate a condition of dynamic equilibrium, see Fig. \ref{fig: economy}
for the different evolution strategy of Samuelson tatonnement \citep{Day1998}
for \emph{nonlinear} Walrasian demand and supply profiles. 

In the \emph{linear} case, let $D(p):=1-\beta p$, $S(p):=\lambda p$,
$\beta,\lambda>0$, rescaled and normalized as $D(0)=1$, $D(1)=1-\beta$,
$S(0)=0$ for $0\le p\le1$, be mappings on the unit square. Then
supply and demand interact (mate) in the market via the shifted nonlinear
qubit \[
f_{DS}(p)=\lambda p(1-\beta p)\]
with a maximum $f_{DS}(p_{m})=\frac{\lambda}{4\beta}$ at $p_{m}=\frac{1}{2\beta}$;
note that at $\beta=1$, $f_{DS}$ reduces to the usual symmetric
form $\lambda p(1-p)$ and at $\beta=\frac{1}{2}$, $p_{m}=1$. Since
we are interested only in the range $\frac{1}{2}\le f_{DS}\le1$ for
possible complex effects, let the slopes of the two opposites be related
by $\beta=0.25\lambda$ for the expected $f_{DS}(1)=0$ at $\lambda=4$.
The market clearing condition $D(p^{*})=S(p^{*})$ at $p^{*}=\frac{1}{\beta+\lambda}=\frac{4}{5\lambda}$
\emph{apparently does not have any significance in the interactive
evolution of $p_{t+1}=f_{\textrm{{DS}}}(p_{t})$ with fixed point}
$p_{\textrm{{fp}}}=\frac{\lambda-1}{\beta\lambda}=\frac{4(\lambda-1)}{\lambda^{2}}$,
except at the uninteresting {}``solid-state'' $\lambda=1.25$ for
$p^{*}=p_{*}$ (see Fig. \ref{fig: bifur-phase}(a)). The time evolution
of the $p_{m}$-shifted, demand-supply qubit \begin{equation}
f_{DS}(p)=\lambda p(1-0.25\lambda p),\quad\beta=0.25\lambda\label{eq: DS}\end{equation}
is similar to the symmetric $\lambda p(1-p)$, except for a right-shift
of $p_{m}$ for $2\le\lambda<4$. 

The identification of demand $D$ with mandatory heat rejection $Q$
(by $E$) and of supply $S$ with heat generated $q$ (by $P$) requires
some elucidation, recall footnotes \ref{fn: capital} and \ref{fn: culture}.
While the supply correspondence $S\Leftrightarrow q(T):=\alpha(T)Q_{h}$
in this positive-negative, auto-feedback loop responsible for a competitive
market {}``capitalist'' philosophy is fairly obvious, the demand
analogy with $Q(T):=(T/T_{h})Q_{h}$ is based on the argument that
the confrontation of $Q$ and $q$, bestows on $Q$ a collective \textquotedblleft{}cultural
demand\textquotedblright{} that is met by individualistic \textquotedblleft{}supply\textquotedblright{}
of $q$ in a bidirectional loop that sustains, and is sustained by
each other, in the overall context of the whole. This collective and
collaborative consumer demand complements, preserves, and nourishes
the individualistic competitive supply $q$ that constitutes the capitalist
base of the fi{}rm%
\footnote{An example should clarify. The amount of sugar released in the blood
by a supply of the \textquotedblleft{}competitive capital\textquotedblright{}
of energy input --- if unchecked leading to heat death --- is effectively
countered in bidirectional homestatic demand of the \textquotedblleft{}collaborative
culture\textquotedblright{} of insulin --- that remaining unchecked
can only lead to a cold demise.%
}.

Putting $Q(T)=q(T)$ for equality of demand and supply in Eq. (\ref{eq: alpha}),
gives \begin{align}
T_{\pm} & =\frac{1}{2}\left(T_{h}-T_{c}\pm\sqrt{T_{h}^{2}-2T_{h}T_{c}+5T_{c}^{2}}\right)\label{eq: econ T}\\
 & =(403.21,-223.21),\nonumber \end{align}
to be compared with the holistic $T_{\pm}=(406.09,161.18)$ of Eq.
(\ref{eq: T_plus(a)}), with the limits \begin{align*}
T_{\pm} & =\begin{cases}
(T_{h},0), & T_{c}=0\\
(T_{h},-T_{h}), & T_{c}=T_{h},\end{cases}\end{align*}
that are inconsistent with the holistic condition $\iota=\alpha$:
the static equilibrium of supply and demand, as noted earlier and
in contrast with the Samuelson tatonnement of Fig. \ref{fig: economy},
is possibly only a linear manifestation of economic complex holism. 

The remarkable correspondence of this evolutionary profi{}le with
the logistic qubit interaction is far too pronounced to be dismissed
as incidental. In situations as in the Prisoner\textquoteright{}s
Dilemma for example, the agents are infact not free to take unilateral
decisions but are in entangled holistic states of competitive collaboration
with an accomplice --- the two (unfi{}lled) unstable fi{}xed points
of fi{}gure (d) --- with the four possible outcomes of footnote \ref{fn: PD}
denoted by the (fi{}lled) stable fi{}xed points, leading to the iterated
dilemma corresponding to the converged holism of (d). When the entanglement
is weak (linear) however, it is possible to consider the dilemma in
terms of the Bell states in the base $(\left|\left\uparrow \right\uparrow \right\rangle +\left|\left\downarrow \right\downarrow \right\rangle )/\sqrt{2}$
resulting in the Nash equilibrium $(\left\downarrow \right\downarrow )$.
Carrying this type of reasoning a step further, it is conceivable
that globalization has effectively transformed the world economy into
a single-celled monolith from its complex multi-cellular form, with
the inevitable consequence that it is incapable of any further self-organization
to a meaningful homeostatic form. 

What is the economic analogy to the thermodynamics of open complex
systems of Fig. \ref{fig: economy}? We suggest that economic profi{}t
\begin{equation}
\pi(Y)=R-C(Y)\label{eq: profit}\end{equation}
as the difference between total revenue $R$ and total investment
$C$, with $Y$ the output of the economy, corresponds to irreversibility
$\iota(T)$ of Eq. (\ref{eq: iota}a) that constitutes the foundation
of chanoxity. With the specifi{}c mappings \begin{align}
R & :=W_{\textrm{{rev}}}(T_{c})=\frac{T_{h}-T_{c}}{T_{h}}\, Q_{h}\nonumber \\
C(Y) & :=W(T)=\frac{T_{h}-T}{T_{h}}\, Q_{h}\label{eq: mappings}\\
Y & :=T\nonumber \end{align}
$Q_{h}$, the total infrastructural resources needed for the sustenance
of a civil society that can support the consumer-fi{}rm interaction
assumed to be suitably normalized, $\iota=\alpha$ holism requires
from Eq. (\ref{eq: T_plus(a)}) the very specifi{}c $R\lyxmathsym{\textminus}C$
relationship \sublabon{equation} \begin{equation}
C=\frac{3R-R\sqrt{5-4R}}{2(1+R)},\quad C(T)\triangleq1-\frac{T}{T_{h}},\, R(T_{c})\triangleq1-\frac{T_{c}}{T_{h}}<\frac{5}{4}\label{eq: cost}\end{equation}
completely solves the economic holistic problem determining the \textquotedblleft{}output\textquotedblright{}
$T$, the profi{}t being given by \begin{equation}
\pi=\frac{2R^{2}-R\left(1-\sqrt{5-4R}\right)}{2(1+R)}\label{eq: Profit}\end{equation}
\sublaboff{equation}Any (unutilized) profi{}t unavailable for the
benefi{}t of the system can only increase the entropy of the universe
by Eq. (\ref{eq: free-energy(a)}). Uninhibited maximization of profi{}t
therefore corresponds to the Second Law dead-state of maximum entropy
of turbulence, anarchy, and chaos %
\footnote{Adopting the point of view that the Second Law maximum entropy forward
state of dissipation, degradation, and waste comprises \textquotedblleft{}bad\textquotedblright{}
while its opposite of enforced constructivism, usefulness and order
defi{}nes \textquotedblleft{}good\textquotedblright{}, the inescapable
synthesis of our analysis is that Nature discards the high-entropy
\textquotedblleft{}bad\textquotedblright{} to make way for the low-entropy
\textquotedblleft{}good\textquotedblright{} in its dynamical quest
of life. Paradoxically either, on its own, spells \textquotedblleft{}death\textquotedblright{}
and only their judiciously engineered intermingling can support and
sustain Nature.%
}. 

In orthodox neoclassical economics, there are two main kinds of recognized
economic thinking --- micro-economics that deals with small-scale
economic activities such as that of the individual or company, and
macroeconomics which is the study of the entire economy in terms of
the total amount of goods and services produced, total income earned,
the level of employment of productive resources, and the general behavior
of prices. Mesoeconomics argues that there are important structures
which are not refl{}ected in the attributes of supply and demand curves,
nor in the large economic measures of infl{}ation, Gross Domestic
Product, the unemployment rate, and other aggregate demand and savings
measures. The argument is that the intermediate scale creates effects
which need to be described using different measurements, mathematical
formalisms and ideas. ChaNoXity represents a specifi{}c manifestation
of this philosophical platform, the correspondences of Eq. (\ref{eq: mappings})
leading to the uncomfortable yet unavoidable diagnosis that the present
social imbroglio --- triggered by arguably ideologically motivated
economic skulduggery --- arises from the predictions of these models
that \textquotedblleft{}are\textquoteright{}nt even wrong, they are
simply non-existent\textquotedblright{} \citep{Farmer2009}. However,
since $\pi=0$ corresponds to the reversible $\iota=0$ quasi-static
dead state, Nature\textquoteright{}s future can only be ensured at
the expense of its past: meaningful survival of the present depends
on a careful and intentioned balance of the forward and backward arrows
through the environmental resources of the system. 

Nature is in fact a delicately balanced nonlinear complex of \textquotedblleft{}capital\textquotedblright{}
and \textquotedblleft{}culture\textquotedblright{} representing the
arrows of individualism and collectivism respectively.

\section*{\textsf{\large The }\textsf{\textsl{\large Creative Destruction }}\textsf{\large of
Mesoeconomics}\textsf{\textsl{\large{} }}}

Economics, as the social science that examines how individuals use
limited or scarce resources in satisfying their unlimited wants, is
dominated by mainstream neoclassical economics that is plainly reductionist
in nature and fi{}ercely micro-individualistic, with societal macro-collectivism
appearing merely as an aggregation of the former, motivated by unashamed
aspiration of unlimited desires. Collectivism is mostly an assumed
axiomatic imposition on the structure, without any inquiry on whether
such predetermined equilibria do in- deed ever exist. There is also
the lurking suspicion that this version of modern economics \textquotedblleft{}is
sick. Economics has increasingly become an intellectual game played
for its own sake and not for its practical consequences for understanding
the economic world. Economists have converted the subject into a sort
of social mathematics in which analytical rigour is everything and
practical relevance is nothing\textquotedblright{} \citep{Blaug1997}
that \textquotedblleft{}bears testimony to a triumph of ideology over
science\textquotedblright{} \citep{Stiglitz2002}. The over-arching
camoufl{}age of mathematical sophistry is the \textquotedblleft{}essence
of neoclassical economics, its response to criticism, and its remarkable
capacity to turn explanatory failure into theoretical triumph\textquotedblright{}
\citep{Arnsperger2005}. \textquotedblleft{}What happened to the economics
profession?\textquotedblright{} inquires Paul Krugman \citep{Krugman2009},
\textquotedblleft{}And where does it go from here?\textquotedblright{}
He believes that \textquotedblleft{}the economics profession went
astray because economists, as a group, mistook beauty, clad in impressive-looking
mathematics, for truth (with) a vision of capitalism as a perfect
or nearly perfect system $\cdots$ in which rational individuals interact
in perfect markets.\textquotedblright{} 

That this \textquotedblleft{}massive failure of the dominant economic
model\textquotedblright{} cannot be understood, explained or remedied
in its current incarnation is the considered opinion of many, even
if this does not represent the mainstream. Kurt Dopfer et al. \citep{Dopfer2004},
following Schumpeter\textquoteright{}s Legacy \citep{Dopfer2006},
analyse evolutionary economics in a new perspec- tive of what they
categorize Micro-Meso-Macro: \textquotedblleft{}Our use of meso is
more in the ontological, and therefore analytical, sense rather than
in its classifi{}catory sense. In our view, a meso is a thing that
is made of complex other things (micro) and is an element of higher
order things (macro). Meso is not in the intermediate sense of either
classifi{}cation or analysis of disequilibrium market structures,
but rather in the specifi{}c sense of identifying and conceptualizing
the dynamical building blocks of an economic system.\textquotedblright{} 

Companies that once revolutionized and dominated new industries ---
for example, Xerox in copiers or Polaroid in instant photography ---
have had their profi{}ts fall and their dominance vanish as rivals
launched improved designs or cut manufacturing costs. Any company
that has achieved a strong position in the markets through its use
of new inventory-management, marketing, and personnel-management techniques,
can use its resulting lower prices to compete with older or smaller
companies. Just as older behemoths were eventually undone by more
innovative competitors, these trend-setters will face the same eventuality.
The seemingly once dominant leaders may well fi{}nd themselves antiquated
through a process of \textquotedblleft{}creative destruction\textquotedblright{}
\citep{Schumpeter2008} that \textquotedblleft{}incessantly revolutionizes
the economic structure, incessantly destroying the old one, incessantly
creating a new one\textquotedblright{}. This mutation from within
is indeed the guiding doctrine of complex adaptive systems that constitutes
the quintessence of our engine-pump philosophy. Thus the birth and
death of stable-unstable fi{}xed points in Figs. \ref{fig: 2-4-8}
and \ref{fig: economy} encapsulating the \textquotedblleft{}incessant
mutation from within\textquotedblright{} of creative destruction,
embraces the view of economics as the \textquotedblleft{}science that
studies the causes and consequences of the behaviour of many individuals
dealing with commodities in a macroscopic system\textquotedblright{}
\citep{Dopfer2006}. Evolutionary economics as an inquiry into the
question of how economic activities of many individuals are coordinated
and change over time, examining the dialectics of competitively collaborating
human interaction of \textquotedblleft{}creative destruction\textquotedblright{},
is a signifi{}cant step towards the understanding of \textquotedblleft{}wild
markets\textquotedblright{} and real phenomena. This of course is
conspicuously absent from neoclassical analysis that \textquotedblleft{}begins
and ends with the individual, and sadly, there is barely any role
to anything which is a refl{}ection of the collective.\textquotedblright{} 

According to the Schumpeterian vision of \textquotedblleft{}the evolutionary
response to the thermodynamic challenge is knowledge. $\cdots$ The
hallmark of knowledge is that it can generate new knowledge which
in turn generates new knowledge and so forth, self-perpetuating a
continuous path of cumulated knowledge growth\textquotedblright{}
needed to counter the inevitable entropic loss, the agent responsible
for introducing change through novelty is an \textquotedblleft{}entrepreneur\textquotedblright{}.
The entrepreneur participates in the evolution not just passively
(as his neoclassical counterpart) but more importantly in an active
fashion, initiating a positive-negative feedback that results not
just in a mere extension of a previous structure but through the novelty
of emergence and self-organization at the \textquotedblleft{}generic
level\textquotedblright{}. By engaging actively in the economic process,
\textquotedblleft{}various building blocks are added one after another
to an existing corpus, (which) also implies that the whole structure
and fundamental characteristics of that corpus changes\textquotedblright{}
\citep{Dopfer2006}, compare the stable periodic cycles of Figs. \ref{fig: 2-4-8},
\ref{fig: economy}. 

In economic theory the process that generates the new structures from
the old is called meso signifying a intermediate hybrid of micro and
macro. The Schumpeterian entrepreneur who \textquotedblleft{}carries
out an innovation (micro) that are adopted and imitated by a population
of followers (meso) thereby destroying the existing structure of the
economy (macro) leads to an elementary unit composed of on the one
hand, an idea or generic rule, and on the other many physical actualizations
of it\textquotedblright{} \citep{Dopfer2006} is a physical actualization
of the increasing multifunctional ill-posedness associated with the
time evolution of the logistic map. The bimodal nature of the elementary
unit breaks the traditional micro-macro dichotomy and by introducing
meso leads to a new framework of micro-meso-macro \citep{Dopfer2006}. 

In contrast with the implied linearity of Schumpeter\textquoteright{}s
canvas, the strong nonlinearity inherent in chanoxity \citep{Sengupta2010a}
adds a new dimension to the details of Schumpeter\textquoteright{}s
vision as analysed and critiqued in Dopfer \citep{Dopfer2006}. Extending
this analogy, it is fair to claim that chanoxity constitutes a generic
competitively-collaborating foundation for creative destruction, with
the present collapse of the economic base having been engineered through
a process of gradual decimation of the self-induced feedback mechanisms
that support the existence of open thermodynamic systems. 

Nonlinear self-organization and emergence are fascinating demonstrations
of dynamical homeostasis of opposites, apparently the source and sustenance
of Nature\textquoteright{}s diversity.

\noindent \bibliographystyle{plainnat}
\bibliography{osegu}

\end{document}

%% file: biconv.pdf_t
\begin{picture}(0,0)%
\includegraphics{biconv.pdf}%
\end{picture}%
\setlength{\unitlength}{2960sp}%
\begingroup\makeatletter\ifx\SetFigFont\undefined%
\gdef\SetFigFont#1#2#3#4#5{%
  \reset@font\fontsize{#1}{#2pt}%
  \fontfamily{#3}\fontseries{#4}\fontshape{#5}%
  \selectfont}%
\fi\endgroup%
\begin{picture}(6245,3378)(878,-3155)
\put(5801,-3086){\makebox(0,0)[b]{\smash{{\SetFigFont{9}{10.8}{\familydefault}{\mddefault}{\updefault}{\color[rgb]{0,0,0}(b) Limit: $([-1,0],0) \cup (0,[0,1]) \cup ([0,1],1)$}%
}}}}
\put(2201,-3086){\makebox(0,0)[b]{\smash{{\SetFigFont{9}{10.8}{\familydefault}{\mddefault}{\updefault}{\color[rgb]{0,0,0}(a) Limit: $([-1,0],0) \cup ((0,1],1)$}%
}}}}
\put(3601,-2786){\makebox(0,0)[b]{\smash{{\SetFigFont{9}{10.8}{\familydefault}{\mddefault}{\updefault}{\color[rgb]{0,0,0}1}%
}}}}
\put(1576,-2786){\makebox(0,0)[b]{\smash{{\SetFigFont{9}{10.8}{\familydefault}{\mddefault}{\updefault}{\color[rgb]{0,0,0}$x_{1}$}%
}}}}
\put(2101,-2786){\makebox(0,0)[b]{\smash{{\SetFigFont{9}{10.8}{\familydefault}{\mddefault}{\updefault}{\color[rgb]{0,0,0}$x_{2}$}%
}}}}
\put(2551,-2786){\makebox(0,0)[b]{\smash{{\SetFigFont{9}{10.8}{\familydefault}{\mddefault}{\updefault}{\color[rgb]{0,0,0}$x_{3}$}%
}}}}
\put(3151,-2786){\makebox(0,0)[b]{\smash{{\SetFigFont{9}{10.8}{\familydefault}{\mddefault}{\updefault}{\color[rgb]{0,0,0}$x_{4}$}%
}}}}
\put(2751,-2066){\makebox(0,0)[lb]{\smash{{\SetFigFont{8}{9.6}{\familydefault}{\mddefault}{\updefault}{\color[rgb]{0,0,0}$g$}%
}}}}
\put(2376,-736){\makebox(0,0)[rb]{\smash{{\SetFigFont{9}{10.8}{\familydefault}{\mddefault}{\updefault}{\color[rgb]{0,0,0}$V$}%
}}}}
\put(1001,-586){\makebox(0,0)[rb]{\smash{{\SetFigFont{9}{10.8}{\familydefault}{\mddefault}{\updefault}{\color[rgb]{0,0,0}$f_{x_4}$}%
}}}}
\put(1001,-1186){\makebox(0,0)[rb]{\smash{{\SetFigFont{9}{10.8}{\familydefault}{\mddefault}{\updefault}{\color[rgb]{0,0,0}$f_{x_3}$}%
}}}}
\put(1001,-1636){\makebox(0,0)[rb]{\smash{{\SetFigFont{9}{10.8}{\familydefault}{\mddefault}{\updefault}{\color[rgb]{0,0,0}$f_{x_2}$}%
}}}}
\put(1001,-2161){\makebox(0,0)[rb]{\smash{{\SetFigFont{9}{10.8}{\familydefault}{\mddefault}{\updefault}{\color[rgb]{0,0,0}$f_{x_1}$}%
}}}}
\put(1001,-136){\makebox(0,0)[rb]{\smash{{\SetFigFont{9}{10.8}{\familydefault}{\mddefault}{\updefault}{\color[rgb]{0,0,0}1}%
}}}}
\put(951,-2786){\makebox(0,0)[b]{\smash{{\SetFigFont{9}{10.8}{\familydefault}{\mddefault}{\updefault}{\color[rgb]{0,0,0}0}%
}}}}
\put(3426,-911){\makebox(0,0)[rb]{\smash{{\SetFigFont{9}{10.8}{\familydefault}{\mddefault}{\updefault}{\color[rgb]{0,0,0}$U$}%
}}}}
\put(1801,-211){\makebox(0,0)[b]{\smash{{\SetFigFont{9}{10.8}{\familydefault}{\mddefault}{\updefault}{\color[rgb]{0,0,0}$\cdots$}%
}}}}
\put(5026,-211){\makebox(0,0)[b]{\smash{{\SetFigFont{9}{10.8}{\familydefault}{\mddefault}{\updefault}{\color[rgb]{0,0,0}$\cdots$}%
}}}}
\put(4376,-136){\makebox(0,0)[rb]{\smash{{\SetFigFont{9}{10.8}{\familydefault}{\mddefault}{\updefault}{\color[rgb]{0,0,0}1}%
}}}}
\put(4676,-2786){\makebox(0,0)[b]{\smash{{\SetFigFont{9}{10.8}{\familydefault}{\mddefault}{\updefault}{\color[rgb]{0,0,0}$x_{1}$}%
}}}}
\put(6976,-2786){\makebox(0,0)[b]{\smash{{\SetFigFont{9}{10.8}{\familydefault}{\mddefault}{\updefault}{\color[rgb]{0,0,0}1}%
}}}}
\put(4951,-2786){\makebox(0,0)[b]{\smash{{\SetFigFont{9}{10.8}{\familydefault}{\mddefault}{\updefault}{\color[rgb]{0,0,0}$x_{2}$}%
}}}}
\put(4351,-2786){\makebox(0,0)[rb]{\smash{{\SetFigFont{9}{10.8}{\familydefault}{\mddefault}{\updefault}{\color[rgb]{0,0,0}0}%
}}}}
\put(5176,-2786){\makebox(0,0)[b]{\smash{{\SetFigFont{9}{10.8}{\familydefault}{\mddefault}{\updefault}{\color[rgb]{0,0,0}$x_{3}$}%
}}}}
\put(1651,-1336){\makebox(0,0)[rb]{\smash{{\SetFigFont{9}{10.8}{\familydefault}{\mddefault}{\updefault}{\color[rgb]{0,0,0}$f_2$}%
}}}}
\put(1301,-411){\makebox(0,0)[lb]{\smash{{\SetFigFont{9}{10.8}{\familydefault}{\mddefault}{\updefault}{\color[rgb]{0,0,0}$f_{10}$}%
}}}}
\put(6041,-626){\makebox(0,0)[lb]{\smash{{\SetFigFont{9}{10.8}{\familydefault}{\mddefault}{\updefault}{\color[rgb]{0,0,0}$g$}%
}}}}
\put(2336,-1286){\makebox(0,0)[rb]{\smash{{\SetFigFont{9}{10.8}{\familydefault}{\mddefault}{\updefault}{\color[rgb]{0,0,0}$f_1$}%
}}}}
\put(4651,-736){\makebox(0,0)[lb]{\smash{{\SetFigFont{9}{10.8}{\familydefault}{\mddefault}{\updefault}{\color[rgb]{0,0,0}$f_{10}$}%
}}}}
\put(5476,-2786){\makebox(0,0)[b]{\smash{{\SetFigFont{9}{10.8}{\familydefault}{\mddefault}{\updefault}{\color[rgb]{0,0,0}$x_{4}$}%
}}}}
\put(5966,-1236){\makebox(0,0)[lb]{\smash{{\SetFigFont{9}{10.8}{\familydefault}{\mddefault}{\updefault}{\color[rgb]{0,0,0}$f_1$}%
}}}}
\end{picture}%

%% file: direct-inverse.pdf_t
\begin{picture}(0,0)%
\includegraphics{direct-inverse.pdf}%
\end{picture}%
\setlength{\unitlength}{2763sp}%
\begingroup\makeatletter\ifx\SetFigFont\undefined%
\gdef\SetFigFont#1#2#3#4#5{%
  \reset@font\fontsize{#1}{#2pt}%
  \fontfamily{#3}\fontseries{#4}\fontshape{#5}%
  \selectfont}%
\fi\endgroup%
\begin{picture}(5000,2731)(-99,-1884)
\put(128,514){\makebox(0,0)[b]{\smash{{\SetFigFont{9}{10.8}{\familydefault}{\mddefault}{\updefault}{\color[rgb]{0,0,0}$X^\alpha$}%
}}}}
\put(4674,-1411){\makebox(0,0)[b]{\smash{{\SetFigFont{9}{10.8}{\familydefault}{\mddefault}{\updefault}{\color[rgb]{0,0,0}$X_\beta$}%
}}}}
\put(2492,-1811){\makebox(0,0)[lb]{\smash{{\SetFigFont{9}{10.8}{\familydefault}{\mddefault}{\updefault}{\color[rgb]{1,0,0}Direct/Inductive Limit: \underline{\textsc{Pump}}}%
}}}}
\put(128,-1436){\makebox(0,0)[b]{\smash{{\SetFigFont{9}{10.8}{\familydefault}{\mddefault}{\updefault}{\color[rgb]{0,0,0}$X^\beta$}%
}}}}
\put(1196,-1061){\makebox(0,0)[lb]{\smash{{\SetFigFont{9}{10.8}{\familydefault}{\mddefault}{\updefault}{\color[rgb]{0,0,0}$\xi^\beta$}%
}}}}
\put(3651,264){\makebox(0,0)[rb]{\smash{{\SetFigFont{9}{10.8}{\familydefault}{\mddefault}{\updefault}{\color[rgb]{0,0,0}$\zeta_\alpha$}%
}}}}
\put(3606,-1061){\makebox(0,0)[rb]{\smash{{\SetFigFont{9}{10.8}{\familydefault}{\mddefault}{\updefault}{\color[rgb]{0,0,0}$\zeta_\beta$}%
}}}}
\put(896,-411){\makebox(0,0)[rb]{\smash{{\SetFigFont{9}{10.8}{\familydefault}{\mddefault}{\updefault}{\color[rgb]{0,.69,0}$X_\leftarrow$}%
}}}}
\put(151,-486){\makebox(0,0)[b]{\smash{{\SetFigFont{9}{10.8}{\familydefault}{\mddefault}{\updefault}{\color[rgb]{0,0,0}$\pi^{\beta\alpha}$}%
}}}}
\put(4719,-361){\makebox(0,0)[b]{\smash{{\SetFigFont{9}{10.8}{\familydefault}{\mddefault}{\updefault}{\color[rgb]{0,0,0}$\eta_{\alpha\beta}$}%
}}}}
\put(1201,264){\makebox(0,0)[lb]{\smash{{\SetFigFont{9}{10.8}{\familydefault}{\mddefault}{\updefault}{\color[rgb]{0,0,0}$\xi^\alpha$}%
}}}}
\put(3946,-421){\makebox(0,0)[lb]{\smash{{\SetFigFont{9}{10.8}{\familydefault}{\mddefault}{\updefault}{\color[rgb]{1,0,0}$_{\rightarrow}X$}%
}}}}
\put(4674,564){\makebox(0,0)[b]{\smash{{\SetFigFont{9}{10.8}{\familydefault}{\mddefault}{\updefault}{\color[rgb]{0,0,0}$X_\alpha$}%
}}}}
\put(1651,-411){\makebox(0,0)[b]{\smash{{\SetFigFont{9}{10.8}{\familydefault}{\mddefault}{\updefault}{\color[rgb]{0,0,0}$h$}%
}}}}
\put(4051,-861){\makebox(0,0)[rb]{\smash{{\SetFigFont{9}{10.8}{\familydefault}{\mddefault}{\updefault}{\color[rgb]{0,0,0}$\eta_\beta$}%
}}}}
\put(4026, 39){\makebox(0,0)[rb]{\smash{{\SetFigFont{9}{10.8}{\familydefault}{\mddefault}{\updefault}{\color[rgb]{0,0,0}$\eta_\alpha$}%
}}}}
\put(826,-861){\makebox(0,0)[lb]{\smash{{\SetFigFont{9}{10.8}{\familydefault}{\mddefault}{\updefault}{\color[rgb]{0,0,0}$\pi^\beta$}%
}}}}
\put(876,-36){\makebox(0,0)[lb]{\smash{{\SetFigFont{9}{10.8}{\familydefault}{\mddefault}{\updefault}{\color[rgb]{0,0,0}$\pi^\alpha$}%
}}}}
\put(3201,-386){\makebox(0,0)[b]{\smash{{\SetFigFont{9}{10.8}{\familydefault}{\mddefault}{\updefault}{\color[rgb]{0,0,0}$g$}%
}}}}
\put(2310,-1811){\makebox(0,0)[rb]{\smash{{\SetFigFont{9}{10.8}{\familydefault}{\mddefault}{\updefault}{\color[rgb]{0,.69,0}Inverse/Projective Limit: \underline{\textsc{Engine}}}%
}}}}
\put(2401,664){\makebox(0,0)[b]{\smash{{\SetFigFont{9}{10.8}{\familydefault}{\mddefault}{\updefault}{\color[rgb]{.82,0,.82}\underline{$\alpha \preceq \beta \in \mathbb{D},$ Directed Set}}%
}}}}
\put(2476,-411){\makebox(0,0)[b]{\smash{{\SetFigFont{9}{10.8}{\familydefault}{\mddefault}{\updefault}$X_{\leftrightarrow}$}}}}
\end{picture}%

%% file: direct-inv1.pdf_t
\begin{picture}(0,0)%
\includegraphics{direct-inv1.pdf}%
\end{picture}%
\setlength{\unitlength}{2368sp}%
\begingroup\makeatletter\ifx\SetFigFont\undefined%
\gdef\SetFigFont#1#2#3#4#5{%
  \reset@font\fontsize{#1}{#2pt}%
  \fontfamily{#3}\fontseries{#4}\fontshape{#5}%
  \selectfont}%
\fi\endgroup%
\begin{picture}(7866,4414)(-782,-3218)
\put(2886,-2856){\makebox(0,0)[rb]{\smash{{\SetFigFont{8}{9.6}{\familydefault}{\mddefault}{\updefault}{\color[rgb]{0,0,0}{\it{emergence}}, $P$-concentration, disorder.}%
}}}}
\put(4205,-243){\makebox(0,0)[b]{\smash{{\SetFigFont{8}{9.6}{\familydefault}{\mddefault}{\updefault}{\color[rgb]{0,0,0}$\pi^1$}%
}}}}
\put(3726, 89){\makebox(0,0)[b]{\smash{{\SetFigFont{8}{9.6}{\familydefault}{\mddefault}{\updefault}{\color[rgb]{0,0,0}$(X^1,\mathcal{T}^1)$}%
}}}}
\put(3083,-1386){\makebox(0,0)[b]{\smash{{\SetFigFont{8}{9.6}{\familydefault}{\mddefault}{\updefault}{\color[rgb]{0,0,0}$(X,\mathcal{T})$}%
}}}}
\put(2382,289){\makebox(0,0)[b]{\smash{{\SetFigFont{8}{9.6}{\familydefault}{\mddefault}{\updefault}{\color[rgb]{0,0,0}$(X_{-1},\mathcal{T}_{-1})$}%
}}}}
\put(1428,-1961){\makebox(0,0)[b]{\smash{{\SetFigFont{8}{9.6}{\familydefault}{\mddefault}{\updefault}{\color[rgb]{0,0,0}$(X_{-n},\mathcal{T}_{-n})$}%
}}}}
\put(1922,-1536){\makebox(0,0)[b]{\smash{{\SetFigFont{7}{8.4}{\familydefault}{\mddefault}{\updefault}{\color[rgb]{0,0,0}$\eta_{-1}$}%
}}}}
\put(2886,989){\makebox(0,0)[rb]{\smash{{\SetFigFont{8}{9.6}{\familydefault}{\mddefault}{\updefault}{\color[rgb]{1,0,0}Contractive negative world $\mathfrak{W}$: \underline{\textsc{Pump}}}%
}}}}
\put(5351,-386){\makebox(0,0)[lb]{\smash{{\SetFigFont{8}{9.6}{\familydefault}{\mddefault}{\updefault}{\color[rgb]{0,0,0}$(\Phi,\mathcal{I})=$}%
}}}}
\put(4514,-1486){\makebox(0,0)[b]{\smash{{\SetFigFont{8}{9.6}{\familydefault}{\mddefault}{\updefault}{\color[rgb]{0,0,0}$(X^n,\mathcal{T}^n)$}%
}}}}
\put(4826,-1261){\makebox(0,0)[b]{\smash{{\SetFigFont{8}{9.6}{\familydefault}{\mddefault}{\updefault}{\color[rgb]{0,0,0}$\pi^n$}%
}}}}
\put(1001,-11){\makebox(0,0)[b]{\smash{{\SetFigFont{7}{8.4}{\familydefault}{\mddefault}{\updefault}{\color[rgb]{0,0,0}$\eta_{-n}$}%
}}}}
\put(5103,-761){\makebox(0,0)[b]{\smash{{\SetFigFont{8}{9.6}{\familydefault}{\mddefault}{\updefault}{\color[rgb]{0,0,0}$X_\leftarrow$}%
}}}}
\put(4138,-1011){\makebox(0,0)[b]{\smash{{\SetFigFont{8}{9.6}{\familydefault}{\mddefault}{\updefault}{\color[rgb]{0,0,0}$\pi^{kl}$}%
}}}}
\put(1931,-561){\makebox(0,0)[b]{\smash{{\SetFigFont{8}{9.6}{\familydefault}{\mddefault}{\updefault}{\color[rgb]{0,0,0}$\eta_{kl}$}%
}}}}
\put(3276,-3136){\makebox(0,0)[lb]{\smash{{\SetFigFont{8}{9.6}{\familydefault}{\mddefault}{\updefault}{\color[rgb]{0,0,0}Convergence in the real world.}%
}}}}
\put(2886,-3136){\makebox(0,0)[rb]{\smash{{\SetFigFont{8}{9.6}{\familydefault}{\mddefault}{\updefault}{\color[rgb]{0,0,0}Convergence in the dual neg-world.}%
}}}}
\put(3276,-2856){\makebox(0,0)[lb]{\smash{{\SetFigFont{8}{9.6}{\familydefault}{\mddefault}{\updefault}{\color[rgb]{0,0,0}{\it{self-organization}}, $E$-dissipation, order.}%
}}}}
\put(3276,-2261){\makebox(0,0)[lb]{\smash{{\SetFigFont{8}{9.6}{\familydefault}{\mddefault}{\updefault}{\color[rgb]{0,.69,0}\textsc{Registration/Deexcitation}, $0<t<\infty$}%
}}}}
\put(531,-661){\makebox(0,0)[rb]{\smash{{\SetFigFont{8}{9.6}{\familydefault}{\mddefault}{\updefault}{\color[rgb]{0,0,0}$\bigcup\,(X_{-k},\mathcal{T}_{-k})$}%
}}}}
\put(5351,-1211){\makebox(0,0)[lb]{\smash{{\SetFigFont{8}{9.6}{\familydefault}{\mddefault}{\updefault}{\color[rgb]{0,0,0}coarsest initial}%
}}}}
\put(531,-936){\makebox(0,0)[rb]{\smash{{\SetFigFont{8}{9.6}{\familydefault}{\mddefault}{\updefault}{\color[rgb]{0,0,0}"hot" $T_h.\; \mathcal{F}$}%
}}}}
\put(5351,-661){\makebox(0,0)[lb]{\smash{{\SetFigFont{8}{9.6}{\familydefault}{\mddefault}{\updefault}{\color[rgb]{0,0,0}$\bigcap\,(X^k,\mathcal{T}^k)$}%
}}}}
\put(5351,-936){\makebox(0,0)[lb]{\smash{{\SetFigFont{8}{9.6}{\familydefault}{\mddefault}{\updefault}{\color[rgb]{0,0,0}"cold" $T_c. \;\mathcal{I}$}%
}}}}
\put(531,-1211){\makebox(0,0)[rb]{\smash{{\SetFigFont{8}{9.6}{\familydefault}{\mddefault}{\updefault}{\color[rgb]{0,0,0}finest final}%
}}}}
\put(531,-386){\makebox(0,0)[rb]{\smash{{\SetFigFont{8}{9.6}{\familydefault}{\mddefault}{\updefault}{\color[rgb]{0,0,0}$(\Phi^{\times},\mathcal{F})=$}%
}}}}
\put(531,-1486){\makebox(0,0)[rb]{\smash{{\SetFigFont{8}{9.6}{\familydefault}{\mddefault}{\updefault}{\color[rgb]{0,0,0}topology}%
}}}}
\put(5901,-1486){\makebox(0,0)[lb]{\smash{{\SetFigFont{8}{9.6}{\familydefault}{\mddefault}{\updefault}{\color[rgb]{0,0,0}topology}%
}}}}
\put(3276,989){\makebox(0,0)[lb]{\smash{{\SetFigFont{8}{9.6}{\familydefault}{\mddefault}{\updefault}{\color[rgb]{0,.69,0}\underline{\textsc{Engine}}: Expansive real world $W$}%
}}}}
\put(2886,689){\makebox(0,0)[rb]{\smash{{\SetFigFont{8}{9.6}{\familydefault}{\mddefault}{\updefault}{\color[rgb]{0,0,0}Gravitational, \emph{exclusion}, negative feedback}%
}}}}
\put(3276,689){\makebox(0,0)[lb]{\smash{{\SetFigFont{8}{9.6}{\familydefault}{\mddefault}{\updefault}{\color[rgb]{0,0,0}Cosmological, \emph{inclusion}, positive feedback}%
}}}}
\put(2886,-2261){\makebox(0,0)[rb]{\smash{{\SetFigFont{8}{9.6}{\familydefault}{\mddefault}{\updefault}{\color[rgb]{1,0,0}\textsc{Preparation/Excitation}, $-\infty<t<0$}%
}}}}
\put(3276,-2576){\makebox(0,0)[lb]{\smash{{\SetFigFont{8}{9.6}{\familydefault}{\mddefault}{\updefault}{\color[rgb]{0,0,0}Forward-Inverse Projective System}%
}}}}
\put(2886,-2576){\makebox(0,0)[rb]{\smash{{\SetFigFont{8}{9.6}{\familydefault}{\mddefault}{\updefault}{\color[rgb]{0,0,0}Backward-Direct Inductive System}%
}}}}
\put(3111,-621){\makebox(0,0)[b]{\smash{{\SetFigFont{8}{9.6}{\familydefault}{\mddefault}{\updefault}{\color[rgb]{0,0,0}$X_\leftrightarrow$}%
}}}}
\put(3076,-896){\makebox(0,0)[b]{\smash{{\SetFigFont{8}{9.6}{\familydefault}{\mddefault}{\updefault}{\color[rgb]{0,0,0}$T$}%
}}}}
\put(526,-1711){\makebox(0,0)[b]{\smash{{\SetFigFont{8}{9.6}{\familydefault}{\mddefault}{\updefault}{\color[rgb]{0,0,0}$_{\rightarrow}X$}%
}}}}
\end{picture}%

%% file: engine-pump.pdf_t
\begin{picture}(0,0)%
\includegraphics{engine-pump.pdf}%
\end{picture}%
\setlength{\unitlength}{3355sp}%
\begingroup\makeatletter\ifx\SetFigFont\undefined%
\gdef\SetFigFont#1#2#3#4#5{%
  \reset@font\fontsize{#1}{#2pt}%
  \fontfamily{#3}\fontseries{#4}\fontshape{#5}%
  \selectfont}%
\fi\endgroup%
\begin{picture}(6522,2548)(308,-1600)
\put(5251,664){\makebox(0,0)[b]{\smash{{\SetFigFont{9}{10.8}{\familydefault}{\mddefault}{\updefault}{\color[rgb]{0,0,0}$T_h$}%
}}}}
\put(5676,-161){\makebox(0,0)[b]{\smash{{\SetFigFont{9}{10.8}{\familydefault}{\mddefault}{\updefault}{\color[rgb]{0,0,0}$Q$}%
}}}}
\put(5801,-1011){\makebox(0,0)[b]{\smash{{\SetFigFont{9}{10.8}{\familydefault}{\mddefault}{\updefault}{\color[rgb]{0,0,0}$q_c$}%
}}}}
\put(5326,-486){\makebox(0,0)[rb]{\smash{{\SetFigFont{9}{10.8}{\familydefault}{\mddefault}{\updefault}{\color[rgb]{0,0,0}$T_c$}%
}}}}
\put(5551,429){\makebox(0,0)[b]{\smash{{\SetFigFont{9}{10.8}{\familydefault}{\mddefault}{\updefault}{\color[rgb]{0,0,0}1}%
}}}}
\put(6426,-861){\makebox(0,0)[lb]{\smash{{\SetFigFont{9}{10.8}{\familydefault}{\mddefault}{\updefault}{\color[rgb]{0,0,0}$W$}%
}}}}
\put(5926,139){\makebox(0,0)[b]{\smash{{\SetFigFont{9}{10.8}{\familydefault}{\mddefault}{\updefault}{\color[rgb]{0,0,0}$Q_h$}%
}}}}
\put(6101,-786){\makebox(0,0)[b]{\smash{{\SetFigFont{9}{10.8}{\familydefault}{\mddefault}{\updefault}{\color[rgb]{0,0,0}$q$}%
}}}}
\put(5301, 64){\makebox(0,0)[rb]{\smash{{\SetFigFont{9}{10.8}{\familydefault}{\mddefault}{\updefault}{\color[rgb]{0,0,0}$T$}%
}}}}
\put(6251,-386){\makebox(0,0)[lb]{\smash{{\SetFigFont{9}{10.8}{\familydefault}{\mddefault}{\updefault}{\color[rgb]{0,0,0}$W_{\textrm{rev}}-W$}%
}}}}
\put(6801,-1311){\makebox(0,0)[b]{\smash{{\SetFigFont{9}{10.8}{\familydefault}{\mddefault}{\updefault}{\color[rgb]{0,0,0}$v$}%
}}}}
\put(5951,-1361){\makebox(0,0)[b]{\smash{{\SetFigFont{9}{10.8}{\familydefault}{\mddefault}{\updefault}{\color[rgb]{0,0,0}(b)}%
}}}}
\put(2906,-411){\makebox(0,0)[b]{\smash{{\SetFigFont{9}{10.8}{\familydefault}{\mddefault}{\updefault}{\color[rgb]{0,0,0}$q$}%
}}}}
\put(3963,-411){\makebox(0,0)[b]{\smash{{\SetFigFont{9}{10.8}{\familydefault}{\mddefault}{\updefault}{\color[rgb]{0,0,0}$q_c$}%
}}}}
\put(3412,-311){\makebox(0,0)[b]{\smash{{\SetFigFont{11}{13.2}{\familydefault}{\mddefault}{\updefault}{\color[rgb]{0,0,0}$P$}%
}}}}
\put(3412,114){\makebox(0,0)[b]{\smash{{\SetFigFont{9}{10.8}{\familydefault}{\mddefault}{\updefault}{\color[rgb]{0,0,0}demonic pump}%
}}}}
\put(2636,-1261){\makebox(0,0)[b]{\smash{{\SetFigFont{9}{10.8}{\familydefault}{\mddefault}{\updefault}{\color[rgb]{0,0,0}$=Q_h(1-T/T_h)$}%
}}}}
\put(3412,314){\makebox(0,0)[b]{\smash{{\SetFigFont{9}{10.8}{\familydefault}{\mddefault}{\updefault}{\color[rgb]{0,0,0}Maxwell}%
}}}}
\put(2551,-1536){\makebox(0,0)[lb]{\smash{{\SetFigFont{9}{10.8}{\familydefault}{\mddefault}{\updefault}{\color[rgb]{0,0,0}\textsc{Concentration}}%
}}}}
\put(1401,564){\makebox(0,0)[b]{\smash{{\SetFigFont{9}{10.8}{\familydefault}{\mddefault}{\updefault}{\color[rgb]{0,0,0}$TS=\iota W_{\textrm{rev}}=W_{\textrm{rev}}-W$}%
}}}}
\put(1937,-436){\makebox(0,0)[b]{\smash{{\SetFigFont{9}{10.8}{\familydefault}{\mddefault}{\updefault}{\color[rgb]{0,0,0}$Q$}%
}}}}
\put(2400,-336){\makebox(0,0)[b]{\smash{{\SetFigFont{11}{13.2}{\familydefault}{\mddefault}{\updefault}{\color[rgb]{0,0,0}$T$}%
}}}}
\put(969,-436){\makebox(0,0)[b]{\smash{{\SetFigFont{9}{10.8}{\familydefault}{\mddefault}{\updefault}{\color[rgb]{0,0,0}$Q_h$}%
}}}}
\put(550,-336){\makebox(0,0)[rb]{\smash{{\SetFigFont{9}{10.8}{\familydefault}{\mddefault}{\updefault}{\color[rgb]{0,0,0}$T_h$}%
}}}}
\put(1387,-311){\makebox(0,0)[b]{\smash{{\SetFigFont{11}{13.2}{\familydefault}{\mddefault}{\updefault}{\color[rgb]{0,0,0}$E$}%
}}}}
\put(701,-1261){\makebox(0,0)[b]{\smash{{\SetFigFont{9}{10.8}{\familydefault}{\mddefault}{\updefault}{\color[rgb]{0,0,0}(a)}%
}}}}
\put(2251,-1536){\makebox(0,0)[rb]{\smash{{\SetFigFont{9}{10.8}{\familydefault}{\mddefault}{\updefault}{\color[rgb]{0,0,0}\textsc{Dispersion}}%
}}}}
\put(4301,-336){\makebox(0,0)[lb]{\smash{{\SetFigFont{9}{10.8}{\familydefault}{\mddefault}{\updefault}{\color[rgb]{0,0,0}$T_c$}%
}}}}
\put(4901,789){\makebox(0,0)[rb]{\smash{{\SetFigFont{9}{10.8}{\familydefault}{\mddefault}{\updefault}{\color[rgb]{0,0,0}$p$}%
}}}}
\put(2400,-936){\makebox(0,0)[b]{\smash{{\SetFigFont{9}{10.8}{\familydefault}{\mddefault}{\updefault}{\color[rgb]{0,0,0}$W=(1-\iota)W_{\textrm{rev}}$}%
}}}}
\put(3434,-761){\makebox(0,0)[lb]{\smash{{\SetFigFont{9}{10.8}{\familydefault}{\mddefault}{\updefault}{\color[rgb]{0,0,0}$W_P$}%
}}}}
\end{picture}%

%% file: Engine-Pump.pdf_t
\begin{picture}(0,0)%
\includegraphics{Engine-Pump.pdf}%
\end{picture}%
\setlength{\unitlength}{2763sp}%
\begingroup\makeatletter\ifx\SetFigFont\undefined%
\gdef\SetFigFont#1#2#3#4#5{%
  \reset@font\fontsize{#1}{#2pt}%
  \fontfamily{#3}\fontseries{#4}\fontshape{#5}%
  \selectfont}%
\fi\endgroup%
\begin{picture}(7634,7439)(11,-6803)
\put(201,289){\makebox(0,0)[rb]{\smash{{\SetFigFont{8}{9.6}{\familydefault}{\mddefault}{\updefault}{\color[rgb]{0,0,0}$10^3$}%
}}}}
\put(201,-2861){\makebox(0,0)[rb]{\smash{{\SetFigFont{8}{9.6}{\familydefault}{\mddefault}{\updefault}{\color[rgb]{0,0,0}$-10^3$}%
}}}}
\put(2851,-1061){\makebox(0,0)[lb]{\smash{{\SetFigFont{8}{9.6}{\familydefault}{\mddefault}{\updefault}{\color[rgb]{0,0,0}$20$K}%
}}}}
\put(301,-451){\makebox(0,0)[lb]{\smash{{\SetFigFont{8}{9.6}{\familydefault}{\mddefault}{\updefault}{\color[rgb]{0,0,0}$T_h=480$}%
}}}}
\put(176,-3336){\makebox(0,0)[rb]{\smash{{\SetFigFont{8}{9.6}{\familydefault}{\mddefault}{\updefault}{\color[rgb]{0,0,0}$4$}%
}}}}
\put(3576,-6711){\makebox(0,0)[b]{\smash{{\SetFigFont{8}{9.6}{\familydefault}{\mddefault}{\updefault}{\color[rgb]{0,0,0}$4$}%
}}}}
\put(201,-4536){\makebox(0,0)[rb]{\smash{{\SetFigFont{8}{9.6}{\familydefault}{\mddefault}{\updefault}{\color[rgb]{0,0,0}$1$}%
}}}}
\put(351,-6711){\makebox(0,0)[rb]{\smash{{\SetFigFont{8}{9.6}{\familydefault}{\mddefault}{\updefault}{\color[rgb]{0,0,0}$-4$}%
}}}}
\put(2551,-411){\makebox(0,0)[lb]{\smash{{\SetFigFont{8}{9.6}{\familydefault}{\mddefault}{\updefault}{\color[rgb]{0,0,0}$T_c=300$}%
}}}}
\put(7276,-3101){\makebox(0,0)[b]{\smash{{\SetFigFont{8}{9.6}{\familydefault}{\mddefault}{\updefault}{\color[rgb]{0,0,0}$4$}%
}}}}
\put(6026,-3061){\makebox(0,0)[b]{\smash{{\SetFigFont{8}{9.6}{\familydefault}{\mddefault}{\updefault}{\color[rgb]{0,0,0}$\alpha_+$}%
}}}}
\put(4001,-3101){\makebox(0,0)[rb]{\smash{{\SetFigFont{8}{9.6}{\familydefault}{\mddefault}{\updefault}{\color[rgb]{0,0,0}$-4$}%
}}}}
\put(2301,-5446){\makebox(0,0)[rb]{\smash{{\SetFigFont{8}{9.6}{\familydefault}{\mddefault}{\updefault}{\color[rgb]{0,0,0}$20$K}%
}}}}
\put(3301,-6361){\makebox(0,0)[b]{\smash{{\SetFigFont{8}{9.6}{\familydefault}{\mddefault}{\updefault}{\color[rgb]{0,0,0}(IV)}%
}}}}
\put(3076,-4686){\makebox(0,0)[b]{\smash{{\SetFigFont{8}{9.6}{\familydefault}{\mddefault}{\updefault}{\color[rgb]{0,0,0}(I)}%
}}}}
\put(3001,-696){\makebox(0,0)[lb]{\smash{{\SetFigFont{8}{9.6}{\familydefault}{\mddefault}{\updefault}{\color[rgb]{0,0,0}$W$ (I)}%
}}}}
\put(6276, 89){\makebox(0,0)[lb]{\smash{{\SetFigFont{8}{9.6}{\familydefault}{\mddefault}{\updefault}{\color[rgb]{0,0,0}$T_c=600$}%
}}}}
\put(6276,-211){\makebox(0,0)[lb]{\smash{{\SetFigFont{8}{9.6}{\familydefault}{\mddefault}{\updefault}{\color[rgb]{0,0,0}$T_c=480$}%
}}}}
\put(5026,-1961){\makebox(0,0)[rb]{\smash{{\SetFigFont{8}{9.6}{\familydefault}{\mddefault}{\updefault}{\color[rgb]{0,0,0}$T_c=-150$}%
}}}}
\put(4301,-926){\makebox(0,0)[lb]{\smash{{\SetFigFont{8}{9.6}{\familydefault}{\mddefault}{\updefault}{\color[rgb]{0,0,0}$T_c=480$}%
}}}}
\put(351, 39){\makebox(0,0)[lb]{\smash{{\SetFigFont{8}{9.6}{\familydefault}{\mddefault}{\updefault}{\color[rgb]{0,0,0}$\mathfrak{W}$ (II)}%
}}}}
\put(5826,-711){\makebox(0,0)[lb]{\smash{{\SetFigFont{8}{9.6}{\familydefault}{\mddefault}{\updefault}{\color[rgb]{0,0,0}$150$}%
}}}}
\put(4926,-461){\makebox(0,0)[rb]{\smash{{\SetFigFont{8}{9.6}{\familydefault}{\mddefault}{\updefault}{\color[rgb]{0,0,0}$T_h=480$}%
}}}}
\put(3701,-4811){\makebox(0,0)[lb]{\smash{{\SetFigFont{8}{9.6}{\familydefault}{\mddefault}{\updefault}{\color[rgb]{0,0,0}$\alpha$}%
}}}}
\put(976,-236){\makebox(0,0)[rb]{\smash{{\SetFigFont{8}{9.6}{\familydefault}{\mddefault}{\updefault}{\color[rgb]{0,0,0}$-\alpha T_h$}%
}}}}
\put(1976,489){\makebox(0,0)[lb]{\smash{{\SetFigFont{7}{8.4}{\familydefault}{\mddefault}{\updefault}{\color[rgb]{0,0,0}$T$}%
}}}}
\put(5686,489){\makebox(0,0)[lb]{\smash{{\SetFigFont{8}{9.6}{\familydefault}{\mddefault}{\updefault}{\color[rgb]{0,0,0}$T$}%
}}}}
\put(1976,-3136){\makebox(0,0)[lb]{\smash{{\SetFigFont{8}{9.6}{\familydefault}{\mddefault}{\updefault}{\color[rgb]{0,0,0}$\iota$}%
}}}}
\put(7401,-1186){\makebox(0,0)[lb]{\smash{{\SetFigFont{8}{9.6}{\familydefault}{\mddefault}{\updefault}{\color[rgb]{0,0,0}$\alpha$}%
}}}}
\put(3701,-1186){\makebox(0,0)[lb]{\smash{{\SetFigFont{8}{9.6}{\familydefault}{\mddefault}{\updefault}{\color[rgb]{0,0,0}$\alpha$}%
}}}}
\put(2026,114){\makebox(0,0)[lb]{\smash{{\SetFigFont{8}{9.6}{\familydefault}{\mddefault}{\updefault}{\color[rgb]{0,0,0}$450$}%
}}}}
\put(2876,-1661){\makebox(0,0)[lb]{\smash{{\SetFigFont{8}{9.6}{\familydefault}{\mddefault}{\updefault}{\color[rgb]{0,0,0}$T_c=0$}%
}}}}
\put(4876,-1736){\makebox(0,0)[rb]{\smash{{\SetFigFont{8}{9.6}{\familydefault}{\mddefault}{\updefault}{\color[rgb]{0,0,0}$T_c=150$}%
}}}}
\put(2876,-1936){\makebox(0,0)[lb]{\smash{{\SetFigFont{8}{9.6}{\familydefault}{\mddefault}{\updefault}{\color[rgb]{0,0,0}$(1-\alpha)T_h$}%
}}}}
\put(2951,-2361){\makebox(0,0)[lb]{\smash{{\SetFigFont{8}{9.6}{\familydefault}{\mddefault}{\updefault}{\color[rgb]{0,0,0}$\mathfrak{W}$ (IV)}%
}}}}
\put(4551, 89){\makebox(0,0)[b]{\smash{{\SetFigFont{8}{9.6}{\familydefault}{\mddefault}{\updefault}{\color[rgb]{0,0,0}$600$}%
}}}}
\put(6751,-2061){\makebox(0,0)[b]{\smash{{\SetFigFont{8}{9.6}{\familydefault}{\mddefault}{\updefault}{\color[rgb]{0,0,0}$-150$}%
}}}}
\put(926,-2486){\makebox(0,0)[lb]{\smash{{\SetFigFont{8}{9.6}{\familydefault}{\mddefault}{\updefault}{\color[rgb]{0,0,0}$T_+$}%
}}}}
\put(201,-4936){\makebox(0,0)[rb]{\smash{{\SetFigFont{8}{9.6}{\familydefault}{\mddefault}{\updefault}{\color[rgb]{0,0,0}$0$}%
}}}}
\put(201,-1336){\makebox(0,0)[rb]{\smash{{\SetFigFont{8}{9.6}{\familydefault}{\mddefault}{\updefault}{\color[rgb]{0,0,0}$0$}%
}}}}
\put(6726,-811){\makebox(0,0)[b]{\smash{{\SetFigFont{8}{9.6}{\familydefault}{\mddefault}{\updefault}{\color[rgb]{0,0,0}$T_c=0$}%
}}}}
\put(6651,-3101){\makebox(0,0)[b]{\smash{{\SetFigFont{8}{9.6}{\familydefault}{\mddefault}{\updefault}{\color[rgb]{0,0,0}(c)}%
}}}}
\put(1101,-3861){\makebox(0,0)[rb]{\smash{{\SetFigFont{8}{9.6}{\familydefault}{\mddefault}{\updefault}{\color[rgb]{0,0,0}$T_c=300$}%
}}}}
\put(801,-6261){\makebox(0,0)[b]{\smash{{\SetFigFont{8}{9.6}{\familydefault}{\mddefault}{\updefault}{\color[rgb]{0,0,0}600}%
}}}}
\put(926,-2736){\makebox(0,0)[lb]{\smash{{\SetFigFont{8}{9.6}{\familydefault}{\mddefault}{\updefault}{\color[rgb]{0,0,0}$T_-$}%
}}}}
\put(226,-2111){\makebox(0,0)[rb]{\smash{{\SetFigFont{8}{9.6}{\familydefault}{\mddefault}{\updefault}{\color[rgb]{0,0,0}$-T_h$}%
}}}}
\put(451,-3586){\makebox(0,0)[b]{\smash{{\SetFigFont{8}{9.6}{\familydefault}{\mddefault}{\updefault}{\color[rgb]{0,0,0}(II)}%
}}}}
\put(526,-5461){\makebox(0,0)[b]{\smash{{\SetFigFont{8}{9.6}{\familydefault}{\mddefault}{\updefault}{\color[rgb]{0,0,0}(III)}%
}}}}
\put(3926,-6511){\makebox(0,0)[rb]{\smash{{\SetFigFont{8}{9.6}{\familydefault}{\mddefault}{\updefault}{\color[rgb]{0,0,0}$-2$}%
}}}}
\put(6727,-6739){\makebox(0,0)[b]{\smash{{\SetFigFont{8}{9.6}{\familydefault}{\mddefault}{\updefault}{\color[rgb]{0,0,0}$T_h \! \le \! T$}%
}}}}
\put(5626,-6739){\makebox(0,0)[b]{\smash{{\SetFigFont{8}{9.6}{\familydefault}{\mddefault}{\updefault}{\color[rgb]{0,0,0}$0 \! <T \! < \! T_h$}%
}}}}
\put(7351,-5461){\makebox(0,0)[lb]{\smash{{\SetFigFont{8}{9.6}{\familydefault}{\mddefault}{\updefault}{\color[rgb]{0,0,0}$\theta=\frac{T}{T_h}$}%
}}}}
\put(226,-5236){\makebox(0,0)[rb]{\smash{{\SetFigFont{8}{9.6}{\familydefault}{\mddefault}{\updefault}{\color[rgb]{0,0,0}$\iota_-$}%
}}}}
\put(226,-4686){\makebox(0,0)[rb]{\smash{{\SetFigFont{8}{9.6}{\familydefault}{\mddefault}{\updefault}{\color[rgb]{0,0,0}$\iota_+$}%
}}}}
\put(376,-1186){\makebox(0,0)[lb]{\smash{{\SetFigFont{8}{9.6}{\familydefault}{\mddefault}{\updefault}{\color[rgb]{0,0,0}$W$ (III)}%
}}}}
\put(226,-661){\makebox(0,0)[rb]{\smash{{\SetFigFont{8}{9.6}{\familydefault}{\mddefault}{\updefault}{\color[rgb]{0,0,0}$T_+$}%
}}}}
\put(226,-1061){\makebox(0,0)[rb]{\smash{{\SetFigFont{8}{9.6}{\familydefault}{\mddefault}{\updefault}{\color[rgb]{0,0,0}$T_-$}%
}}}}
\put(301,-786){\makebox(0,0)[lb]{\smash{{\SetFigFont{8}{9.6}{\familydefault}{\mddefault}{\updefault}{\color[rgb]{0,0,0}$T_c=300$}%
}}}}
\put(3926,-5611){\makebox(0,0)[rb]{\smash{{\SetFigFont{8}{9.6}{\familydefault}{\mddefault}{\updefault}{\color[rgb]{0,0,0}$-1$}%
}}}}
\put(3926,-3661){\makebox(0,0)[rb]{\smash{{\SetFigFont{8}{9.6}{\familydefault}{\mddefault}{\updefault}{\color[rgb]{0,0,0}$4$}%
}}}}
\put(7326,-5761){\makebox(0,0)[lb]{\smash{{\SetFigFont{8}{9.6}{\familydefault}{\mddefault}{\updefault}{\color[rgb]{0,0,0}$2$}%
}}}}
\put(4726,-3361){\makebox(0,0)[lb]{\smash{{\SetFigFont{8}{9.6}{\familydefault}{\mddefault}{\updefault}{\color[rgb]{0,0,0}$\iota\alpha=\zeta\theta(1-\theta); \zeta=(1-T_c/T_h)^{-1}$}%
}}}}
\put(4976,-5511){\makebox(0,0)[rb]{\smash{{\SetFigFont{8}{9.6}{\familydefault}{\mddefault}{\updefault}{\color[rgb]{0,0,0}$0$}%
}}}}
\put(6101,-5446){\makebox(0,0)[rb]{\smash{{\SetFigFont{8}{9.6}{\familydefault}{\mddefault}{\updefault}{\color[rgb]{0,0,0}$1$}%
}}}}
\put(6776,-4411){\makebox(0,0)[b]{\smash{{\SetFigFont{8}{9.6}{\familydefault}{\mddefault}{\updefault}{\color[rgb]{0,0,0}$T_h=480$}%
}}}}
\put(6901,-6436){\makebox(0,0)[b]{\smash{{\SetFigFont{8}{9.6}{\familydefault}{\mddefault}{\updefault}{\color[rgb]{0,0,0}$150$}%
}}}}
\put(4201,-6286){\makebox(0,0)[b]{\smash{{\SetFigFont{8}{9.6}{\familydefault}{\mddefault}{\updefault}{\color[rgb]{0,0,0}$0$}%
}}}}
\put(4276,-5836){\makebox(0,0)[b]{\smash{{\SetFigFont{8}{9.6}{\familydefault}{\mddefault}{\updefault}{\color[rgb]{0,0,0}(IV)}%
}}}}
\put(6976,-5836){\makebox(0,0)[b]{\smash{{\SetFigFont{8}{9.6}{\familydefault}{\mddefault}{\updefault}{\color[rgb]{0,0,0}(II)}%
}}}}
\put(5626,-4861){\makebox(0,0)[b]{\smash{{\SetFigFont{8}{9.6}{\familydefault}{\mddefault}{\updefault}{\color[rgb]{0,0,0}(I)+(III)}%
}}}}
\put(2341,-6686){\makebox(0,0)[b]{\smash{{\SetFigFont{8}{9.6}{\familydefault}{\mddefault}{\updefault}{\color[rgb]{0,0,0}$\alpha_+$}%
}}}}
\put(1476,-6686){\makebox(0,0)[b]{\smash{{\SetFigFont{8}{9.6}{\familydefault}{\mddefault}{\updefault}{\color[rgb]{0,0,0}$\alpha_-$}%
}}}}
\put(1876,-1461){\makebox(0,0)[rb]{\smash{{\SetFigFont{8}{9.6}{\familydefault}{\mddefault}{\updefault}{\color[rgb]{0,0,0}$0$}%
}}}}
\put(5186,-3061){\makebox(0,0)[b]{\smash{{\SetFigFont{8}{9.6}{\familydefault}{\mddefault}{\updefault}{\color[rgb]{0,0,0}$\alpha_-$}%
}}}}
\put(1436,-3061){\makebox(0,0)[rb]{\smash{{\SetFigFont{8}{9.6}{\familydefault}{\mddefault}{\updefault}{\color[rgb]{0,0,0}$-1$}%
}}}}
\put(2376,-3061){\makebox(0,0)[lb]{\smash{{\SetFigFont{8}{9.6}{\familydefault}{\mddefault}{\updefault}{\color[rgb]{0,0,0}$1$}%
}}}}
\put(4526,-6739){\makebox(0,0)[b]{\smash{{\SetFigFont{8}{9.6}{\familydefault}{\mddefault}{\updefault}{\color[rgb]{0,0,0}$T \! \le \! 0$}%
}}}}
\put(3301,-5386){\makebox(0,0)[b]{\smash{{\SetFigFont{8}{9.6}{\familydefault}{\mddefault}{\updefault}{\color[rgb]{0,0,0}$-150$}%
}}}}
\put( 26,-5736){\makebox(0,0)[lb]{\smash{{\SetFigFont{8}{9.6}{\familydefault}{\mddefault}{\updefault}{\color[rgb]{0,0,0}$\iota_c=-1.6667$}%
}}}}
\put(6751,-5461){\makebox(0,0)[b]{\smash{{\SetFigFont{8}{9.6}{\familydefault}{\mddefault}{\updefault}{\color[rgb]{0,0,0}$T_c \! = \! \mp\infty$}%
}}}}
\put(2951,-3486){\makebox(0,0)[rb]{\smash{{\SetFigFont{8}{9.6}{\familydefault}{\mddefault}{\updefault}{\color[rgb]{0,0,0}$\iota=\alpha$}%
}}}}
\put(5601,-3984){\makebox(0,0)[b]{\smash{{\SetFigFont{8}{9.6}{\familydefault}{\mddefault}{\updefault}{\color[rgb]{0,0,0}$T_c=450$}%
}}}}
\end{picture}%

%% file: 2-phase.pdf_t
\begin{picture}(0,0)%
\includegraphics{2-phase.pdf}%
\end{picture}%
\setlength{\unitlength}{2763sp}%
\begingroup\makeatletter\ifx\SetFigFont\undefined%
\gdef\SetFigFont#1#2#3#4#5{%
  \reset@font\fontsize{#1}{#2pt}%
  \fontfamily{#3}\fontseries{#4}\fontshape{#5}%
  \selectfont}%
\fi\endgroup%
\begin{picture}(7544,7544)(-1371,-10583)
\put(-1124,-3911){\makebox(0,0)[b]{\smash{{\SetFigFont{8}{9.6}{\familydefault}{\mddefault}{\updefault}{\color[rgb]{0,0,0}$P$}%
}}}}
\put(126,-6786){\makebox(0,0)[b]{\smash{{\SetFigFont{8}{9.6}{\familydefault}{\mddefault}{\updefault}{\color[rgb]{0,0,0}$v$}%
}}}}
\put(  1,-6211){\makebox(0,0)[b]{\smash{{\SetFigFont{8}{9.6}{\familydefault}{\mddefault}{\updefault}{\color[rgb]{0,0,0}S-Vap}%
}}}}
\put(151,-5456){\makebox(0,0)[b]{\smash{{\SetFigFont{8}{9.6}{\familydefault}{\mddefault}{\updefault}{\color[rgb]{0,0,0}L-Vap}%
}}}}
\put(826,-4636){\makebox(0,0)[b]{\smash{{\SetFigFont{8}{9.6}{\familydefault}{\mddefault}{\updefault}{\color[rgb]{0,0,0}Vapour}%
}}}}
\put(2876,-5986){\makebox(0,0)[rb]{\smash{{\SetFigFont{8}{9.6}{\familydefault}{\mddefault}{\updefault}{\color[rgb]{0,0,0}$T_c$}%
}}}}
\put(6051,-3461){\makebox(0,0)[rb]{\smash{{\SetFigFont{8}{9.6}{\familydefault}{\mddefault}{\updefault}{\color[rgb]{0,0,0}\underbar{$T_h=500$}}%
}}}}
\put(5776,-6736){\makebox(0,0)[b]{\smash{{\SetFigFont{8}{9.6}{\familydefault}{\mddefault}{\updefault}{\color[rgb]{0,0,0}(b)}%
}}}}
\put(3951,-6586){\makebox(0,0)[rb]{\smash{{\SetFigFont{8}{9.6}{\familydefault}{\mddefault}{\updefault}{\color[rgb]{0,0,0}$0$}%
}}}}
\put(3126,-5851){\makebox(0,0)[rb]{\smash{{\SetFigFont{8}{9.6}{\familydefault}{\mddefault}{\updefault}{\color[rgb]{0,0,0}$300$}%
}}}}
\put(-774,-5136){\makebox(0,0)[b]{\smash{{\SetFigFont{8}{9.6}{\familydefault}{\mddefault}{\updefault}{\color[rgb]{0,0,0}Solid}%
}}}}
\put(2401,-3286){\makebox(0,0)[lb]{\smash{{\SetFigFont{8}{9.6}{\familydefault}{\mddefault}{\updefault}{\color[rgb]{0,0,0}($T_+-T_-$)}%
}}}}
\put(4436,-4286){\makebox(0,0)[rb]{\smash{{\SetFigFont{8}{9.6}{\familydefault}{\mddefault}{\updefault}{\color[rgb]{0,0,0}$\alpha=\eta\zeta$}%
}}}}
\put(931,-3786){\makebox(0,0)[lb]{\smash{{\SetFigFont{8}{9.6}{\familydefault}{\mddefault}{\updefault}{\color[rgb]{1,0,0}Critical Point (CP)}%
}}}}
\put(1531,-5611){\makebox(0,0)[lb]{\smash{{\SetFigFont{8}{9.6}{\familydefault}{\mddefault}{\updefault}{\color[rgb]{0,0,0}$T=T_{cr}$}%
}}}}
\put(-464,-5611){\rotatebox{90.0}{\makebox(0,0)[b]{\smash{{\SetFigFont{7}{8.4}{\familydefault}{\mddefault}{\updefault}{\color[rgb]{0,0,0}S-L}%
}}}}}
\put(1576,-6226){\makebox(0,0)[b]{\smash{{\SetFigFont{8}{9.6}{\familydefault}{\mddefault}{\updefault}{\color[rgb]{0,0,0}$T$}%
}}}}
\put(1211,-5111){\makebox(0,0)[lb]{\smash{{\SetFigFont{8}{9.6}{\familydefault}{\mddefault}{\updefault}{\color[rgb]{0,0,0}$P=$const}%
}}}}
\put(-144,-4566){\makebox(0,0)[b]{\smash{{\SetFigFont{8}{9.6}{\familydefault}{\mddefault}{\updefault}{\color[rgb]{0,0,0}Liquid}%
}}}}
\put(2401,-6511){\makebox(0,0)[b]{\smash{{\SetFigFont{8}{9.6}{\familydefault}{\mddefault}{\updefault}{\color[rgb]{.5,.17,0}Triple Line/Point (TP)}%
}}}}
\put(5901,-9826){\makebox(0,0)[lb]{\smash{{\SetFigFont{8}{9.6}{\familydefault}{\mddefault}{\updefault}{\color[rgb]{0,0,0}$\chi$}%
}}}}
\put(5901,-10236){\makebox(0,0)[lb]{\smash{{\SetFigFont{8}{9.6}{\familydefault}{\mddefault}{\updefault}{\color[rgb]{0,0,0}$T$}%
}}}}
\put(2551,-8079){\makebox(0,0)[rb]{\smash{{\SetFigFont{8}{9.6}{\familydefault}{\mddefault}{\updefault}{\color[rgb]{0,0,0}$X$}%
}}}}
\put(2551,-7111){\makebox(0,0)[rb]{\smash{{\SetFigFont{8}{9.6}{\familydefault}{\mddefault}{\updefault}{\color[rgb]{0,0,0}1}%
}}}}
\put(5787,-9756){\makebox(0,0)[rb]{\smash{{\SetFigFont{8}{9.6}{\familydefault}{\mddefault}{\updefault}{\color[rgb]{0,0,0}4}%
}}}}
\put(3634,-9756){\makebox(0,0)[b]{\smash{{\SetFigFont{8}{9.6}{\familydefault}{\mddefault}{\updefault}{\color[rgb]{0,0,0}3}%
}}}}
\put(4266,-9192){\makebox(0,0)[b]{\smash{{\SetFigFont{8}{9.6}{\familydefault}{\mddefault}{\updefault}{\color[rgb]{0,0,0}(I)}%
}}}}
\put(3634,-10063){\makebox(0,0)[b]{\smash{{\SetFigFont{8}{9.6}{\familydefault}{\mddefault}{\updefault}{\color[rgb]{0,0,0}0}%
}}}}
\put(4933,-10039){\makebox(0,0)[b]{\smash{{\SetFigFont{8}{9.6}{\familydefault}{\mddefault}{\updefault}{\color[rgb]{0,0,0}0}%
}}}}
\put(4787,-10039){\makebox(0,0)[b]{\smash{{\SetFigFont{8}{9.6}{\familydefault}{\mddefault}{\updefault}{\color[rgb]{0,0,0}1}%
}}}}
\put(5828,-10039){\makebox(0,0)[b]{\smash{{\SetFigFont{8}{9.6}{\familydefault}{\mddefault}{\updefault}{\color[rgb]{0,0,0}1}%
}}}}
\put(3418,-9465){\makebox(0,0)[b]{\smash{{\SetFigFont{8}{9.6}{\familydefault}{\mddefault}{\updefault}{\color[rgb]{0,0,0}(L)}%
}}}}
\put(3204,-10063){\makebox(0,0)[b]{\smash{{\SetFigFont{8}{9.6}{\familydefault}{\mddefault}{\updefault}{\color[rgb]{0,0,0}0}%
}}}}
\put(5704,-10039){\makebox(0,0)[b]{\smash{{\SetFigFont{8}{9.6}{\familydefault}{\mddefault}{\updefault}{\color[rgb]{0,0,0}0}%
}}}}
\put(4851,-10401){\makebox(0,0)[b]{\smash{{\SetFigFont{8}{9.6}{\familydefault}{\mddefault}{\updefault}{\color[rgb]{0,0,0}$T_h$}%
}}}}
\put(4851,-9756){\makebox(0,0)[b]{\smash{{\SetFigFont{8}{9.6}{\familydefault}{\mddefault}{\updefault}{\color[rgb]{0,0,0}$\lambda_*$}%
}}}}
\put(3204,-9756){\makebox(0,0)[b]{\smash{{\SetFigFont{8}{9.6}{\familydefault}{\mddefault}{\updefault}{\color[rgb]{0,0,0}2}%
}}}}
\put(2904,-9465){\makebox(0,0)[b]{\smash{{\SetFigFont{8}{9.6}{\familydefault}{\mddefault}{\updefault}{\color[rgb]{0,0,0}(S)}%
}}}}
\put(4266,-9465){\makebox(0,0)[b]{\smash{{\SetFigFont{8}{9.6}{\familydefault}{\mddefault}{\updefault}{\color[rgb]{0,0,0}(L+Vap)}%
}}}}
\put(2601,-9756){\makebox(0,0)[b]{\smash{{\SetFigFont{8}{9.6}{\familydefault}{\mddefault}{\updefault}{\color[rgb]{0,0,0}0}%
}}}}
\put(2601,-10063){\makebox(0,0)[b]{\smash{{\SetFigFont{8}{9.6}{\familydefault}{\mddefault}{\updefault}{\color[rgb]{0,0,0}0}%
}}}}
\put(3651,-10401){\makebox(0,0)[b]{\smash{{\SetFigFont{8}{9.6}{\familydefault}{\mddefault}{\updefault}{\color[rgb]{0,0,0}$T_c$}%
}}}}
\put(3826,-9986){\makebox(0,0)[lb]{\smash{{\SetFigFont{8}{9.6}{\familydefault}{\mddefault}{\updefault}{\color[rgb]{0,0,0}Multi$_\parallel$}%
}}}}
\put(5326,-9465){\makebox(0,0)[b]{\smash{{\SetFigFont{8}{9.6}{\familydefault}{\mddefault}{\updefault}{\color[rgb]{0,0,0}(Vap)}%
}}}}
\put(5201,-9191){\makebox(0,0)[b]{\smash{{\SetFigFont{8}{9.6}{\familydefault}{\mddefault}{\updefault}{\color[rgb]{0,0,0}(ii)}%
}}}}
\put(3426,-9192){\makebox(0,0)[b]{\smash{{\SetFigFont{8}{9.6}{\familydefault}{\mddefault}{\updefault}{\color[rgb]{0,0,0}(III)}%
}}}}
\put(2911,-9192){\makebox(0,0)[b]{\smash{{\SetFigFont{8}{9.6}{\familydefault}{\mddefault}{\updefault}{\color[rgb]{0,0,0}(iv)}%
}}}}
\put(5426,-10436){\makebox(0,0)[b]{\smash{{\SetFigFont{8}{9.6}{\familydefault}{\mddefault}{\updefault}{\color[rgb]{0,0,0}(c)}%
}}}}
\put(3204,-10386){\makebox(0,0)[b]{\smash{{\SetFigFont{8}{9.6}{\familydefault}{\mddefault}{\updefault}{\color[rgb]{0,0,0}$0$}%
}}}}
\put(2601,-10369){\makebox(0,0)[b]{\smash{{\SetFigFont{8}{9.6}{\familydefault}{\mddefault}{\updefault}{\color[rgb]{0,0,0}$-\infty$}%
}}}}
\put(5801,-10369){\makebox(0,0)[b]{\smash{{\SetFigFont{8}{9.6}{\familydefault}{\mddefault}{\updefault}{\color[rgb]{0,0,0}$\infty$}%
}}}}
\put(5901,-9591){\makebox(0,0)[lb]{\smash{{\SetFigFont{8}{9.6}{\familydefault}{\mddefault}{\updefault}{\color[rgb]{0,0,0}$\lambda$}%
}}}}
\put(1276,-6736){\makebox(0,0)[b]{\smash{{\SetFigFont{8}{9.6}{\familydefault}{\mddefault}{\updefault}{\color[rgb]{0,0,0}(a)}%
}}}}
\put(2601,-4511){\makebox(0,0)[rb]{\smash{{\SetFigFont{8}{9.6}{\familydefault}{\mddefault}{\updefault}{\color[rgb]{0,0,0}$400$}%
}}}}
\put(2601,-4136){\makebox(0,0)[rb]{\smash{{\SetFigFont{8}{9.6}{\familydefault}{\mddefault}{\updefault}{\color[rgb]{0,0,0}$600$}%
}}}}
\put(2601,-4886){\makebox(0,0)[rb]{\smash{{\SetFigFont{8}{9.6}{\familydefault}{\mddefault}{\updefault}{\color[rgb]{0,0,0}$200$}%
}}}}
\put(3666,-6336){\makebox(0,0)[rb]{\smash{{\SetFigFont{8}{9.6}{\familydefault}{\mddefault}{\updefault}{\color[rgb]{0,0,0}$100$}%
}}}}
\put(6076,-5911){\makebox(0,0)[b]{\smash{{\SetFigFont{8}{9.6}{\familydefault}{\mddefault}{\updefault}{\color[rgb]{0,0,0}$2$}%
}}}}
\put(4651,-6461){\makebox(0,0)[b]{\smash{{\SetFigFont{8}{9.6}{\familydefault}{\mddefault}{\updefault}{\color[rgb]{0,0,0}$-1$}%
}}}}
\put(5501,-6361){\makebox(0,0)[rb]{\smash{{\SetFigFont{8}{9.6}{\familydefault}{\mddefault}{\updefault}{\color[rgb]{0,0,0}$\alpha$}%
}}}}
\put(3401,-5461){\rotatebox{19.0}{\makebox(0,0)[lb]{\smash{{\SetFigFont{8}{9.6}{\familydefault}{\mddefault}{\updefault}{\color[rgb]{0,0,0}Complex: $\alpha=\iota$}%
}}}}}
\put(2601,-5211){\makebox(0,0)[rb]{\smash{{\SetFigFont{8}{9.6}{\familydefault}{\mddefault}{\updefault}{\color[rgb]{0,0,0}$0$}%
}}}}
\put(2601,-5386){\makebox(0,0)[rb]{\smash{{\SetFigFont{8}{9.6}{\familydefault}{\mddefault}{\updefault}{\color[rgb]{0,0,0}$500$}%
}}}}
\put(5601,-6086){\makebox(0,0)[b]{\smash{{\SetFigFont{8}{9.6}{\familydefault}{\mddefault}{\updefault}{\color[rgb]{0,0,0}$1$}%
}}}}
\put(5136,-6286){\makebox(0,0)[b]{\smash{{\SetFigFont{8}{9.6}{\familydefault}{\mddefault}{\updefault}{\color[rgb]{0,0,0}$0$}%
}}}}
\put(4176,-6661){\makebox(0,0)[b]{\smash{{\SetFigFont{8}{9.6}{\familydefault}{\mddefault}{\updefault}{\color[rgb]{0,0,0}$-2$}%
}}}}
\put(1926,-4336){\makebox(0,0)[b]{\smash{{\SetFigFont{8}{9.6}{\familydefault}{\mddefault}{\updefault}{\color[rgb]{1,0,0}$\alpha_-$}%
}}}}
\put(4201,-6211){\makebox(0,0)[b]{\smash{{\SetFigFont{8}{9.6}{\familydefault}{\mddefault}{\updefault}{\color[rgb]{1,0,0}$\alpha_+$}%
}}}}
\end{picture}%

%% file: 2-4-8_cycles.pdf_t
\begin{picture}(0,0)%
\includegraphics{2-4-8_cycles.pdf}%
\end{picture}%
\setlength{\unitlength}{2960sp}%
\begingroup\makeatletter\ifx\SetFigFont\undefined%
\gdef\SetFigFont#1#2#3#4#5{%
  \reset@font\fontsize{#1}{#2pt}%
  \fontfamily{#3}\fontseries{#4}\fontshape{#5}%
  \selectfont}%
\fi\endgroup%
\begin{picture}(7553,10250)(549,-9626)
\put(7701,-2686){\makebox(0,0)[b]{\smash{{\SetFigFont{9}{10.8}{\familydefault}{\mddefault}{\updefault}{\color[rgb]{0,0,0}1}%
}}}}
\put(4501,-2686){\makebox(0,0)[b]{\smash{{\SetFigFont{9}{10.8}{\familydefault}{\mddefault}{\updefault}{\color[rgb]{0,0,0}0 }%
}}}}
\put(7701,-6146){\makebox(0,0)[b]{\smash{{\SetFigFont{9}{10.8}{\familydefault}{\mddefault}{\updefault}{\color[rgb]{0,0,0}1}%
}}}}
\put(4501,-6146){\makebox(0,0)[b]{\smash{{\SetFigFont{9}{10.8}{\familydefault}{\mddefault}{\updefault}{\color[rgb]{0,0,0}0}%
}}}}
\put(4526,-2991){\makebox(0,0)[rb]{\smash{{\SetFigFont{9}{10.8}{\familydefault}{\mddefault}{\updefault}{\color[rgb]{0,0,0}1}%
}}}}
\put(4526,474){\makebox(0,0)[rb]{\smash{{\SetFigFont{9}{10.8}{\familydefault}{\mddefault}{\updefault}{\color[rgb]{0,0,0}1}%
}}}}
\put(6776,-3791){\makebox(0,0)[rb]{\smash{{\SetFigFont{9}{10.8}{\familydefault}{\mddefault}{\updefault}{\color[rgb]{0,0,0}$x_*$}%
}}}}
\put(6751,-336){\makebox(0,0)[rb]{\smash{{\SetFigFont{9}{10.8}{\familydefault}{\mddefault}{\updefault}{\color[rgb]{0,0,0}$x_*$}%
}}}}
\put(6151,-2361){\makebox(0,0)[b]{\smash{{\SetFigFont{9}{10.8}{\familydefault}{\mddefault}{\updefault}{\color[rgb]{0,0,0}$2^2$-cycle, $N=2$}%
}}}}
\put(6151,-5811){\makebox(0,0)[b]{\smash{{\SetFigFont{9}{10.8}{\familydefault}{\mddefault}{\updefault}{\color[rgb]{0,0,0}$2^3$-cycle, $N=3$}%
}}}}
\put(6151,-5536){\makebox(0,0)[b]{\smash{{\SetFigFont{9}{10.8}{\familydefault}{\mddefault}{\updefault}{\color[rgb]{0,0,0}(c) $\lambda=3.55463$}%
}}}}
\put(6151,-6661){\makebox(0,0)[b]{\smash{{\SetFigFont{9}{10.8}{\familydefault}{\mddefault}{\updefault}{\color[rgb]{0,0,0}(f) $\lambda=3.57$, chaos, maximum}%
}}}}
\put(7501,-1786){\makebox(0,0)[b]{\smash{{\SetFigFont{11}{13.2}{\familydefault}{\mddefault}{\updefault}{\color[rgb]{1,0,0}$(\downarrow)$}%
}}}}
\put(4801,-1861){\makebox(0,0)[b]{\smash{{\SetFigFont{11}{13.2}{\familydefault}{\mddefault}{\updefault}{\color[rgb]{1,0,0}$(\uparrow)$}%
}}}}
\put(7501,-5261){\makebox(0,0)[b]{\smash{{\SetFigFont{9}{10.8}{\familydefault}{\mddefault}{\updefault}{\color[rgb]{1,0,0}$(\downarrow)$}%
}}}}
\put(4576,-9611){\makebox(0,0)[b]{\smash{{\SetFigFont{9}{10.8}{\familydefault}{\mddefault}{\updefault}{\color[rgb]{0,0,0}0.444}%
}}}}
\put(7726,-9611){\makebox(0,0)[b]{\smash{{\SetFigFont{9}{10.8}{\familydefault}{\mddefault}{\updefault}{\color[rgb]{0,0,0}0.452}%
}}}}
\put(851,-6436){\makebox(0,0)[rb]{\smash{{\SetFigFont{9}{10.8}{\familydefault}{\mddefault}{\updefault}{\color[rgb]{0,0,0}1}%
}}}}
\put(4051,-9611){\makebox(0,0)[b]{\smash{{\SetFigFont{9}{10.8}{\familydefault}{\mddefault}{\updefault}{\color[rgb]{0,0,0}1}%
}}}}
\put(3701,-1586){\makebox(0,0)[b]{\smash{{\SetFigFont{11}{13.2}{\familydefault}{\mddefault}{\updefault}{\color[rgb]{1,0,0}$(\downarrow)$}%
}}}}
\put(826,-2686){\makebox(0,0)[b]{\smash{{\SetFigFont{9}{10.8}{\familydefault}{\mddefault}{\updefault}{\color[rgb]{0,0,0}0}%
}}}}
\put(826,474){\makebox(0,0)[rb]{\smash{{\SetFigFont{9}{10.8}{\familydefault}{\mddefault}{\updefault}{\color[rgb]{0,0,0}1}%
}}}}
\put(4026,-2686){\makebox(0,0)[b]{\smash{{\SetFigFont{9}{10.8}{\familydefault}{\mddefault}{\updefault}{\color[rgb]{0,0,0}1}%
}}}}
\put(2991,-401){\makebox(0,0)[rb]{\smash{{\SetFigFont{9}{10.8}{\familydefault}{\mddefault}{\updefault}{\color[rgb]{0,0,0}$x_*$}%
}}}}
\put(2476,-2136){\makebox(0,0)[b]{\smash{{\SetFigFont{9}{10.8}{\familydefault}{\mddefault}{\updefault}{\color[rgb]{0,0,0}(a) $\lambda=3.23607$}%
}}}}
\put(2476,-2361){\makebox(0,0)[b]{\smash{{\SetFigFont{9}{10.8}{\familydefault}{\mddefault}{\updefault}{\color[rgb]{0,0,0}2-cycle, $N=1$}%
}}}}
\put(1251,-1586){\makebox(0,0)[b]{\smash{{\SetFigFont{11}{13.2}{\familydefault}{\mddefault}{\updefault}{\color[rgb]{1,0,0}$(\uparrow)$}%
}}}}
\put(3001,-3881){\makebox(0,0)[rb]{\smash{{\SetFigFont{9}{10.8}{\familydefault}{\mddefault}{\updefault}{\color[rgb]{0,0,0}$x_*$}%
}}}}
\put(2466,-2686){\makebox(0,0)[b]{\smash{{\SetFigFont{9}{10.8}{\familydefault}{\mddefault}{\updefault}{\color[rgb]{0,0,0}0.5}%
}}}}
\put(6151,-2686){\makebox(0,0)[b]{\smash{{\SetFigFont{9}{10.8}{\familydefault}{\mddefault}{\updefault}{\color[rgb]{0,0,0}0.5}%
}}}}
\put(6151,-6136){\makebox(0,0)[b]{\smash{{\SetFigFont{9}{10.8}{\familydefault}{\mddefault}{\updefault}{\color[rgb]{0,0,0}0.5}%
}}}}
\put(2001,-6136){\makebox(0,0)[b]{\smash{{\SetFigFont{9}{10.8}{\familydefault}{\mddefault}{\updefault}{\color[rgb]{0,0,0}0.5}%
}}}}
\put(2476,-5836){\makebox(0,0)[b]{\smash{{\SetFigFont{9}{10.8}{\familydefault}{\mddefault}{\updefault}{\color[rgb]{0,0,0}(d) $\lambda=3.55463$, zoomed}%
}}}}
\put(1651,-2711){\makebox(0,0)[b]{\smash{{\SetFigFont{9}{10.8}{\familydefault}{\mddefault}{\updefault}{\color[rgb]{0,0,0}L}%
}}}}
\put(3226,-2711){\makebox(0,0)[b]{\smash{{\SetFigFont{9}{10.8}{\familydefault}{\mddefault}{\updefault}{\color[rgb]{0,0,0}R}%
}}}}
\put(826,-1036){\makebox(0,0)[rb]{\smash{{\SetFigFont{9}{10.8}{\familydefault}{\mddefault}{\updefault}{\color[rgb]{0,0,0}0.5}%
}}}}
\put(3001,-6161){\makebox(0,0)[b]{\smash{{\SetFigFont{9}{10.8}{\familydefault}{\mddefault}{\updefault}{\color[rgb]{0,0,0}R}%
}}}}
\put(1526,-6161){\makebox(0,0)[b]{\smash{{\SetFigFont{9}{10.8}{\familydefault}{\mddefault}{\updefault}{\color[rgb]{0,0,0}L}%
}}}}
\put(4801,-5261){\makebox(0,0)[b]{\smash{{\SetFigFont{9}{10.8}{\familydefault}{\mddefault}{\updefault}{\color[rgb]{1,0,0}$(\uparrow)$}%
}}}}
\put(5326,-6161){\makebox(0,0)[b]{\smash{{\SetFigFont{9}{10.8}{\familydefault}{\mddefault}{\updefault}{\color[rgb]{0,0,0}L}%
}}}}
\put(6901,-6161){\makebox(0,0)[b]{\smash{{\SetFigFont{9}{10.8}{\familydefault}{\mddefault}{\updefault}{\color[rgb]{0,0,0}R}%
}}}}
\put(4501,-4476){\makebox(0,0)[rb]{\smash{{\SetFigFont{9}{10.8}{\familydefault}{\mddefault}{\updefault}{\color[rgb]{0,0,0}0.5}%
}}}}
\put(1486,-3911){\makebox(0,0)[lb]{\smash{{\SetFigFont{9}{10.8}{\familydefault}{\mddefault}{\updefault}{\color[rgb]{0,0,0}$n=3\;\,4$}%
}}}}
\put(1486,-4111){\makebox(0,0)[lb]{\smash{{\SetFigFont{9}{10.8}{\familydefault}{\mddefault}{\updefault}{\color[rgb]{0,0,0}$n=5\;\,6$}%
}}}}
\put(1486,-3711){\makebox(0,0)[lb]{\smash{{\SetFigFont{9}{10.8}{\familydefault}{\mddefault}{\updefault}{\color[rgb]{0,0,0}$n=1\;\,2$}%
}}}}
\put(1486,-4311){\makebox(0,0)[lb]{\smash{{\SetFigFont{9}{10.8}{\familydefault}{\mddefault}{\updefault}{\color[rgb]{0,0,0}$n=7\;\,8$}%
}}}}
\put(2476,-536){\makebox(0,0)[b]{\smash{{\SetFigFont{9}{10.8}{\familydefault}{\mddefault}{\updefault}{\color[rgb]{1,0,0}$f_{12}$}%
}}}}
\put(826,-5586){\makebox(0,0)[rb]{\smash{{\SetFigFont{9}{10.8}{\familydefault}{\mddefault}{\updefault}{\color[rgb]{.69,0,0}$f_{26}$}%
}}}}
\put(826,-4851){\makebox(0,0)[rb]{\smash{{\SetFigFont{9}{10.8}{\familydefault}{\mddefault}{\updefault}{\color[rgb]{.69,0,0}$f_{48}$}%
}}}}
\put(826,-3086){\makebox(0,0)[rb]{\smash{{\SetFigFont{9}{10.8}{\familydefault}{\mddefault}{\updefault}{\color[rgb]{.69,0,0}$f_{15}$}%
}}}}
\put(826,-3401){\makebox(0,0)[rb]{\smash{{\SetFigFont{9}{10.8}{\familydefault}{\mddefault}{\updefault}{\color[rgb]{.69,0,0}$f_{37}$}%
}}}}
\put(6151,-2086){\makebox(0,0)[b]{\smash{{\SetFigFont{9}{10.8}{\familydefault}{\mddefault}{\updefault}{\color[rgb]{0,0,0}(b) $\lambda=3.49856$}%
}}}}
\put(2476,-6661){\makebox(0,0)[b]{\smash{{\SetFigFont{9}{10.8}{\familydefault}{\mddefault}{\updefault}{\color[rgb]{0,0,0}(e) $\lambda=1$, minimum entropy,}%
}}}}
\put(2476,-6886){\makebox(0,0)[b]{\smash{{\SetFigFont{9}{10.8}{\familydefault}{\mddefault}{\updefault}{\color[rgb]{0,0,0}collective cold death}%
}}}}
\put(6151,-6886){\makebox(0,0)[b]{\smash{{\SetFigFont{9}{10.8}{\familydefault}{\mddefault}{\updefault}{\color[rgb]{0,0,0}entropy, individualistic heat death}%
}}}}
\put(826,-9586){\makebox(0,0)[b]{\smash{{\SetFigFont{9}{10.8}{\familydefault}{\mddefault}{\updefault}{\color[rgb]{0,0,0}0}%
}}}}
\put(4526, 89){\makebox(0,0)[rb]{\smash{{\SetFigFont{9}{10.8}{\familydefault}{\mddefault}{\updefault}{\color[rgb]{.82,0,0}$f_{13}$}%
}}}}
\put(4526,-1186){\makebox(0,0)[rb]{\smash{{\SetFigFont{9}{10.8}{\familydefault}{\mddefault}{\updefault}{\color[rgb]{.82,0,0}$f_{24}$}%
}}}}
\put(7826,-1011){\makebox(0,0)[lb]{\smash{{\SetFigFont{9}{10.8}{\familydefault}{\mddefault}{\updefault}{\color[rgb]{0,0,0}0.5}%
}}}}
\end{picture}%

%% file: S-C-T.pdf_t
\begin{picture}(0,0)%
\includegraphics{S-C-T.pdf}%
\end{picture}%
\setlength{\unitlength}{2565sp}%
\begingroup\makeatletter\ifx\SetFigFont\undefined%
\gdef\SetFigFont#1#2#3#4#5{%
  \reset@font\fontsize{#1}{#2pt}%
  \fontfamily{#3}\fontseries{#4}\fontshape{#5}%
  \selectfont}%
\fi\endgroup%
\begin{picture}(5606,8133)(971,-7853)
\put(3545,121){\makebox(0,0)[b]{\smash{{\SetFigFont{8}{9.6}{\familydefault}{\mddefault}{\updefault}{\color[rgb]{0,0,0}$5$}%
}}}}
\put(1351,-536){\makebox(0,0)[lb]{\smash{{\SetFigFont{8}{9.6}{\familydefault}{\mddefault}{\updefault}{\color[rgb]{0,0,0}$W$: (I)+(III)}%
}}}}
\put(2451,-1211){\makebox(0,0)[lb]{\smash{{\SetFigFont{8}{9.6}{\familydefault}{\mddefault}{\updefault}{\color[rgb]{0,0,0}$C_V\!\triangleq\!\displaystyle \frac{dE}{dT}\!=\!(1-E^2)\tanh^{-2}E$}%
}}}}
\put(3876,-2911){\makebox(0,0)[rb]{\smash{{\SetFigFont{8}{9.6}{\familydefault}{\mddefault}{\updefault}{\color[rgb]{0,0,0}$T$}%
}}}}
\put(5751,-2986){\makebox(0,0)[rb]{\smash{{\SetFigFont{8}{9.6}{\familydefault}{\mddefault}{\updefault}{\color[rgb]{0,0,0}$\mathfrak{W}$: (II)+(IV)}%
}}}}
\put(1026,-1761){\makebox(0,0)[rb]{\smash{{\SetFigFont{8}{9.6}{\familydefault}{\mddefault}{\updefault}{\color[rgb]{0,0,0}$-1$}%
}}}}
\put(1351, 89){\makebox(0,0)[b]{\smash{{\SetFigFont{8}{9.6}{\familydefault}{\mddefault}{\updefault}{\color[rgb]{0,0,0}$\underleftarrow{\textrm{Backward}} \, P \! :$ Gravity}%
}}}}
\put(5786,-3611){\makebox(0,0)[b]{\smash{{\SetFigFont{8}{9.6}{\familydefault}{\mddefault}{\updefault}{\color[rgb]{0,0,0}$\mathfrak{P \! : Concentration}$}%
}}}}
\put(3476,-3361){\makebox(0,0)[rb]{\smash{{\SetFigFont{8}{9.6}{\familydefault}{\mddefault}{\updefault}{\color[rgb]{0,0,0}$-5$}%
}}}}
\put(1351,-2236){\makebox(0,0)[lb]{\smash{{\SetFigFont{8}{9.6}{\familydefault}{\mddefault}{\updefault}{\color[rgb]{0,0,0}$S\!\triangleq\!\displaystyle{\!\int\!\frac{dE}{T}}\!=\!\ln\displaystyle{\frac{2}{\sqrt{1-E^2}}}-E \tanh^{-1}E$}%
}}}}
\put(5786,-2561){\makebox(0,0)[b]{\smash{{\SetFigFont{8}{9.6}{\familydefault}{\mddefault}{\updefault}{\color[rgb]{0,0,0}$\mathfrak{E \! : Dispersion}$}%
}}}}
\put(1301,-911){\makebox(0,0)[b]{\smash{{\SetFigFont{8}{9.6}{\familydefault}{\mddefault}{\updefault}{\color[rgb]{0,0,0}$\underrightarrow{\textrm{Forward}} \, E \! :$ Anti-gravity}%
}}}}
\put(6311,-5812){\makebox(0,0)[lb]{\smash{{\SetFigFont{8}{9.6}{\familydefault}{\mddefault}{\updefault}{\color[rgb]{0,0,0}$r$}%
}}}}
\put(3500,-7784){\makebox(0,0)[b]{\smash{{\SetFigFont{8}{9.6}{\familydefault}{\mddefault}{\updefault}{\color[rgb]{0,0,0}$-1$}%
}}}}
\put(3391,-7286){\makebox(0,0)[rb]{\smash{{\SetFigFont{8}{9.6}{\familydefault}{\mddefault}{\updefault}{\color[rgb]{0,0,0}$\mathfrak{W}$}%
}}}}
\put(986,-7036){\makebox(0,0)[rb]{\smash{{\SetFigFont{8}{9.6}{\familydefault}{\mddefault}{\updefault}{\color[rgb]{0,0,0}$-\ln 2$}%
}}}}
\put(986,-5827){\makebox(0,0)[rb]{\smash{{\SetFigFont{8}{9.6}{\familydefault}{\mddefault}{\updefault}{\color[rgb]{0,0,0}$-5$}%
}}}}
\put(3001,-5386){\makebox(0,0)[rb]{\smash{{\SetFigFont{8}{9.6}{\familydefault}{\mddefault}{\updefault}{\color[rgb]{0,0,0}$C_V=-r^2\textrm{sech}^2(r)$}%
}}}}
\put(3401,-3961){\makebox(0,0)[b]{\smash{{\SetFigFont{8}{9.6}{\familydefault}{\mddefault}{\updefault}{\color[rgb]{0,0,0}1}%
}}}}
\put(3541,-3761){\makebox(0,0)[b]{\smash{{\SetFigFont{8}{9.6}{\familydefault}{\mddefault}{\updefault}{\color[rgb]{0,0,0}$E$}%
}}}}
\put(6076,-1761){\makebox(0,0)[lb]{\smash{{\SetFigFont{8}{9.6}{\familydefault}{\mddefault}{\updefault}{\color[rgb]{0,0,0}1}%
}}}}
\put(6401,-1738){\makebox(0,0)[lb]{\smash{{\SetFigFont{8}{9.6}{\familydefault}{\mddefault}{\updefault}{\color[rgb]{0,0,0}$E$}%
}}}}
\put(6076,-1476){\makebox(0,0)[lb]{\smash{{\SetFigFont{8}{9.6}{\familydefault}{\mddefault}{\updefault}{\color[rgb]{0,0,0}$+ \ln 2$}%
}}}}
\put(6076,-2011){\makebox(0,0)[lb]{\smash{{\SetFigFont{8}{9.6}{\familydefault}{\mddefault}{\updefault}{\color[rgb]{0,0,0}$- \ln 2$}%
}}}}
\put(4501,-6536){\makebox(0,0)[lb]{\smash{{\SetFigFont{8}{9.6}{\familydefault}{\mddefault}{\updefault}{\color[rgb]{0,0,0}$S=\textrm{ln(cosh}(r))\!-\!r\textrm{tanh}(r)$}%
}}}}
\put(1376,-6436){\makebox(0,0)[lb]{\smash{{\SetFigFont{8}{9.6}{\familydefault}{\mddefault}{\updefault}{\color[rgb]{0,0,0}$T\propto\displaystyle{\frac{1}{r}}$}%
}}}}
\put(4301,-4986){\makebox(0,0)[lb]{\smash{{\SetFigFont{8}{9.6}{\familydefault}{\mddefault}{\updefault}{\color[rgb]{0,0,0}$E=\textrm{tanh}(r)$}%
}}}}
\put(3301,-486){\makebox(0,0)[lb]{\smash{{\SetFigFont{8}{9.6}{\familydefault}{\mddefault}{\updefault}{\color[rgb]{0,0,0}$T\!\triangleq\!\displaystyle {\frac{dE}{dS}}\!=\!-\displaystyle{\frac{1}{\tanh^{-1}E}}$}%
}}}}
\end{picture}%

%% file: bifur-phase.pdf_t
\begin{picture}(0,0)%
\includegraphics{bifur-phase.pdf}%
\end{picture}%
\setlength{\unitlength}{2644sp}%
\begingroup\makeatletter\ifx\SetFigFont\undefined%
\gdef\SetFigFont#1#2#3#4#5{%
  \reset@font\fontsize{#1}{#2pt}%
  \fontfamily{#3}\fontseries{#4}\fontshape{#5}%
  \selectfont}%
\fi\endgroup%
\begin{picture}(9324,4974)(589,-4423)
\put(5626,-1811){\makebox(0,0)[rb]{\smash{{\SetFigFont{8}{9.6}{\familydefault}{\mddefault}{\updefault}{\color[rgb]{0,0,0}$200$}%
}}}}
\put(5626,-1361){\makebox(0,0)[rb]{\smash{{\SetFigFont{8}{9.6}{\familydefault}{\mddefault}{\updefault}{\color[rgb]{0,0,0}$400$}%
}}}}
\put(8701,-136){\makebox(0,0)[lb]{\smash{{\SetFigFont{8}{9.6}{\familydefault}{\mddefault}{\updefault}{\color[rgb]{0,0,0}\underbar{$T_h=500$}}%
}}}}
\put(5626,-886){\makebox(0,0)[rb]{\smash{{\SetFigFont{8}{9.6}{\familydefault}{\mddefault}{\updefault}{\color[rgb]{0,0,0}$600$}%
}}}}
\put(4726,-2511){\makebox(0,0)[rb]{\smash{{\SetFigFont{8}{9.6}{\familydefault}{\mddefault}{\updefault}{\color[rgb]{0,0,0}4}%
}}}}
\put(2141,-2511){\makebox(0,0)[b]{\smash{{\SetFigFont{8}{9.6}{\familydefault}{\mddefault}{\updefault}{\color[rgb]{0,0,0}3}%
}}}}
\put(2141,-2836){\makebox(0,0)[b]{\smash{{\SetFigFont{8}{9.6}{\familydefault}{\mddefault}{\updefault}{\color[rgb]{0,0,0}0}%
}}}}
\put(3701,-2811){\makebox(0,0)[b]{\smash{{\SetFigFont{8}{9.6}{\familydefault}{\mddefault}{\updefault}{\color[rgb]{0,0,0}0}%
}}}}
\put(3526,-2811){\makebox(0,0)[b]{\smash{{\SetFigFont{8}{9.6}{\familydefault}{\mddefault}{\updefault}{\color[rgb]{0,0,0}1}%
}}}}
\put(4776,-2811){\makebox(0,0)[b]{\smash{{\SetFigFont{8}{9.6}{\familydefault}{\mddefault}{\updefault}{\color[rgb]{0,0,0}1}%
}}}}
\put(5101,-2336){\makebox(0,0)[lb]{\smash{{\SetFigFont{8}{9.6}{\familydefault}{\mddefault}{\updefault}{\color[rgb]{0,0,0}$\lambda$}%
}}}}
\put(1881,-2201){\makebox(0,0)[b]{\smash{{\SetFigFont{8}{9.6}{\familydefault}{\mddefault}{\updefault}{\color[rgb]{0,0,0}(L)}%
}}}}
\put(1626,-2836){\makebox(0,0)[b]{\smash{{\SetFigFont{8}{9.6}{\familydefault}{\mddefault}{\updefault}{\color[rgb]{0,0,0}0}%
}}}}
\put(4626,-2811){\makebox(0,0)[b]{\smash{{\SetFigFont{8}{9.6}{\familydefault}{\mddefault}{\updefault}{\color[rgb]{0,0,0}0}%
}}}}
\put(5101,-2976){\makebox(0,0)[lb]{\smash{{\SetFigFont{8}{9.6}{\familydefault}{\mddefault}{\updefault}{\color[rgb]{0,0,0}$\iota$}%
}}}}
\put(3601,-3186){\makebox(0,0)[b]{\smash{{\SetFigFont{8}{9.6}{\familydefault}{\mddefault}{\updefault}{\color[rgb]{0,0,0}1}%
}}}}
\put(3601,-3511){\makebox(0,0)[b]{\smash{{\SetFigFont{8}{9.6}{\familydefault}{\mddefault}{\updefault}{\color[rgb]{0,0,0}$T_h$}%
}}}}
\put(901,-3486){\makebox(0,0)[b]{\smash{{\SetFigFont{8}{9.6}{\familydefault}{\mddefault}{\updefault}{\color[rgb]{0,0,0}$-\infty$}%
}}}}
\put(2141,-3161){\makebox(0,0)[b]{\smash{{\SetFigFont{8}{9.6}{\familydefault}{\mddefault}{\updefault}{\color[rgb]{0,0,0}$0$}%
}}}}
\put(4701,-3486){\makebox(0,0)[b]{\smash{{\SetFigFont{8}{9.6}{\familydefault}{\mddefault}{\updefault}{\color[rgb]{0,0,0}$\infty$}%
}}}}
\put(5101,-2616){\makebox(0,0)[lb]{\smash{{\SetFigFont{8}{9.6}{\familydefault}{\mddefault}{\updefault}{\color[rgb]{0,0,0}$\chi$}%
}}}}
\put(4701,-3136){\makebox(0,0)[b]{\smash{{\SetFigFont{8}{9.6}{\familydefault}{\mddefault}{\updefault}{\color[rgb]{0,0,0}$\infty$}%
}}}}
\put(3601,-2511){\makebox(0,0)[b]{\smash{{\SetFigFont{8}{9.6}{\familydefault}{\mddefault}{\updefault}{\color[rgb]{0,0,0}$\lambda_*$}%
}}}}
\put(1626,-2511){\makebox(0,0)[b]{\smash{{\SetFigFont{8}{9.6}{\familydefault}{\mddefault}{\updefault}{\color[rgb]{0,0,0}2}%
}}}}
\put(1266,-2201){\makebox(0,0)[b]{\smash{{\SetFigFont{8}{9.6}{\familydefault}{\mddefault}{\updefault}{\color[rgb]{0,0,0}(S)}%
}}}}
\put(2876,-2201){\makebox(0,0)[b]{\smash{{\SetFigFont{8}{9.6}{\familydefault}{\mddefault}{\updefault}{\color[rgb]{0,0,0}(L+Vap)}%
}}}}
\put(901,-2511){\makebox(0,0)[b]{\smash{{\SetFigFont{8}{9.6}{\familydefault}{\mddefault}{\updefault}{\color[rgb]{0,0,0}0}%
}}}}
\put(901,-2836){\makebox(0,0)[b]{\smash{{\SetFigFont{8}{9.6}{\familydefault}{\mddefault}{\updefault}{\color[rgb]{0,0,0}0}%
}}}}
\put(5101,-3321){\makebox(0,0)[lb]{\smash{{\SetFigFont{8}{9.6}{\familydefault}{\mddefault}{\updefault}{\color[rgb]{0,0,0}$T$}%
}}}}
\put(2141,-3511){\makebox(0,0)[b]{\smash{{\SetFigFont{8}{9.6}{\familydefault}{\mddefault}{\updefault}{\color[rgb]{0,0,0}$T_c$}%
}}}}
\put(901,-3136){\makebox(0,0)[b]{\smash{{\SetFigFont{8}{9.6}{\familydefault}{\mddefault}{\updefault}{\color[rgb]{0,0,0}$-\infty$}%
}}}}
\put(5101,-3661){\makebox(0,0)[lb]{\smash{{\SetFigFont{8}{9.6}{\familydefault}{\mddefault}{\updefault}{\color[rgb]{0,0,0}$r$}%
}}}}
\put(4701,-3861){\makebox(0,0)[b]{\smash{{\SetFigFont{8}{9.6}{\familydefault}{\mddefault}{\updefault}{\color[rgb]{0,0,0}$0_+$}%
}}}}
\put(976,-3861){\makebox(0,0)[b]{\smash{{\SetFigFont{8}{9.6}{\familydefault}{\mddefault}{\updefault}{\color[rgb]{0,0,0}$0_-$}%
}}}}
\put(1626,-3511){\makebox(0,0)[b]{\smash{{\SetFigFont{8}{9.6}{\familydefault}{\mddefault}{\updefault}{\color[rgb]{0,0,0}$0$}%
}}}}
\put(4201,-2201){\makebox(0,0)[b]{\smash{{\SetFigFont{8}{9.6}{\familydefault}{\mddefault}{\updefault}{\color[rgb]{0,0,0}(Vap)}%
}}}}
\put(4051,-1911){\makebox(0,0)[b]{\smash{{\SetFigFont{8}{9.6}{\familydefault}{\mddefault}{\updefault}{\color[rgb]{0,0,0}(ii)}%
}}}}
\put(3626,-3861){\makebox(0,0)[b]{\smash{{\SetFigFont{8}{9.6}{\familydefault}{\mddefault}{\updefault}{\color[rgb]{0,0,0}$r_M$}%
}}}}
\put(2326,-2811){\makebox(0,0)[lb]{\smash{{\SetFigFont{8}{9.6}{\familydefault}{\mddefault}{\updefault}{\color[rgb]{0,0,0}Multi$_\parallel$}%
}}}}
\put(1876,-1911){\makebox(0,0)[b]{\smash{{\SetFigFont{8}{9.6}{\familydefault}{\mddefault}{\updefault}{\color[rgb]{0,0,0}(III)}%
}}}}
\put(2876,-1911){\makebox(0,0)[b]{\smash{{\SetFigFont{8}{9.6}{\familydefault}{\mddefault}{\updefault}{\color[rgb]{0,0,0}(I)}%
}}}}
\put(1276,-1911){\makebox(0,0)[b]{\smash{{\SetFigFont{8}{9.6}{\familydefault}{\mddefault}{\updefault}{\color[rgb]{0,0,0}(iv)}%
}}}}
\put(2151,-3861){\makebox(0,0)[b]{\smash{{\SetFigFont{8}{9.6}{\familydefault}{\mddefault}{\updefault}{\color[rgb]{0,0,0}$r_\Lambda$}%
}}}}
\put(826,-361){\makebox(0,0)[rb]{\smash{{\SetFigFont{8}{9.6}{\familydefault}{\mddefault}{\updefault}{\color[rgb]{0,0,0}$X$}%
}}}}
\put(851,314){\makebox(0,0)[rb]{\smash{{\SetFigFont{8}{9.6}{\familydefault}{\mddefault}{\updefault}{\color[rgb]{0,0,0}1}%
}}}}
\put(2876,-3836){\makebox(0,0)[b]{\smash{{\SetFigFont{8}{9.6}{\familydefault}{\mddefault}{\updefault}{\color[rgb]{0,0,0}$\rho_N$}%
}}}}
\put(1666,-3861){\makebox(0,0)[b]{\smash{{\SetFigFont{8}{9.6}{\familydefault}{\mddefault}{\updefault}{\color[rgb]{0,0,0}$\pm\infty$}%
}}}}
\put(1626,-3136){\makebox(0,0)[b]{\smash{{\SetFigFont{8}{9.6}{\familydefault}{\mddefault}{\updefault}{\color[rgb]{0,0,0}$\iota_c$}%
}}}}
\put(2101,-3736){\makebox(0,0)[rb]{\smash{{\SetFigFont{8}{9.6}{\familydefault}{\mddefault}{\updefault}{\color[rgb]{0,0,0}$\leftarrow$}%
}}}}
\put(3676,-3736){\makebox(0,0)[lb]{\smash{{\SetFigFont{8}{9.6}{\familydefault}{\mddefault}{\updefault}{\color[rgb]{0,0,0}$\rightarrow$}%
}}}}
\put(2926,-3536){\makebox(0,0)[lb]{\smash{{\SetFigFont{8}{9.6}{\familydefault}{\mddefault}{\updefault}{\color[rgb]{0,0,0}$\leftarrow \! \rho_M$}%
}}}}
\put(2351,-3536){\makebox(0,0)[lb]{\smash{{\SetFigFont{8}{9.6}{\familydefault}{\mddefault}{\updefault}{\color[rgb]{0,0,0}$\rho_{\Lambda} \! \rightarrow$}%
}}}}
\put(6851,-3461){\makebox(0,0)[rb]{\smash{{\SetFigFont{8}{9.6}{\familydefault}{\mddefault}{\updefault}{\color[rgb]{0,0,0}$100$}%
}}}}
\put(6226,-2886){\makebox(0,0)[rb]{\smash{{\SetFigFont{8}{9.6}{\familydefault}{\mddefault}{\updefault}{\color[rgb]{0,0,0}$300$}%
}}}}
\put(8999,-3461){\makebox(0,0)[rb]{\smash{{\SetFigFont{8}{9.6}{\familydefault}{\mddefault}{\updefault}{\color[rgb]{0,0,0}$\alpha$}%
}}}}
\put(5874,-2986){\makebox(0,0)[rb]{\smash{{\SetFigFont{8}{9.6}{\familydefault}{\mddefault}{\updefault}{\color[rgb]{0,0,0}$T_c$}%
}}}}
\put(5626,-2136){\makebox(0,0)[rb]{\smash{{\SetFigFont{8}{9.6}{\familydefault}{\mddefault}{\updefault}{\color[rgb]{0,0,0}$0$}%
}}}}
\put(7426,-3836){\makebox(0,0)[b]{\smash{{\SetFigFont{8}{9.6}{\familydefault}{\mddefault}{\updefault}{\color[rgb]{0,0,0}$-2$}%
}}}}
\put(5601,-2336){\makebox(0,0)[rb]{\smash{{\SetFigFont{8}{9.6}{\familydefault}{\mddefault}{\updefault}{\color[rgb]{0,0,0}$500$}%
}}}}
\put(5851,-61){\makebox(0,0)[b]{\smash{{\SetFigFont{8}{9.6}{\familydefault}{\mddefault}{\updefault}{\color[rgb]{0,0,0}$(T_+-T_-)$}%
}}}}
\put(9676,-2961){\makebox(0,0)[b]{\smash{{\SetFigFont{8}{9.6}{\familydefault}{\mddefault}{\updefault}{\color[rgb]{0,0,0}$2$}%
}}}}
\put(7076,-3736){\makebox(0,0)[b]{\smash{{\SetFigFont{8}{9.6}{\familydefault}{\mddefault}{\updefault}{\color[rgb]{0,0,0}$0$}%
}}}}
\put(7251,-3011){\rotatebox{22.5}{\makebox(0,0)[lb]{\smash{{\SetFigFont{8}{9.6}{\familydefault}{\mddefault}{\updefault}{\color[rgb]{0,0,0}Complex: $\alpha=\iota$}%
}}}}}
\put(7901,-2536){\rotatebox{22.5}{\makebox(0,0)[lb]{\smash{{\SetFigFont{8}{9.6}{\familydefault}{\mddefault}{\updefault}{\color[rgb]{0,0,0}$W$, (I)}%
}}}}}
\put(9001,-2386){\rotatebox{19.0}{\makebox(0,0)[b]{\smash{{\SetFigFont{8}{9.6}{\familydefault}{\mddefault}{\updefault}{\color[rgb]{0,0,0}(III)}%
}}}}}
\put(7726,-1811){\rotatebox{22.5}{\makebox(0,0)[rb]{\smash{{\SetFigFont{8}{9.6}{\familydefault}{\mddefault}{\updefault}{\color[rgb]{0,0,0}CP $\alpha_-, r=0:$ BH}%
}}}}}
\put(8701,-2911){\rotatebox{22.5}{\makebox(0,0)[rb]{\smash{{\SetFigFont{8}{9.6}{\familydefault}{\mddefault}{\updefault}{\color[rgb]{0,0,0}TP $\alpha_+, r=\infty:$ BB}%
}}}}}
\put(8026,-3586){\makebox(0,0)[b]{\smash{{\SetFigFont{8}{9.6}{\familydefault}{\mddefault}{\updefault}{\color[rgb]{0,0,0}$-1$}%
}}}}
\put(8591,-3386){\makebox(0,0)[b]{\smash{{\SetFigFont{8}{9.6}{\familydefault}{\mddefault}{\updefault}{\color[rgb]{0,0,0}$0$}%
}}}}
\put(9101,-3161){\makebox(0,0)[b]{\smash{{\SetFigFont{8}{9.6}{\familydefault}{\mddefault}{\updefault}{\color[rgb]{0,0,0}$1$}%
}}}}
\put(7837,-1111){\makebox(0,0)[rb]{\smash{{\SetFigFont{8}{9.6}{\familydefault}{\mddefault}{\updefault}{\color[rgb]{0,0,0}$\alpha=\eta\zeta$}%
}}}}
\put(9426,-1786){\makebox(0,0)[rb]{\smash{{\SetFigFont{8}{9.6}{\familydefault}{\mddefault}{\updefault}{\color[rgb]{0,0,0}$\mathfrak{W}$, (II) $\cup$ (IV)}%
}}}}
\end{picture}%

%% file: Penrose-Grav.pdf_t
\begin{picture}(0,0)%
\includegraphics{Penrose-Grav.pdf}%
\end{picture}%
\setlength{\unitlength}{2565sp}%
\begingroup\makeatletter\ifx\SetFigFont\undefined%
\gdef\SetFigFont#1#2#3#4#5{%
  \reset@font\fontsize{#1}{#2pt}%
  \fontfamily{#3}\fontseries{#4}\fontshape{#5}%
  \selectfont}%
\fi\endgroup%
\begin{picture}(7549,4469)(117,-3608)
\put(7651,-1486){\makebox(0,0)[lb]{\smash{{\SetFigFont{9}{10.8}{\familydefault}{\mddefault}{\updefault}{\color[rgb]{0,0,0}(b): $\mathfrak{W}$}%
}}}}
\put(7651,164){\makebox(0,0)[lb]{\smash{{\SetFigFont{9}{10.8}{\familydefault}{\mddefault}{\updefault}{\color[rgb]{0,0,0}(a): $W$}%
}}}}
\put(7651,-2986){\makebox(0,0)[lb]{\smash{{\SetFigFont{9}{10.8}{\familydefault}{\mddefault}{\updefault}{\color[rgb]{0,0,0}(c): $W$}%
}}}}
\end{picture}%

%% file: E-P-econ_1.pdf_t
\begin{picture}(0,0)%
\includegraphics{E-P-econ_1.pdf}%
\end{picture}%
\setlength{\unitlength}{2763sp}%
\begingroup\makeatletter\ifx\SetFigFont\undefined%
\gdef\SetFigFont#1#2#3#4#5{%
  \reset@font\fontsize{#1}{#2pt}%
  \fontfamily{#3}\fontseries{#4}\fontshape{#5}%
  \selectfont}%
\fi\endgroup%
\begin{picture}(8819,6894)(129,-5658)
\put(3412,-311){\makebox(0,0)[b]{\smash{{\SetFigFont{9}{10.8}{\familydefault}{\mddefault}{\updefault}{\color[rgb]{0,0,0}$P$}%
}}}}
\put(2400,-936){\makebox(0,0)[b]{\smash{{\SetFigFont{8}{9.6}{\familydefault}{\mddefault}{\updefault}{\color[rgb]{0,0,0}$W\!=\!(1-\iota)W_{\textrm{rev}}$}%
}}}}
\put(3412,114){\makebox(0,0)[b]{\smash{{\SetFigFont{8}{9.6}{\familydefault}{\mddefault}{\updefault}{\color[rgb]{0,0,0}demonic pump}%
}}}}
\put(3412,314){\makebox(0,0)[b]{\smash{{\SetFigFont{8}{9.6}{\familydefault}{\mddefault}{\updefault}{\color[rgb]{0,0,0}Maxwell}%
}}}}
\put(2551,-1536){\makebox(0,0)[lb]{\smash{{\SetFigFont{8}{9.6}{\familydefault}{\mddefault}{\updefault}{\color[rgb]{0,0,0}\textsc{\underbar{Concentration}}}%
}}}}
\put(2400,-336){\makebox(0,0)[b]{\smash{{\SetFigFont{9}{10.8}{\familydefault}{\mddefault}{\updefault}{\color[rgb]{0,0,0}$T$}%
}}}}
\put(969,-511){\makebox(0,0)[b]{\smash{{\SetFigFont{8}{9.6}{\familydefault}{\mddefault}{\updefault}{\color[rgb]{0,0,0}$Q_h$}%
}}}}
\put(576,-336){\makebox(0,0)[rb]{\smash{{\SetFigFont{8}{9.6}{\familydefault}{\mddefault}{\updefault}{\color[rgb]{0,0,0}$T_h$}%
}}}}
\put(1387,-311){\makebox(0,0)[b]{\smash{{\SetFigFont{9}{10.8}{\familydefault}{\mddefault}{\updefault}{\color[rgb]{0,0,0}$E$}%
}}}}
\put(2276,-1536){\makebox(0,0)[rb]{\smash{{\SetFigFont{8}{9.6}{\familydefault}{\mddefault}{\updefault}{\color[rgb]{0,0,0}\textsc{\underbar{Dispersion}}}%
}}}}
\put(4351,839){\makebox(0,0)[b]{\smash{{\SetFigFont{8}{9.6}{\familydefault}{\mddefault}{\updefault}{\color[rgb]{0,0,0}$U$}%
}}}}
\put(6851,-1861){\makebox(0,0)[b]{\smash{{\SetFigFont{8}{9.6}{\familydefault}{\mddefault}{\updefault}{\color[rgb]{0,0,0}(b)}%
}}}}
\put(2401,-1861){\makebox(0,0)[b]{\smash{{\SetFigFont{8}{9.6}{\familydefault}{\mddefault}{\updefault}{\color[rgb]{0,0,0}(a)}%
}}}}
\put(6751,-661){\makebox(0,0)[rb]{\smash{{\SetFigFont{8}{9.6}{\familydefault}{\mddefault}{\updefault}{\color[rgb]{0,0,0}\textsc{\underbar{Disp}}}%
}}}}
\put(6480,243){\makebox(0,0)[rb]{\smash{{\SetFigFont{7}{8.4}{\familydefault}{\mddefault}{\updefault}{\color[rgb]{0,0,0}$D$}%
}}}}
\put(6851,990){\makebox(0,0)[b]{\smash{{\SetFigFont{8}{9.6}{\familydefault}{\mddefault}{\updefault}{\color[rgb]{0,0,0}$T$}%
}}}}
\put(8221,189){\makebox(0,0)[b]{\smash{{\SetFigFont{8}{9.6}{\familydefault}{\mddefault}{\updefault}{\color[rgb]{0,0,0}$(P, T_c)$}%
}}}}
\put(8221,-427){\makebox(0,0)[b]{\smash{{\SetFigFont{8}{9.6}{\familydefault}{\mddefault}{\updefault}{\color[rgb]{0,0,0}Profit}%
}}}}
\put(8221,-607){\makebox(0,0)[b]{\smash{{\SetFigFont{8}{9.6}{\familydefault}{\mddefault}{\updefault}{\color[rgb]{0,0,0}Maximation}%
}}}}
\put(5441,-607){\makebox(0,0)[b]{\smash{{\SetFigFont{8}{9.6}{\familydefault}{\mddefault}{\updefault}{\color[rgb]{0,0,0}Maximation}%
}}}}
\put(6851,625){\makebox(0,0)[b]{\smash{{\SetFigFont{8}{9.6}{\familydefault}{\mddefault}{\updefault}{\color[rgb]{0,0,0}Goods' Markets}%
}}}}
\put(5137,831){\makebox(0,0)[b]{\smash{{\SetFigFont{8}{9.6}{\familydefault}{\mddefault}{\updefault}{\color[rgb]{0,0,0}$U$}%
}}}}
\put(5441,-427){\makebox(0,0)[b]{\smash{{\SetFigFont{8}{9.6}{\familydefault}{\mddefault}{\updefault}{\color[rgb]{0,0,0}Utility}%
}}}}
\put(6851,-1443){\makebox(0,0)[b]{\smash{{\SetFigFont{8}{9.6}{\familydefault}{\mddefault}{\updefault}{\color[rgb]{0,0,0}Factors Markets}%
}}}}
\put(7821,-1461){\makebox(0,0)[lb]{\smash{{\SetFigFont{7}{8.4}{\familydefault}{\mddefault}{\updefault}{\color[rgb]{0,0,0}$D^\prime$}%
}}}}
\put(7201,243){\makebox(0,0)[lb]{\smash{{\SetFigFont{7}{8.4}{\familydefault}{\mddefault}{\updefault}{\color[rgb]{0,0,0}$S$}%
}}}}
\put(6851,-1061){\makebox(0,0)[b]{\smash{{\SetFigFont{8}{9.6}{\familydefault}{\mddefault}{\updefault}{\color[rgb]{0,0,0}$(1-\iota)$}%
}}}}
\put(5901,-1461){\makebox(0,0)[rb]{\smash{{\SetFigFont{7}{8.4}{\familydefault}{\mddefault}{\updefault}{\color[rgb]{0,0,0}$S^\prime$}%
}}}}
\put(301,539){\makebox(0,0)[lb]{\smash{{\SetFigFont{8}{9.6}{\familydefault}{\mddefault}{\updefault}{\color[rgb]{0,0,0}$T_{\textrm{R}}S\!=\!\iota W_{\textrm{rev}}\!=\!W_{\textrm{rev}}\!-\!W$}%
}}}}
\put(4301,-336){\makebox(0,0)[lb]{\smash{{\SetFigFont{8}{9.6}{\familydefault}{\mddefault}{\updefault}{\color[rgb]{0,0,0}$T_c$}%
}}}}
\put(1926,-1261){\makebox(0,0)[lb]{\smash{{\SetFigFont{8}{9.6}{\familydefault}{\mddefault}{\updefault}{\color[rgb]{0,0,0}$=\!Q_h(1-T/T_h)$}%
}}}}
\put(2201,-486){\makebox(0,0)[rb]{\smash{{\SetFigFont{8}{9.6}{\familydefault}{\mddefault}{\updefault}{\color[rgb]{0,0,0}$Q$}%
}}}}
\put(2926,-326){\makebox(0,0)[b]{\smash{{\SetFigFont{10}{12.0}{\familydefault}{\mddefault}{\updefault}{\color[rgb]{1,0,0}$(\downarrow)$}%
}}}}
\put(1876,-326){\makebox(0,0)[b]{\smash{{\SetFigFont{10}{12.0}{\familydefault}{\mddefault}{\updefault}{\color[rgb]{1,0,0}$(\uparrow)$}%
}}}}
\put(5441,189){\makebox(0,0)[b]{\smash{{\SetFigFont{8}{9.6}{\familydefault}{\mddefault}{\updefault}{\color[rgb]{0,0,0}$(E, T_h)$}%
}}}}
\put(7151,-136){\makebox(0,0)[b]{\smash{{\SetFigFont{10}{12.0}{\familydefault}{\mddefault}{\updefault}{\color[rgb]{1,0,0}$(\downarrow)$}%
}}}}
\put(6551,-136){\makebox(0,0)[b]{\smash{{\SetFigFont{10}{12.0}{\familydefault}{\mddefault}{\updefault}{\color[rgb]{1,0,0}$(\uparrow)$}%
}}}}
\put(6301,-311){\makebox(0,0)[rb]{\smash{{\SetFigFont{8}{9.6}{\familydefault}{\mddefault}{\updefault}{\color[rgb]{0,0,0}$Q$}%
}}}}
\put(7426,-311){\makebox(0,0)[lb]{\smash{{\SetFigFont{8}{9.6}{\familydefault}{\mddefault}{\updefault}{\color[rgb]{0,0,0}$q$}%
}}}}
\put(5441,-136){\makebox(0,0)[b]{\smash{{\SetFigFont{8}{9.6}{\familydefault}{\mddefault}{\updefault}{\color[rgb]{0,0,0}\textsc{\underbar{Consumer, $C$}}}%
}}}}
\put(8221,-136){\makebox(0,0)[b]{\smash{{\SetFigFont{8}{9.6}{\familydefault}{\mddefault}{\updefault}{\color[rgb]{0,0,0}\textsc{\underbar{Firm, $F$}}}%
}}}}
\put(5251,-3436){\makebox(0,0)[b]{\smash{{\SetFigFont{10}{12.0}{\familydefault}{\mddefault}{\updefault}{\color[rgb]{1,0,0}$(\downarrow)$}%
}}}}
\put(6701,-2461){\makebox(0,0)[b]{\smash{{\SetFigFont{8}{9.6}{\familydefault}{\mddefault}{\updefault}{\color[rgb]{0,0,0}\textsc{\underbar{Economic Holism}}: $2^2$-cycle}%
}}}}
\put(3901,-3661){\makebox(0,0)[b]{\smash{{\SetFigFont{10}{12.0}{\familydefault}{\mddefault}{\updefault}{\color[rgb]{1,0,0}$(\uparrow)$}%
}}}}
\put(2551,-5486){\makebox(0,0)[b]{\smash{{\SetFigFont{8}{9.6}{\familydefault}{\mddefault}{\updefault}{\color[rgb]{0,0,0}(c)}%
}}}}
\put(2601,-2711){\makebox(0,0)[b]{\smash{{\SetFigFont{8}{9.6}{\familydefault}{\mddefault}{\updefault}{\color[rgb]{0,0,0}$p_{t+1}\!=\!p_t+0.40[D(p_t)\!-\!S(p_t)]$}%
}}}}
\put(781,-5411){\makebox(0,0)[b]{\smash{{\SetFigFont{8}{9.6}{\familydefault}{\mddefault}{\updefault}{\color[rgb]{0,0,0}0}%
}}}}
\put(4388,-5411){\makebox(0,0)[b]{\smash{{\SetFigFont{8}{9.6}{\familydefault}{\mddefault}{\updefault}{\color[rgb]{0,0,0}3}%
}}}}
\put(714,-2236){\makebox(0,0)[b]{\smash{{\SetFigFont{8}{9.6}{\familydefault}{\mddefault}{\updefault}{\color[rgb]{0,0,0}3}%
}}}}
\put(6702,-5486){\makebox(0,0)[b]{\smash{{\SetFigFont{8}{9.6}{\familydefault}{\mddefault}{\updefault}{\color[rgb]{0,0,0}(d)}%
}}}}
\put(7499,-4411){\makebox(0,0)[b]{\smash{{\SetFigFont{9}{10.8}{\familydefault}{\mddefault}{\updefault}{\color[rgb]{1,0,0}$(\uparrow)$}%
}}}}
\put(6701,-2711){\makebox(0,0)[b]{\smash{{\SetFigFont{8}{9.6}{\familydefault}{\mddefault}{\updefault}{\color[rgb]{0,0,0}$p_{t+1}\!=\!p_t+0.43[D(p_t)-S(p_t)]$}%
}}}}
\put(8438,-5411){\makebox(0,0)[b]{\smash{{\SetFigFont{7}{8.4}{\familydefault}{\mddefault}{\updefault}{\color[rgb]{0,0,0}3}%
}}}}
\put(4860,-2236){\makebox(0,0)[b]{\smash{{\SetFigFont{7}{8.4}{\familydefault}{\mddefault}{\updefault}{\color[rgb]{0,0,0}3}%
}}}}
\put(4902,-5411){\makebox(0,0)[b]{\smash{{\SetFigFont{7}{8.4}{\familydefault}{\mddefault}{\updefault}{\color[rgb]{0,0,0}0}%
}}}}
\put(6961,-661){\makebox(0,0)[lb]{\smash{{\SetFigFont{8}{9.6}{\familydefault}{\mddefault}{\updefault}{\color[rgb]{0,0,0}\textsc{\underbar{Conc}}}%
}}}}
\put(2621,-461){\makebox(0,0)[lb]{\smash{{\SetFigFont{8}{9.6}{\familydefault}{\mddefault}{\updefault}{\color[rgb]{0,0,0}$q$}%
}}}}
\put(3926,-461){\makebox(0,0)[b]{\smash{{\SetFigFont{8}{9.6}{\familydefault}{\mddefault}{\updefault}{\color[rgb]{0,0,0}$q_c$}%
}}}}
\put(1101,-3511){\makebox(0,0)[b]{\smash{{\SetFigFont{10}{12.0}{\familydefault}{\mddefault}{\updefault}{\color[rgb]{1,0,0}$(\downarrow)$}%
}}}}
\put(2601,-2461){\makebox(0,0)[b]{\smash{{\SetFigFont{8}{9.6}{\familydefault}{\mddefault}{\updefault}{\color[rgb]{0,0,0}\textsc{\underbar{Economic Holism}}: $2$-cycle}%
}}}}
\put(3434,-761){\makebox(0,0)[lb]{\smash{{\SetFigFont{8}{9.6}{\familydefault}{\mddefault}{\updefault}{\color[rgb]{0,0,0}$W_P$}%
}}}}
\end{picture}%